\documentclass[traditabstract]{aa}  
\usepackage[colorlinks=true,allcolors=blue]{hyperref}
\usepackage{graphicx}
\usepackage{txfonts}
\usepackage{color}
\usepackage{multirow}
\usepackage{xcolor}
\usepackage[normalem]{ulem}

\begin{document} 

\title{Bound or blown: the fate of hot gas in galaxy groups
}
\author{
R. Seppi\inst{\ref{unige}}\thanks{E-mail: riccardo.seppi@unige.ch} \and
D. Eckert\inst{\ref{unige}} \and
J. Schaye\inst{\ref{leidenobs}} \and
J. Braspenning\inst{\ref{MPH}} \and
M. Schaller\inst{\ref{leidenobs}, \ref{Lorentz}} \and
B. D. Oppenheimer\inst{\ref{Colorado}} \and
E. O'Sullivan\inst{\ref{cfa}} \and
F. Gastaldello\inst{\ref{inafMI}} \and
L. Lovisari\inst{\ref{inafMI}. \ref{cfa}} \and
M. A. Bourne\inst{\ref{cfar}, \ref{kavli}} \and
M. Sun\inst{\ref{alabama}} \and
A. Finoguenov\inst{\ref{helsinki}} \and
H. Khalil\inst{\ref{helsinki}} \and 
G. Gozaliasl\inst{\ref{helsinki},\ref{aalto}} \and
K. Kolokythas\inst{\ref{rhodes},\ref{SARAO}, \ref{inafbo_radio}} \and
Y. E. Bahar\inst{\ref{inafBO}} \and
R. Santra\inst{\ref{tata}}
}

\institute{
Department of Astronomy, University of Geneva, Ch. d’Ecogia 16, CH-1290 Versoix, Switzerland \label{unige} \and
Leiden Observatory, Leiden University, PO Box 9513, 2300 RA Leiden, the Netherlands \label{leidenobs} \and
Max-Planck-Institut f\"{u}r Astronomie, K\"{o}nigstuhl 17, D-69117 Heidelberg, Germany \label{MPH} \and
Lorentz Institute for Theoretical Physics, Leiden University, PO box 9506, 2300 RA Leiden, the Netherlands \label{Lorentz} \and
University of Colorado, Center for Astrophysics and Space Astronomy, 389 UCB, Boulder, CO 80309, USA \label{Colorado} \and
Center for Astrophysics | Harvard \& Smithsonian, 60 Garden Street, Cambridge, MA 02138, USA \label{cfa} \and
INAF, Istituto di Astrofisica Spaziale e Fisica Cosmica di Milano, via A. Corti 12, 20133 Milano, Italy \label{inafMI} \and 
Centre for Astrophysics Research, Department of Physics, Astronomy and Mathematics, University of Hertfordshire, College Lane, Hatfield AL10 9AB, UK \label{cfar} \and 
Kavli Institute for Cosmology, University of Cambridge, Madingley Road, Cambridge, CB3 0HA, UK \label{kavli} \and
Department of Physics and Astronomy, University of Alabama in
Huntsville, Huntsville, AL35899, USA \label{alabama} \and
Department of Physics,  University of Helsinki, Gustaf H\"{a}llstr\"{o}min katu 2A, Helsinki, FI-00014, Finland \label{helsinki} \and
Department of Computer Science, Aalto University, PO Box 15400, Espoo, FI-00076, Finland \label{aalto} \and
Centre for Radio Astronomy Techniques and Technologies, Department of Physics and Electronics, Rhodes University,
PO Box 94, Makhanda 6140, South Africa \label{rhodes} \and
South African Radio Astronomy Observatory, Black River Park North, 2 Fir St, Cape Town 7925, South Africa \label{SARAO} \and
INAF – Istituto di Radioastronomia, via Gobetti 101, I–40129 Bologna, Italy \label{inafbo_radio} \and
INAF, Osservatorio di Astrofisica e Scienza dello Spazio, via Piero Gobetti 93/3, I-40129 Bologna, Italy \label{inafBO} \and
International Centre for Theoretical Sciences, Tata Institute of Fundamental Research,Bangalore 560089, India \label{tata}
}

\date{Accepted XXX. Received YYY; in original form ZZZ}

\titlerunning{The fate of the hot gas in galaxy groups}
\authorrunning{Seppi, Eckert, Schaye et al.}

\abstract{
The impact of AGN feedback on the hot gas content of galaxy groups remains a key uncertainty in galaxy formation and its connection to the large scale structure of the Universe.  
%
We aim to compare the XMM-Newton Group AGN Project (X-GAP) sample to the hydrodynamical FLAMINGO simulations, which span a wide range of AGN feedback prescriptions. 
%
We construct X-GAP analogues by forward-modelling the full selection function, including detection and observational systematics, and generate end-to-end XMM-Newton mock observations analysed consistently with the data. We study multiple observables, including the $L$--$T$ and $M_{\rm gas}$--$T$ relations, number of groups, mean temperature, and velocity dispersion, accounting for their covariance. 
The forward model accurately recovers input luminosities, gas masses, and core-excised temperatures for regular systems, enabling direct comparison in observable space.
The normalisation of the scaling relations is the best discriminator between feedback models, while cosmic variance introduces $>20\%$ fluctuations in the number of detected systems, making counts alone a weak discriminator. Models with intermediate feedback strength provide the best agreement with X-GAP, with the $\mathrm{f_{gas}}-2\sigma$ model yielding the lowest tension of only $0.8\sigma$, while the most extreme feedback scenario ($\mathrm{f_{gas}}-8\sigma$) is ruled out at $>4\sigma$. 
Our results indicate that the thermodynamic properties of galaxy groups favour feedback stronger than the fiducial FLAMINGO calibration, but disfavour the most ejective models. This highlights the importance of combining forward modelling and multi-observable constraints to probe the fate of hot baryons in low-mass haloes.
}

\keywords{Galaxies: groups - X-rays: galaxies: clusters -  Galaxies: clusters: intracluster medium - Surveys -  Cosmology: large-scale structure of Universe - Methods: data analysis}
\maketitle

\section{Introduction}

Understanding galaxy formation requires a comprehensive view of how baryons are accreted, transformed, and redistributed within dark matter haloes across cosmic time \citep[][]{ Wechsler2018ARA&A_galhalo}. While only a small fraction of baryons condense into stars, the majority remains in diffuse gas phases, making the thermal and spatial distribution of baryons a key tracer of feedback and accretion processes in structure formation in the large scale structure (LSS) of the Universe \citep[][]{Fukugita1998ApJ_baryonbudget, Fukugita2004ApJ...616..643F, Ayromlou2023MNRAS_baryonsTNG}.

Galaxy groups, the most common hosts of galaxies in the Universe and occupying the intermediate mass regime between isolated galaxies and rich clusters \citep[][]{Tinker2008}, are particularly sensitive to baryonic processes \citep[][]{Mulchaey2000ARA&A_groupsprop, Giodini2009ApJGroups, Eckmiller2011A&A_chandragroups}. With halo masses of $M_{\rm 500c}\sim10^{13}$--$10^{14}$ M$_\odot$\footnote{$M_{\rm 500c}$ is the mass enclosed within $R_{\rm 500c}$, i.e. the radius encompassing an average density that is 500 times larger than the critical density of the universe, $\rho_c = 3H^2(z)/8\pi G$, at a given redshift z.}, their relatively shallow potential makes them especially susceptible to non-gravitational heating from supernovae and active galactic nuclei \citep[AGN,][]{Padovani2017A&A_agnreview}, allowing feedback to strongly modify their gas content and thermodynamic structure \citep[][]{McNamara2007ARA&A_gasAGN, McCarthy2010MNRAS_agngrps, LeBrun2014MNRAS_cosmowols, Gaspari2020NatAs_feedback}.

Understanding the thermodynamic state of the hot gas in galaxy groups is crucial for constraining models of galaxy formation and evolution. X-ray and Sunyaev-Zel'dovich (SZ) observations show that the gas in groups is less dense, more extended, and has higher entropy than expected from purely gravitational collapse, due to the role of AGN feedback in redistributing baryons \citep[][]{Sun2009ApJ_grpsChandra, Pratt2009A&A_rexcess, Johnson2009_xmmgrps, Stott2012MNRAS_agnXMM, Ettori2013arXiv_grpscluathena, Lovisari2015A&A_grps, Gozaliasl2019MNRAS_chandra}. However, the precise impact of feedback on the intra-group medium (IGrM), and the extent to which simulations reproduce these effects, remain open questions \citep[][]{Ponman1999Natur.397..135P, Finoguenov2002ApJ_grps, Oppenheimer2021Univ....7..209O, Bahar2024_erositagrps}.

In recent years, cosmological hydrodynamical simulations have advanced substantially through increasingly sophisticated sub-grid models for star formation, metal enrichment, and AGN feedback. The latter is now recognised as a key ingredient to reproduce several observed properties, including the suppression of cooling flows in cluster cores \citep[][]{Brighenti2006ApJ...643..120B, Fabian2012ARA&Areview, Hlavacek-Larrondo2022hxga.book_agnfeedback}, the quenching of star formation \citep[][]{Silk1998A&A_FEG_SMBH, BoothSchaye2009MNRAS_feedback, Pillepich2018MNRAS.473.4077P}, and the scaling relations between black holes and their host galaxies \citep[][]{Magorrian1998AJ....115.2285M, Kormendy2013ARA&A_BHcoevo, Sahu2019ApJ_BHscalrel, Habouzit2021MNRAS_BHmass_Mstar}. Modern simulations adopt different AGN feedback implementations. For example, cosmo-OWLS \citep[][]{LeBrun2014MNRAS_cosmowols}, EAGLE \citep[][]{Schaye2015MNRAS.446..521S}, BAHAMAS \citep[][]{McCarthy2017MNRASbahams}, FLAMINGO \citep[][]{Schaye2023MNRAS_flamingo}, and COLIBRE \citep[][]{Schaye2025arXiv_colibre} use isotropic thermal energy injection linked to gas accretion onto the supermassive black hole \citep[SMBH,][]{BoothSchaye2009MNRAS_feedback}. Other models include radio-mode bubbles (Illustris, FABLE; \citealt[][]{Vogelsberger2014MNRAS_illustris, Henden2018MNRAS_FABLE}), mechanical or radiative feedback channels (MAGNETICUM; \citealt[][]{Steinborn2015MNRAS_agnmagnet}), or kinetic feedback schemes (HorizonAGN, IllustrisTNG, SIMBA; \citealt[][]{Dubois2016MNRAS_horizonAGN, Weinberger2017MNRAS_agnTNG, Dave2019MNRAS_simba}). Some simulations additionally implement directional jet feedback, including SIMBA, selected FLAMINGO runs, and hybrid AGN feedback models in COLIBRE \citep[][]{Husko2025arXiv_colibreAGN}. While these simulations are calibrated to reproduce the galaxy stellar mass function (GSMF) and, in some cases, properties of massive clusters, their predictions for lower-mass haloes remain sensitive to the details of feedback modelling. Galaxy groups therefore provide a powerful test-bed for assessing the realism of baryonic physics in simulations, that is closely tied to the efficiency with which feedback expels baryons from dark matter haloes. \citep[][]{Oppenheimer2021Univ....7..209O, Eckert2021_review, Gastaldello2021Univ....7..208G, Lovisari2021Univ_review, Eckert2025A&A_4436}.

Multiple studies suggest substantial ejection of hot gas in galaxy groups, although its strength remains uncertain \citep[][]{Gastaldello2007ApJ_fgas_grps, Sun2009ApJ_grpsChandra, Rasmussen2009MNRAS_grps, Gonzalez2013ApJ_fbaryon, Lagana2013A&A_baryons_grpsclu, Sanderson2013MNRAS_baryondeficit, Akino2022PASJ...74..175A}. More recently, stacking analyses of eROSITA data around optically selected systems have inferred gas fractions of $\lesssim5\%$ for haloes in the $10^{13.5}$--$10^{14.5}$ M$_\odot$ range, below the predictions of most hydrodynamical simulations \citep[][]{Popesso2024arXiv_simsSTACKS}. Independent constraints from entropy profiles and kinetic SZ (kSZ) measurements also indicate low baryon fractions in groups \citep[][]{Molendi2025A&A_entropy, Siegel2025arXiv_kszFGAS, Roper2025arXiv_ksz}. In particular, analyses combining ACT, DESI, and eROSITA data favour stronger AGN feedback models, including the most ejective FLAMINGO variants \citep[][]{Siegel2025arXiv_kszFGAS, Siegel2025_baryonification}. However, recent XRISM observations suggest lower velocity dispersions and kinetic pressure than expected, potentially indicating that some feedback models may be overly ejective \citep[][]{XRISM2025_veldisp_feedback}. Similarly, analyses combining Planck and DES data find weaker power-spectrum suppression than predicted by simulations \citep[][]{Xu2025arXiv_DES_Planck_weakfeedback}, highlighting ongoing tensions that depend sensitively on the adopted feedback prescription \citep[e.g. including jets,][]{Bourne2023Galax_feedback}. By analysing the hot atmospheres of galaxy groups with a forward-modelling pipeline that reproduces the observational selection and observable measurements, we perform a direct comparison between simulations and data, thereby constraining the strength of baryon expulsion from massive haloes.

The XMM-Newton Group AGN Project \citep[X-GAP,][]{Eckert2024_xgap} is a sample of 49 galaxy groups observed with XMM-Newton, selected from the All-Sky Extended Sources project \citep[AXES,][]{Damsted2024_axes, Khalil2024A&A_AXES} using X-ray data from the ROSAT all-sky survey (RASS) and optical information from the Sloan Digital Sky Survey (SDSS), based on optically identified groups \citep[][]{Tempel2017A&A_sdssFOF}. X-GAP includes peculiar systems such as SDSSTG-4436, which shows a high entropy floor and a kinked density profile, providing valuable insight into the cumulative impact of AGN activity across cosmic time \citep[][]{Eckert2025A&A_4436}, as well as sources with extended radio emission \citep[][]{Santra2026arXiv_XGAPradio}. 
In a previous work, we forward-modelled the selection process using end-to-end simulations that reproduce the detection scheme \citep{Seppi2025A&A_selfunc}. \citet{Eckert2025arXiv_letter} presented a first comparison between the thermodynamic properties of X-GAP groups and predictions from the hydrodynamical FLAMINGO simulations \citep[][]{Schaye2023MNRAS_flamingo, Kugel2023MNRAS.526.6103K}, which provide an important benchmark thanks to multiple runs with different feedback implementations.

In this work, we take a further step compared to \cite{Eckert2025arXiv_letter} by generating and analysing detailed end-to-end XMM-Newton mock observations of X-GAP-like groups selected from FLAMINGO. We use the selection function from \cite{Seppi2025A&A_selfunc} to identify an X-GAP like sample in FLAMINGO, which allows a direct one-to-one comparison with simulations accounting for selection effects in real observations.
This framework enables a stringent test of whether state of the art hydrodynamical models reproduce observations of galaxy groups in the nearby Universe.
This paper is organised as follows. Sect. \ref{sec:data_sims} describes the data and simulations, Sect. \ref{sec:XMM_mock} the forward modelling and analysis of the XMM-Newton mocks, Sect. \ref{sec:results} presents the results, Sect. \ref{sec:discussion} discusses them in the context of recent complementary measurements, and Sect. \ref{sec:conclusion} provides a summary and outlook. We adopt the FLAMINGO cosmology, i.e. a $\Lambda$CDM model with parameters from the Dark Energy Survey plus external constraints \citep[DES Y3,][]{Abbott2022PhRvD_DESY3}: $H_0=68.1$ km s$^{-1}$ Mpc$^{-1}$, $\Omega_{\rm M}=0.306$, and $\Omega_{\Lambda}=0.694$.

\section{Data and Simulations}
\label{sec:data_sims}
In this section we describe the data and simulations used in this work: the X-GAP group sample, the FLAMINGO simulation suite, and the selection of simulated X-GAP analogues therein.

\subsection{X-GAP and its selection}
\label{subsec:X-GAP}

X-GAP\footnote{\url{https://www.astro.unige.ch/xgap}} \citep{Eckert2024_xgap} is an XMM-Newton large programme targeting galaxy groups (10$^{13}$--10$^{14}$ M$\odot$) to map IGrM properties out to $R_{\rm 500c}$. By comparing observed integrated quantities ($T$, $L$, $M_{\rm gas}$), gas fractions, and thermodynamic profiles to hydrodynamical simulations, the project aims to constrain AGN feedback models. The sample originates from the AXES group catalogue \citep{Damsted2024_axes}, which cross-matched wavelet-detected ROSAT sources \citep{Kaefer2019A&A...628A..43K} with optically detected groups from SDSS galaxies ($r < 17.77$; \citealt{Tempel2017A&A_sdssFOF}, $z < 0.05$). In \cite{Seppi2025A&A_selfunc}, we quantified AXES completeness and purity using end-to-end simulations of a full-sky light cone. We derived a robust selection function model based directly on observables, X-ray flux ($F_{\rm X}$), galaxy velocity dispersion ($\sigma_{\rm v,T}$), and redshift (z), rather than indirect halo properties. It reads:
\begin{align}
    P_{\rm det}(F_{\rm X}, z, \sigma_{\rm v,T}) =\ (1 &+ \exp[-\alpha_{\rm Fx}(\log_{\rm 10}F_{\rm X} - F_{\rm X,0}) \ + \nonumber \\
    & - \alpha_{\rm \sigma v} \times \log_{\rm 10}\sigma_{\rm v,T} + \alpha_{\rm z} \times \log_{10}z ])^{-1},
    \label{eq:selfunc_model}
\end{align}
where $F_{\rm X,0}$, $\alpha_{\rm Fx}$, $\alpha_{\rm \sigma v}$, and $\alpha_{\rm z}$ are model parameters.
We use the median point for each best fit parameter in the posterior chains from \cite{Seppi2025A&A_selfunc}.

X-GAP is designed to mitigate the problems in the optical identification of galaxy groups, such as projection effects and merging of unvirialised systems \citep[][]{Robotham2011MNRAS_gamagrps, Old2014MNRAS_massrec, Marini2025A&A_optical} with purer X-ray selection based on the IGrM. The trade-off is that large X-ray surveys tend to be shallow, limiting the group identification to low redshift. Indeed, \cite{Seppi2025A&A_selfunc} showed that the cross-match with X-ray data serves as a cleaning step of the optical candidates, increasing purity but reducing completeness. The X-ray detection via wavelet filters is particularly beneficial for identifying the outskirts of galaxy groups, as it is not sensitive to the shape of the surface brightness profile, thereby reducing biases related to the presence of a cool core \citep[][]{Eckert2011A&A...526A..79E, Seppi2021A&A_MF}. This is especially critical at low masses, while high-mass, brighter systems are typically detected regardless of dynamical state, lower sensitivity at lower masses can limit detection to centrally peaked systems, introducing significant biases into the selection function modelling.

\subsection{FLAMINGO}

FLAMINGO\footnote{\url{https://flamingo.strw.leidenuniv.nl}} is a large suite of cosmological hydrodynamical simulations \citep[][]{Schaye2023MNRAS_flamingo}. It features various box sizes and resolutions. FLAMINGO is calibrated to reproduce the z=0 GSMF and the gas mass fraction of low-redshift galaxy clusters \citep[][]{Kugel2023MNRAS.526.6103K}. In this work we use the high-resolution 1 Gpc box, i.e. with 3600$^3$ dark matter particles and gas particles mass (m$_g$) of 10$^8$ M$_\odot$ (here after L1\_m8), as a benchmark for the sample selection. In a second step, we use the intermediate resolution, with 2$\times$1800$^3$ particles and m$_g$ of 10$^9$ M$_\odot$ (L1\_m9), to test the various implementations of baryonic physics. The L1\_m9 set occupies the lower-resolution/large volume parameter space compared to smaller simulations, such as EAGLE \citep[][]{Schaye2015MNRAS.446..521S}, TNG \citep[][]{Pillepich2018MNRAS.473.4077P}, or SIMBA \citep[][]{Dave2019MNRAS_simba}. Gas particle masses are not strictly fixed, as they incorporate mass and metals by neighbouring particles through outflows and feedback \citep{Schaye2023MNRAS_flamingo}.
On top of the fiducial hydro model, FLAMINGO provides calibrations to lower GSMF (M$_\star$-$\sigma$), and to different gas fractions in clusters (f$_{\rm gas} \pm N\sigma$). In particular, the $N\sigma$ label indicates the shift in the gas fraction used for calibration: more negative values (e.g. $-2\sigma$) correspond to stronger AGN feedback, which further depletes the gas content of haloes and reduces the gas fraction at a fixed mass. Stellar and AGN feedback parameters are jointly tuned to reproduce the chosen sets of observables \citep[][]{Kugel2023MNRAS.526.6103K}.

L1\_m9 includes different AGN feedback implementations: thermal, isotropic energy injection \citep{BoothSchaye2009MNRAS_feedback}, and kinetic feedback with AGN jets \citep{Husko2022MNRAS_agnmodel}. FLAMINGO models feedback using a black hole (BH) energy reservoir that stores a fraction of the accreted rest-mass energy. 
In the thermal model, energy is injected once it exceeds a threshold corresponding to heating a neighbouring gas particle by $\Delta T_{\rm AGN}$ \citep[see Sect. 2.3.6 in][]{Schaye2023MNRAS_flamingo}. This parameter sets the temperature jump and thus regulates the burstiness of feedback events. In the jet model, the threshold corresponds to $m_g v_{\rm jet}^2$, and once reached, two particles are kicked within a cone aligned with the BH spin at velocity $v_{\rm jet}$, injecting kinetic energy in a collimated outflow. Both $\Delta T_{\rm AGN}$ and $v_{\rm jet}$ are free calibration parameters.

Haloes are identified with HBT-HERONS \citep{ForouharMoreno2025MNRAS_HBT}, which locates centres at the most bound particle, and halo properties are computed with the Spherical Overdensity and Aperture Processor \citep[SOAP;][]{McGibbon2025JOSS_SOAP}.

\subsection{Group sample selection}
\label{subsubsec:flamingo_groupsel}

X-GAP groups are identified from SDSS galaxies with r-band magnitude $r<17.77$. At very low redshifts $z<0.02$, it includes galaxies with stellar masses below $10^9$ M$_\odot$, unresolved in L1\_m9. We therefore use the higher-resolution L1\_m8 run to properly resolve the X-GAP galaxy population. 

We do not explicitly model spectroscopic incompleteness: the SDSS main galaxy sample is highly complete, with the dominant missing fraction of about $5\%$ arising from fiber collisions \citep[fibers can not be placed within 55 arcsec of each other,][]{Strauss2002AJ_sdss, Eisenstein2011AJ_sdssdr12, Alam2015ApJS_sdss3}. For the nearby groups considered here, subtending large angular extents on the sky, this effect is expected to be modest.
We provide full details for the $m_r$ modelling in FLAMINGO in Appendix \ref{appendix:Ngalsel}.

To select X-GAP analogues in FLAMINGO we use a single snapshot at $z=0$, which extensively covers the proper distance within $z<0.05$ corresponding to about 250 Mpc. Therefore, there is no need to concatenate different snapshots. Starting from the halo catalogue, we place an observer at the box centre and construct a light cone by converting Cartesian coordinates $(X,Y,Z)$ to $(\mathrm{RA},\mathrm{DEC},z)$ following the geometry of \cite{Seppi2025A&A_selfunc}. \\
To predict X-ray fluxes for FLAMINGO haloes we use the true X-ray luminosities derived from the emissivity of gas particles within $r_{\rm 500c}$ \citep[][]{Braspenning2024MNRAS_flamingo}. A key methodological difference concerns the luminosity definition: in FLAMINGO it is computed as a spherical (3D) integral of the gas emissivity within $r_{\rm 500c}$ \citep{Braspenning2024MNRAS_flamingo}, whereas in X-GAP \citep{Eckert2024_xgap} and in our selection modelling \citep{Seppi2025A&A_selfunc} $L_{\rm 500c}$ is obtained from a cylindrical (projected) integral of the emission measure within a projected radius $R_{\rm 500c}$. The cylindrical definition includes additional emission along the line of sight, implying $L_{\rm 500c}^{\rm cyl} > L_{\rm 500c}^{\rm sph}$ for the same gas distribution.

Computing $L_{\rm 500c}^{\rm cyl}$ for every single halo in the FLAMINGO boxes is computationally expensive. Therefore, to place the two on a common footing, we derive a multiplicative conversion using $\beta$-model gas density profiles. We denote by $n_e n_H \sim n_{\rm e,0}^2\left[1+\left(\frac{r}{r_c}\right)^2\right]^{-3\beta}$ the 3D emission measure density, while $EM(R)$ represents the line-of-sight emission measure (also known as emission integral). We compute both the spherical and cylindrical integrals:
\begin{align}
    L_{\rm X}^{\rm sph} =& \int_0^{r_{\rm 500c}} 4\pi r^2 n_e n_H(r)\, \Lambda(T,Z)\, dr \nonumber \\
    EM(R) =&\, 2\int_{R}^{\infty}\frac{n_e n_H(r)}{\sqrt{r^2 - R^2}}\,dr \nonumber \\
    L_{\rm X}^{\rm cyl} =& \int_0^{R_{\rm 500c}} 2\pi R\, EM(R)\, \Lambda(T,Z)\, dR,
    \label{eq:spherical_cylindircal_integrals}
\end{align}
where $\Lambda$ is the cooling function that depends on temperature and metallicity \citep[see][]{Eckert2020_pyproffit}.
We interpolate the cumulative sum of the integrals to R$_{\rm 500c}$ and evaluate the ratio $C_{\rm cyl}$ = L$_{\rm 500c}^{\rm cyl}$/L$_{\rm 500c}^{\rm sph}$ over a grid in slope $\beta$ and core radius r$_C$. X-GAP groups are well described by $\beta$-models with 
$\beta$ around 0.45 and 
r$_C$ below 0.1$\times$R$_{\rm 500c}$.
Exploring slopes between 0.4 and 0.5, and core radii below 0.05 on a fine grid we obtain a median correction of $C_{\rm cyl}$=1.16 with a scatter of 0.06.

Because feedback variants in FLAMINGO may modify the ICM profile shape (weaker feedback producing peakier cores and stronger feedback flatter distributions), we adopt a conservative conversion between spherical and cylindrical luminosities. We assume a log-normal distribution with median 0.1 dex ($C_{\rm cyl}\approx1.26$) and scatter 0.06 dex, truncated at $C_{\rm cyl}=1$ since the cylindrical luminosity must exceed the spherical one. For each halo we draw $C_{\rm cyl}$ and compute $L_{\rm 500c}^{\rm cyl} = C_{\rm cyl}\, L_{\rm 500c}^{\rm sph}$. We verified that this correction does not significantly bias the reconstructed luminosity (see Appendix \ref{appendix:LinLout}).
$C_{\rm cyl}$ maps FLAMINGO luminosities to the cylindrical aperture used by X-GAP while allowing for plausible profile variations. X-ray fluxes are then computed including a K-correction \citep[see][]{Seppi2025A&A_selfunc}. Hereafter we refer to cylindrical luminosities as $L_{\rm 500c}$.

For each halo we identify member subhaloes, project their velocities along the line of sight to obtain peculiar velocities, and compute the velocity dispersion. The light cone construction, flux computation, and velocity-dispersion estimation follow \citet{Seppi2025A&A_selfunc}.

We then apply the X-GAP selection using rejection sampling: for each halo we compute the detection probability (Eq.~\ref{eq:selfunc_model}) and retain the object if this exceeds a random number drawn in $[0,1]$. Additional X-GAP cuts are imposed: (i) $0.02<z<0.05$, (ii) $R_{\rm 500c}<15$ arcmin to ensure the system fits within the XMM-Newton field of view (FoV), and (iii) number of galaxies $N_{\rm gal,SDSS}\ge8$.

The second condition translates into an upper limit on the ROSAT luminosity at fixed redshift because $R_{\rm 500c}$ was originally estimated from RASS luminosities \citep[Eq. 1 in][]{Eckert2025arXiv_letter}. We therefore calibrate the relation between $L_{\rm 500c}$ and the ROSAT luminosity in the 0.5--2.0 keV band using X-GAP groups, in the form $\log_{10} L_{\rm 500c} = A + B\,\log_{10} L_{\rm ROSAT} + \mathcal{N}(0,\sigma_{\rm int})$, finding $B=0.70$, $A=0.51$, and $\sigma_{\rm int}=0.45$. This relation is applied to convert FLAMINGO luminosities before imposing the luminosity--redshift cut.

Finally, the requirement $N_{\rm gal,SDSS}\ge8$ cannot be applied directly because the number of members identified by the optical FoF algorithm in observations does not necessarily perfectly correspond to the intrinsic subhalo population of the parent dark matter halo. We therefore use the simulations developed in \citet{Seppi2025A&A_selfunc} to calibrate the mapping between true and detected membership, enabling a forward-modelled implementation of this cut. Details are provided in Appendix \ref{appendix:Ngalsel}.

\subsection{Selection Uncertainties}
\label{subsec:cosmic_variance}

\begin{figure}
    \centering
    \includegraphics[width=\columnwidth]{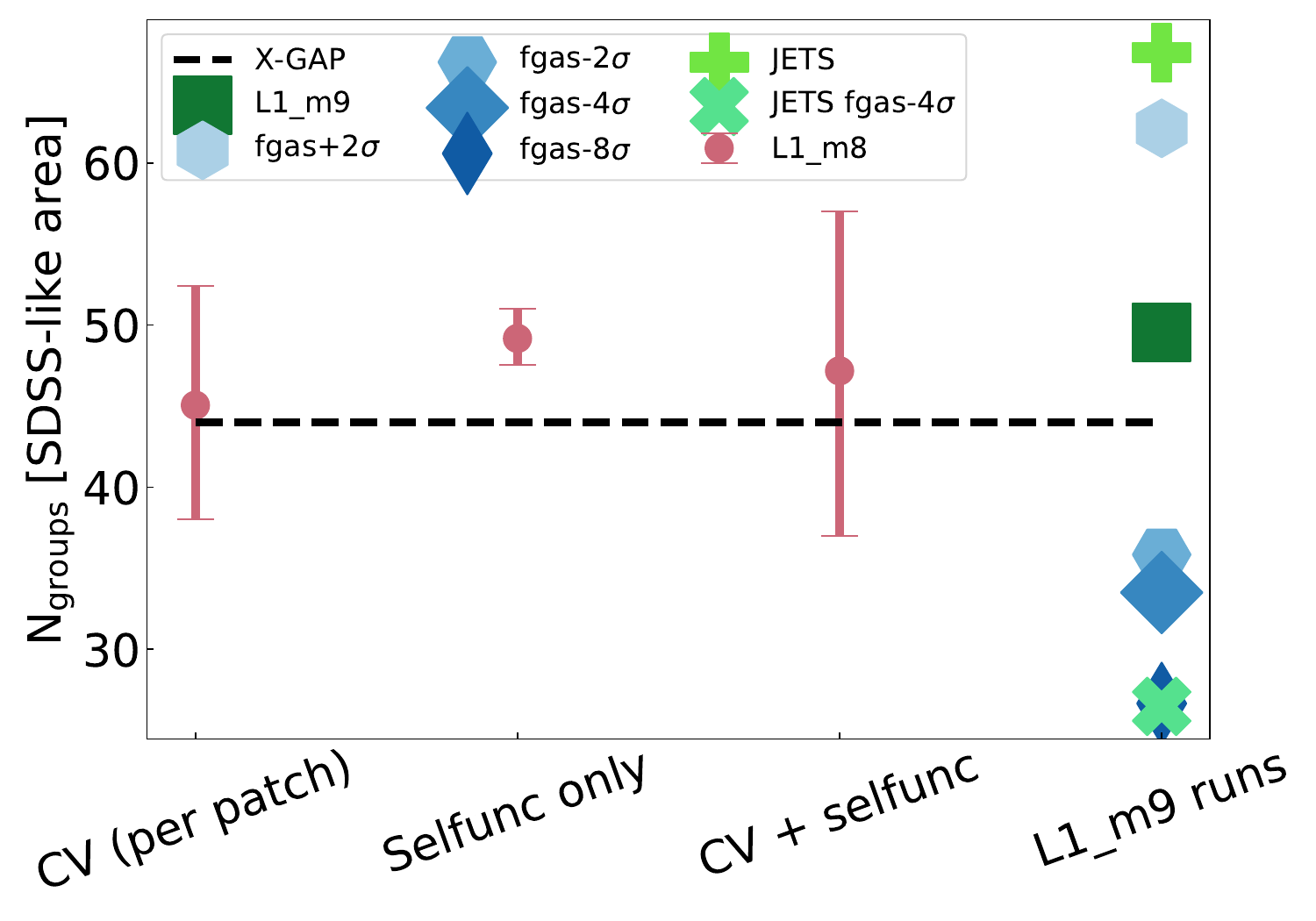}
    \includegraphics[width=\columnwidth]{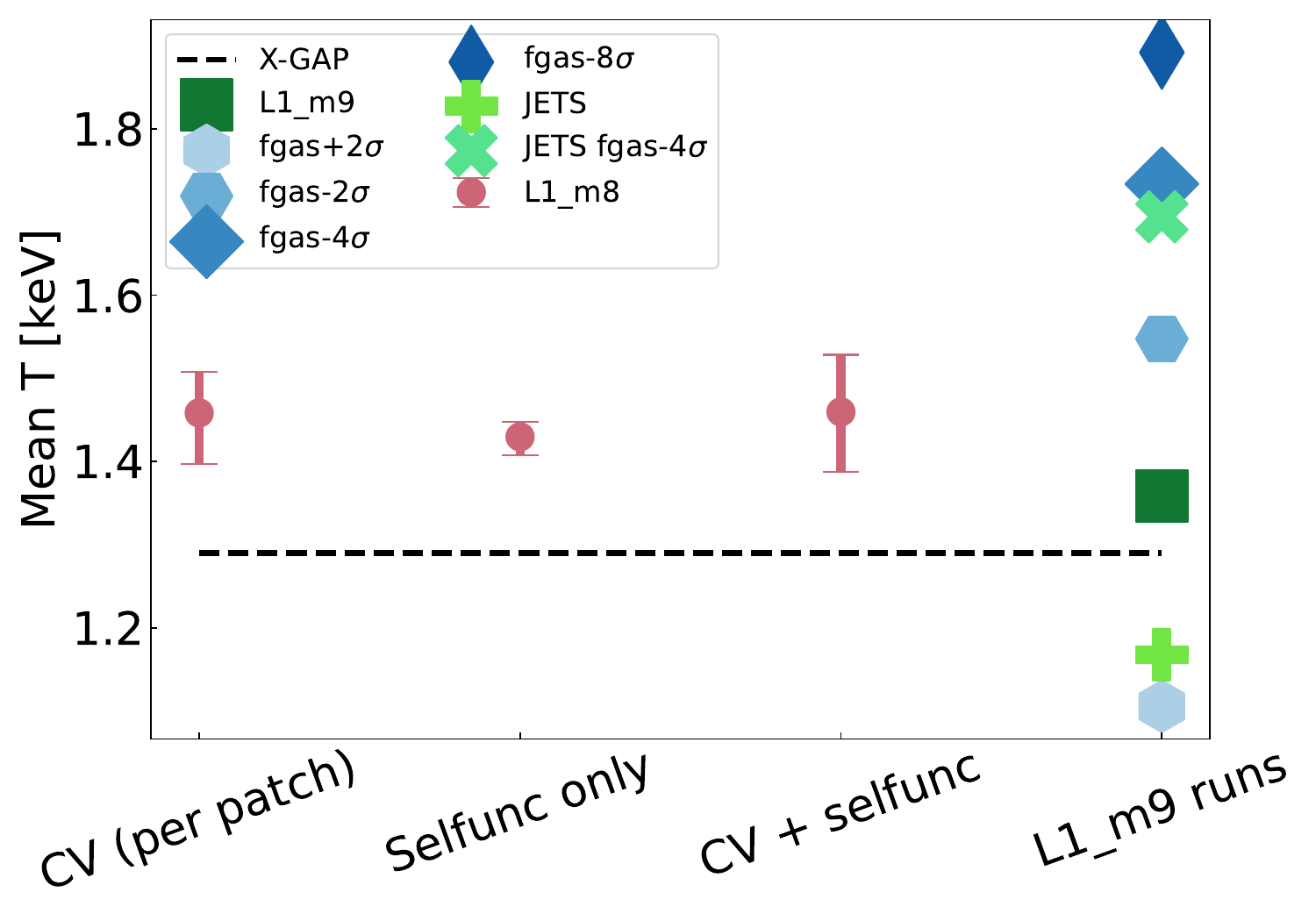}
    \caption{Expected properties of an X-GAP-like sample selected from different FLAMINGO models. The L1\_m8 includes tests for cosmic variance (CV) and uncertainties on the selection function. The top panel shows the number of groups within an SDSS-like area of 7430 deg$^2$, the bottom one shows the median temperature of the selected sample. The latter is a promising discriminator between FLAMINGO models.}
    \label{fig:Number_of_groups}
\end{figure}

We evaluate possible sources of systematics in our selected samples: cosmic variance, and uncertainties on the selection function. Their impact on the total number of selected groups and on the median temperature of the sample is shown in Fig. \ref{fig:Number_of_groups}.

\subsubsection{Cosmic variance}

To estimate the impact of cosmic variance, we generate nine independent light cones by placing the observer at different positions within the simulation box of L1\_m8. One observer is positioned at the centre of the box (500, 500, 500) Mpc, while the remaining eight are located at the centres of the eight sub-octants of the box, i.e., the midpoints of each corner cube of size 500 Mpc. This tiling strategy ensures that each light cone probes a unique volume. Given that the centres of the sub-octants are separated by 500 Mpc, the light cones originating from adjacent observers do not overlap within a radial distance of 250 Mpc, corresponding to approximately $z=0.05$ in the \cite{Abbott2022PhRvD_DESY3} cosmology, covering the redshift range relevant for the X-GAP sample without spatial redundancy. In addition, we divide each full-sky light cone into four quadrants, splitting the full RA, DEC plane in half. Then we count the number of haloes within an area sub-selected from each of the four quadrants that matches the SDSS survey area. To do this, we find a maximum value of DEC according to $\text{DEC}_{\rm max} = \arcsin{\frac{A_{\rm SDSS}}{\Delta_{\rm RA}}}$, where $A_{\rm SDSS}=7430 \times (\pi/180)$ sr, and $\Delta_{\rm RA}=\pi$ rad. We find DEC$_{\rm max}$=46.09 deg. We obtain a combination of four SDSS-like sky areas for each of the nine light cones, for a total of 36 independent areas to quantify cosmic variance (in combination with Poisson noise) in our selected sample.

The mean number of objects expected within an SDSS-like area is 45.1, with 16th and 84th percentiles of 38.0 and 52.4 (standard deviation 9.3), in excellent agreement with the 44 groups in X-GAP. This prediction depends on cosmology, in particular on $S_8 = \sigma_8\sqrt{\Omega_{\rm M}/0.3}$, which sets the halo abundance, and on $H_0$, which affects both the survey volume and the X-ray fluxes through the luminosity distance. As shown in Appendix~\ref{appendix:cosmo}, this changes the predicted counts by $\sim$2 groups, subdominant compared to cosmic variance. 
The impact of cosmic variance on the temperature distribution is smaller, with mean 1.46 keV, 16th and 84th percentiles of 1.40 and 1.51 keV, and a standard deviation of 0.06 keV.

\subsubsection{Uncertainty on the selection function}
The best fit parameters on the selection function model from \cite{Seppi2025A&A_selfunc} have error bars which naturally propagate as an uncertainty on the selected sample. To quantify such an effect, we randomly draw 20 models from the posterior chains, which allows us to properly account for the correlation between different parameters. We use a single area in one individual light cone to neglect cosmic variance and isolate the effect due to uncertainty in the selection function.

We obtain a mean number of 49 groups, with 16th and 84th percentiles at 47.5 and 50.9, for a standard deviation of 2.6. The temperature distribution is even less sensitive to uncertainties on the selection function: we find a mean of 1.43 keV, with 16th and 84th percentiles at 1.41 and 1.45, for a standard deviation of 0.02. We find that this effect is subdominant with respect to cosmic variance concerning both the number of groups and the temperature distribution of the selected sample. 

We finally combine cosmic variance and selection function uncertainty by drawing 20 random selection function models in each sky area of the nine light cones, which gives us a total of 720 experiments to evaluate their combined effect. We obtain a total number (mean temperature) with respective 16th and 84th percentiles of 47, 37, 57 (1.46, 1.39, 1.53 keV), with a standard deviation of 10.2 (0.07 keV). In both cases the fluctuations are dominated by cosmic variance.

\subsection{Hydro runs}
\label{subsec:hydro_runs}

\begin{figure}
    \centering
    \includegraphics[width=\columnwidth]{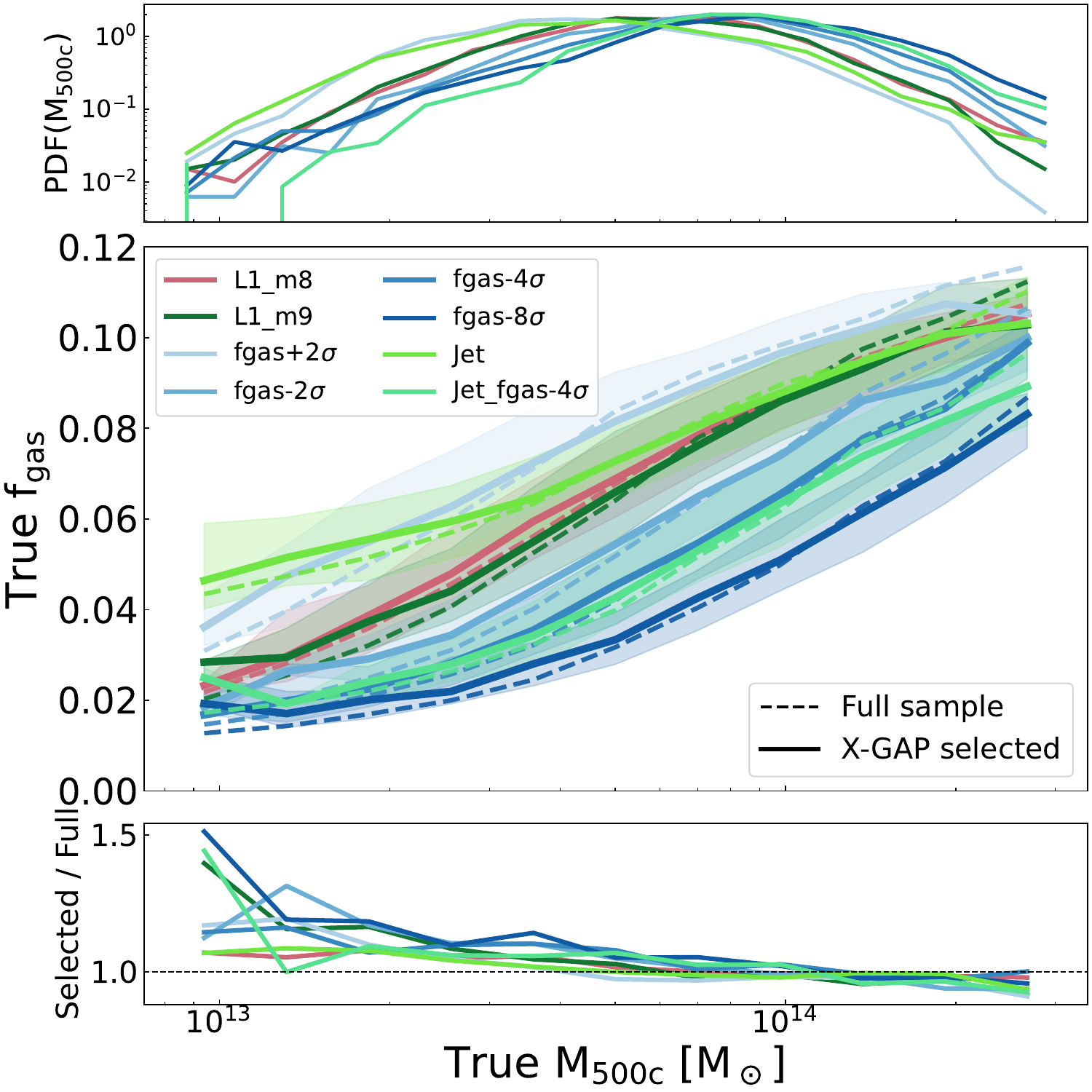}
    \caption{Gas fraction as a function of mass for the selected systems (solid lines and shaded areas) and the full sample (dashed lines). Both are true input quantities. The top panel shows the mass distribution of the selected systems: skewed to higher masses for strong feedback models. The bottom panel denotes the ratio between the gas fraction in the X-GAP-like selected sample and the whole population.}
    \label{fig:fgas}
\end{figure}

After modelling the N$_{\rm gal}$ selection in L1\_m8 (see Sect. \ref{subsubsec:flamingo_groupsel}), we identify corresponding haloes across the L1\_m9 FLAMINGO runs, which share the same initial conditions with L1\_m8, allowing consistent tracking of individual systems. We match each selected group to its counterpart within 1 Mpc, and replace the galaxy populations in lower-resolution runs with those from L1\_m8, which fully resolve X-GAP members. This ensures accurate velocity dispersions and a consistent application of the N$_{\rm gal}\geq 8$ selection.
This approach is justified as FLAMINGO simulations are calibrated to reproduce the observed GSMF, yielding consistent galaxy populations across feedback variants, while AGN feedback changes do not significantly affect galaxy properties at fixed resolution \citep[see Fig. 9 in][]{Schaye2023MNRAS_flamingo}. While the L1\_m8 exhibits a slight offset from L1\_m9 at the high-mass end \citep[see Fig. 8 in][]{Schaye2023MNRAS_flamingo}, this discrepancy is comparable to the systematic uncertainties arising from cosmic variance in the observational data. In addition, these deviations primarily affect the massive cluster regime, rather than the galaxy group scales analysed in this work. We exclude runs calibrated to different GSMFs.

For each group, we compute the detection probability using Eq. \ref{eq:selfunc_model} based on flux, redshift, and velocity dispersion (from L1\_m8), and obtain X-GAP-like catalogues for all simulations.

We compare the samples selected for different model runs in FLAMINGO to the systematics tests from Sect. \ref{subsec:cosmic_variance} in the last columns of Fig. \ref{fig:Number_of_groups}. 
The number of selected groups depends on the feedback implementation. In an SDSS-like area, the model with strong jets provides the lowest number of selected groups with 27. The Jet model produces the most with 64, followed by the model with weak AGN $f_{\rm gas}+2\sigma$ at 60. The particle resolution does not affect this statistic: the number of groups in the low resolution run L1\_m9 is 47, the same as in the higher resolution L1\_m8. Given the uncertainty due to the combination of cosmic variance and selection function, the number of groups alone is not a great discriminator between different models, that are all encompassed within about 1.5$\sigma$ compared to the fiducial L1\_m8.
Conversely, the temperature distribution shows smaller fluctuations, with the $f_{\rm gas}+2\sigma$ ($f_{\rm gas}-8\sigma$) at about 1.0 (1.8) keV being incompatible with the fiducial model at about 2.9 (5.8)$\sigma$. We conclude that temperature is an observable that potentially can discriminate FLAMINGO models compared to the real X-GAP sample (see also Sect. \ref{sec:results}).

We evaluate the distribution of selected systems in the $f_{\rm gas}$--M$_{\rm 500c}$ plane. The total gas mass is computed by summing the masses of all hot gas elements (see Sect. \ref{subsec:xmmsim}) within the true R$_{\rm 500c}$. The resulting relation is shown in Fig. \ref{fig:fgas}, where solid lines and shaded regions indicate the selected samples, while dashed lines show the full halo populations. The top panel shows the mass distribution of the selected systems. As feedback strength increases, the selection shifts towards higher masses, since more massive haloes are required to retain enough baryons to produce X-ray luminosities that satisfy the X-GAP selection. The lower panel presents the ratio between the gas fractions of the selected and full samples. At the low-mass end the ratio exceeds unity due to incompleteness for faint systems, with deviations larger than 10\% below $3\times10^{13}$ M$_\odot$ \citep[][]{Seppi2025A&A_selfunc}. In this regime the X-GAP selection preferentially includes systems that lie above the mean $f_{\rm gas}$--M$_{\rm 500c}$ relation (i.e. up-scattered systems). Conversely, at the high-mass end the upper luminosity cut (imposed to ensure R$_{\rm 500c}$ < 15$'$) results in a preferential selection of systems below the mean relation, with an impact below 5\% at $2\times10^{14}$ M$_\odot$. Overall, the X-GAP sample is broadly representative of the underlying group population, particularly near the typical mass scale of $(5$--$8)\times10^{13}$ M$_\odot$. We obtain similar results with a study of the impact of selection of the scaling relation between X-ray luminosity and temperature (see Appendix \ref{subsec:inout} and Fig. \ref{fig:LTselection}).

Out of all the selected systems, we randomly sub-select 50 groups independently for each run, in order to have a statistical power that is similar to X-GAP. These are the systems we use for our full forward modelling. 
The various models show different gas fractions at fixed mass: the same selection function yields systems with different gas content as a function of mass based on the feedback model. Indeed this means that the gas content of the selected groups can discriminate between the FLAMINGO models \citep[][]{Kugel2023MNRAS.526.6103K}. For example, at M$_{\rm 500c}$=7$\times$10$^{13}$ M$_\odot$ the typical gas fraction changes by up to a factor of 2.3, with a value of 0.089 for $f_{\rm gas}+2\sigma$ and 0.038 for $f_{\rm gas}-8\sigma$. Milder values are shown by the fiducial L1\_m8 with 0.076 and $f_{\rm gas}-2\sigma$ with 0.061. The fiducial L1\_m9 is in excellent agreement with L1\_m8 at 0.075. This is expected, because they are calibrated on the same target, and it confirms that particle resolution does not affect our result here.

The gas fraction has often been used in the context of AGN feedback, as stronger gas ejection would deplete haloes, therefore reducing the gas fraction at fixed halo mass \citep[][]{Eckert2021_review}. Simulations, including FLAMINGO, have used the gas fraction as an anchor to calibrate their feedback model \citep[][]{Schaye2023MNRAS_flamingo}. 
However, directly interpreting the observed f$_{\rm gas}$--M relation is challenging due to the inherent coupling between gas mass, total halo mass, and their integration volume. In observations, f$_{\rm gas}$ is integrated within a radius derived from the estimated halo mass, which is itself subject to uncertainties such as hydrostatic bias and projection effects. Consequently, errors in M$_{\rm 500c}$ propagate into the determination of the integration radius of the gas mass, introducing non-trivial covariance in the f$_{\rm gas}$--M$_{\rm 500c}$ relation. Observed trends thus reflect a combination of physical feedback effects and these measurement systematics, complicating comparisons between simulations and different observational datasets.

\section{XMM-Newton forward modelling}
\label{sec:XMM_mock}
\begin{figure*}
    \centering
    \includegraphics[trim=0 4.5cm 0 4.5cm, clip, width=2\columnwidth]{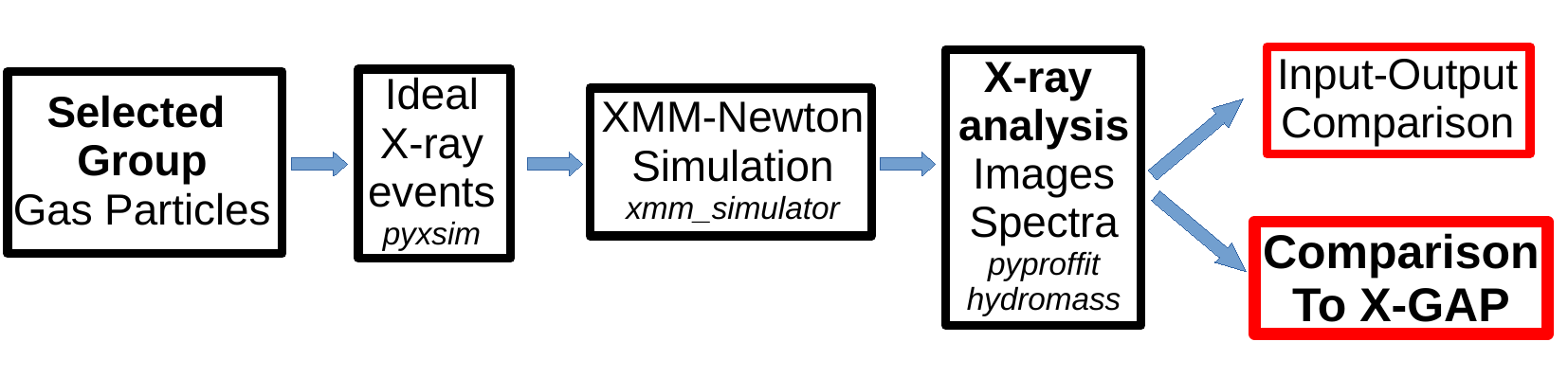}
    \caption{Workflow for the end-to-end XMM-Newton simulation of FLAMINGO galaxy groups down to the direct comparison with X-GAP.}
    \label{fig:workflow}
\end{figure*}

In this section we describe the generation and analysis of XMM-Newton mock data starting from the selected galaxy groups. The complete workflow is depicted in Fig. \ref{fig:workflow}.

\subsection{The XMM mocks}
\label{subsec:xmmsim}

We collect the gas particles in 3D within two times r$_{\rm 200c}$ of each halo and isolate hot gas particles hotter than 3$\times$10$^{5}$ K and null star formation rates. Cold gas and star forming particles are not expected to contribute to the X-ray emission of the IGrM. We also exclude gas particles that were heated by an AGN event in the last 100 Myr \citep[comparable to a typical AGN duty cycle,][]{Martini2004cbhg_qsoLT, Vantyghem2014MNRAS_dutycycle}, they are affected by numerical artifacts and unphysical temperatures and did not yet mix with the surrounding medium in a physical way. This makes the cores more realistic and typically removes less than 0.5$\%$ of the total luminosity. We assume that the hot gas is in collisional ionization equilibrium and each hot gas particle emits following an \texttt{APEC} model \citep[][]{Smith2001ApJapec}. We do not introduce any assumption (such as the typical 0.3 Z$_\odot$) on the average metallicity of the gas, we rather use the individual Z value for each particle. We use the SPH smoothed metallicity values readily available in the FLAMINGO gas particle data. We also account for the individual abundances of the following elements that are tracked in FLAMINGO: C, N, O, Ne, Mg, Si, and Fe. For non-tracked elements (S and Ca), we scale their solar contributions according to the total metallicity of particle Z/Z$_\odot$. This approach, which allows for non-solar abundance ratios, follows the methodology detailed in \citet{Braspenning2024MNRAS_flamingo}. This follows the same recipe as the one described in Sect. 2.2-2.4 of \cite{Braspenning2024MNRAS_flamingo}, who used \texttt{cloudy} to compute cooling rates. We verified in Appendix \ref{subsec:inout} that any intrinsic difference between \texttt{cloudy} and \texttt{APEC} does not affect our results.
We estimate the X-ray spectrum of each particle using the \texttt{pyXSIM} software \citep[][]{ZuHone2016ascl_pyxsim, Biffi2013MNRAS_phox}. We generate idealised photon lists with a large effective area  of 10\ 000 cm$^2$ and exposure time of 2 Ms. To account for galactic absorption, we fix the neutral hydrogen column density to $N_H=2\times$10$^{20}$ cm$^{-2}$. This value represents the typical median column density across the SDSS footprint, which is primarily located at high Galactic latitudes where $N_H$ is relatively low.

To simulate realistic XMM-Newton images, we use the \texttt{xmm\_simulator}\footnote{\url{github.com/domeckert/xmm_simulator}} package presented in \cite{Seppi2026arXiv260303440S}. We developed a new interface to this software to generate realistic mock XMM-Newton event lists from idealised photon simulations, generated for example with \texttt{pyXSIM}. The interface accounts for the desired exposure time and the spatially varying effective area across the XMM field of view. As a first step, we convolve the idealised events with the instrument's point spread function (PSF), modelled as a King profile. For each photon, we sample a radial offset from the PSF cumulative distribution function and a random azimuthal angle in [0, 2$\pi$], which together define a blurred photon position.
The three XMM detectors MOS1, MOS2, and PN are processed independently. Events falling outside each camera's footprint are discarded. We then compute, for each pixel, the fraction of idealised photons that would actually be observed by the observational setup, given the ratios of effective areas and exposure times. This accounts for the varying instrument response across the FoV. A Poisson draw using a binomial distribution is used to select a subset of events accordingly.
Finally, the redistribution matrix function (RMF) is applied: for each event, we sample a detected energy from its true energy using the appropriate RMF-derived probability distribution. The resulting event list closely mimics real XMM-Newton data products in format and instrumental characteristics. The \texttt{xmm\_simulator} includes a realistic simulation of the X-ray foreground and background, with contributions from the local hot bubble, the galactic halo, and the cosmic X-ray background. We also simulate individual AGN in the FoV \citep[see][for details]{Seppi2026arXiv260303440S}. We do not specifically add X-ray binaries. They would only show up as a hard tail in the spectrum of the inner most bin in our analysis, and their contribution is expected to be around 10$^{40}$ erg/s \citep[][]{Boroson2011ApJ_XRBs}, negligible in the soft X-rays compared to the hot gas luminosity in the regime of galaxy groups.

\subsection{X-ray data analysis}

We analyse the mock XMM-Newton images following the standard X-GAP procedures described in \cite{Eckert2025A&A_4436}, starting from the count image, exposure map, and particle background map produced by the \texttt{xmm\_simulator}. Point sources are detected with the \texttt{ewavelet} task using wavelet filtering. Compared to \cite{Seppi2026arXiv260303440S}, we additionally perform a hard-band (2--7 keV) detection to identify heavily obscured AGN. PN-to-MOS count ratios of 3.42 and 2.74 are applied in the soft (0.7--1.2 keV) and hard (2--7 keV) bands, respectively. After visual inspection, the detections from both bands are combined and used to construct a mask for point sources.

Temperature profiles are derived from spectra extracted in 12 logarithmically spaced annuli centred on the X-ray peak, following \cite{Rossetti2024A&A_TxprofCXM}. The background spectrum is taken from the outermost annulus (12.5--15 arcmin) and jointly modelled with a simulated RASS background generated with the same parameters as in \texttt{xmm\_simulator}. The background model includes three components: the local hot bubble (kT = 0.11 keV), the Galactic halo, and the cosmic X-ray background described by a power law with photon index 1.46. RASS spectra are rescaled by a cross-calibration factor of 0.12 \citep[][]{Rossetti2024A&A_TxprofCXM}. In each annulus the group emission is modelled as absorbed thermal plasma (\texttt{tbabs$\times$APEC}), fixing the hydrogen column density to $2\times10^{20}$ cm$^{-2}$ as assumed in the simulation, while fitting for temperature and metallicity.

\subsection{X-ray modelling}
\label{subsubsec:thermo_profs}

We extracted surface brightness profiles from the 0.7-1.2 keV mock XMM-Newton EPIC images in bins of 3 arcsec using the \texttt{pyproffit} package \citep[][]{Eckert2020_pyproffit}, accounting for the telescope vignetting and PSF. We use the measured 2D surface brightness and temperature profile to infer deprojected 3D thermodynamical profiles assuming that the systems are spherically symmetric with the \texttt{hydromass} package \citep[we use the FORW model, see][]{Eckert2022_hydromass}. It allows linking the pressure, temperature, and density profiles under the assumption of hydrostatic equilibrium to jointly model the dark matter halo mass profile. We compute integrated quantities within the estimated R$_{\rm 500c}$ and the core-excised ones by excluding the inner 0.15$\times$R$_{\rm 500c}$. The gas mass is obtained by integrating the gas number density profile weighted by the mean molecular weight and proton mass \citep[][]{Eckert2020_pyproffit, Seppi2026arXiv260303440S}. Any bias in the estimation of R$_{\rm 500c}$ would impact the integration radius to measure integrated quantities. This is expected to be negligible for X-ray luminosity and temperature, where most of the signal comes from the inner regions, but it is important for gas mass, that mostly lies in the outskirts (see Sect. \ref{subsubsec:MgasT}). The integrated temperature follows the surface brightness weighting scheme from \citet{Lovisari2024A&A_CXMATETX}, that has been shown to faithfully reproduce spectroscopic-like temperature \citep[][]{Mazzotta2004MNRAS_SL, Rasia2014_Txstruct}. We also refer the reader to \cite{Eckert2025A&A_4436} for more details on the X-ray data analysis of X-GAP groups, which we implement in the same way on our mock groups from FLAMINGO.

\begin{figure}
    \sidecaption
    \includegraphics[width=\columnwidth]{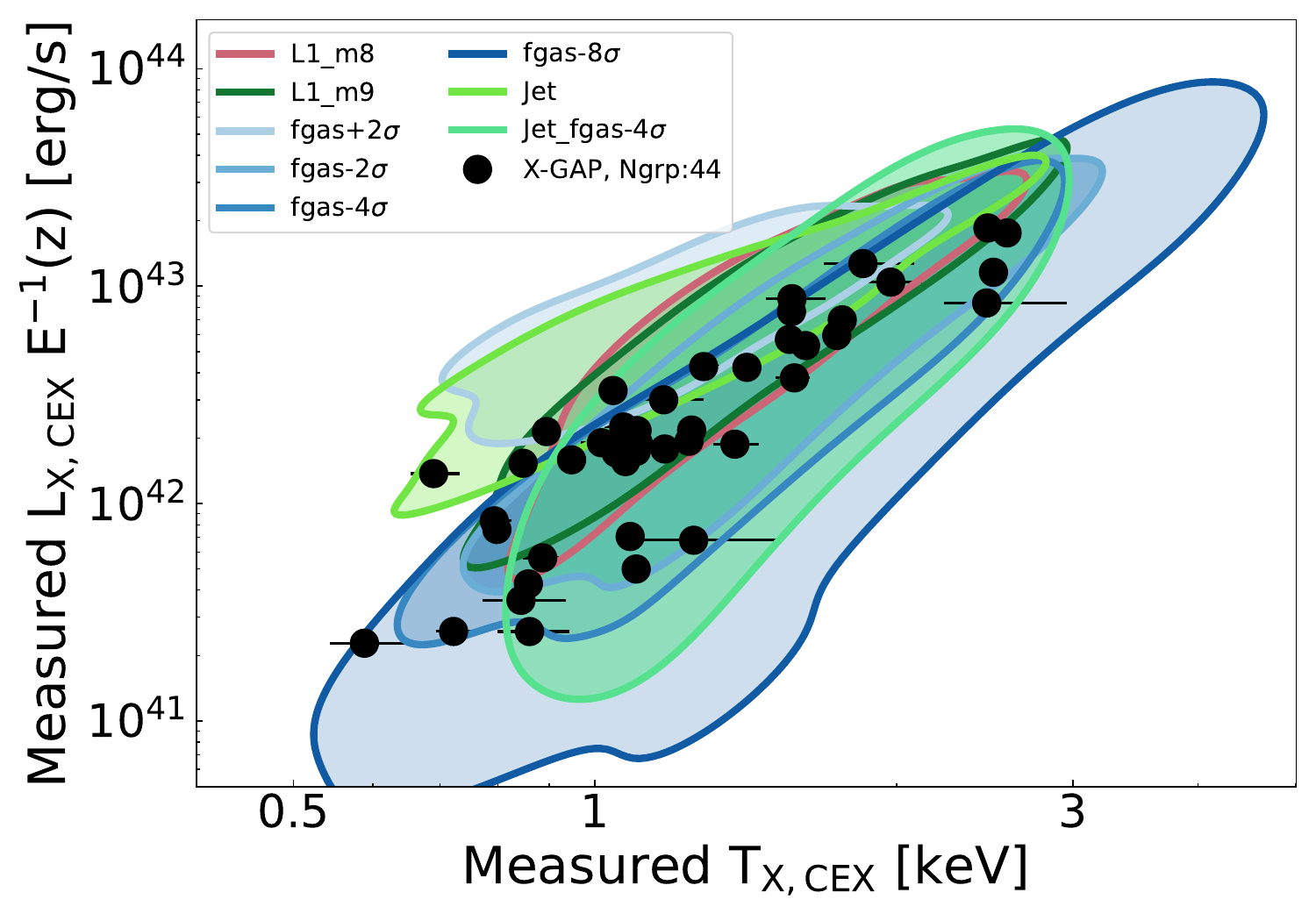}
\includegraphics[width=\columnwidth]{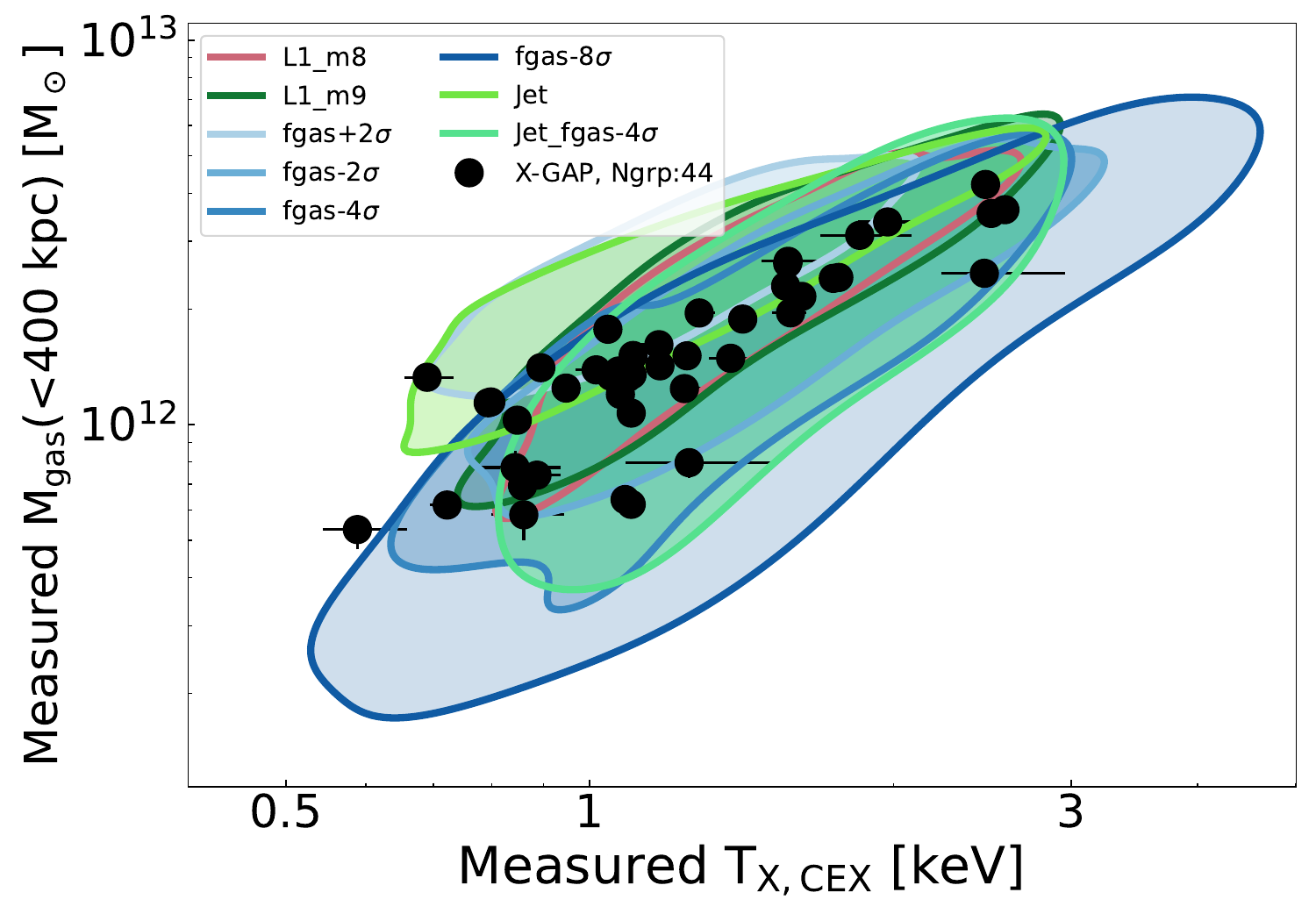}
    \caption{Comparison between X-GAP and the FLAMINGO models using the scaling relations between observables: core-excised X-ray luminosity and core-excised temperature (top panel) and gas mass and core-excised temperature (bottom panel). The contours represent the 2$\sigma$ distribution for each FLAMINGO model estimated with a Gaussian KDE.}
    \label{fig:LT_comparison}
\end{figure}

\section{Results}
\label{sec:results}

First, we verified that our X-ray analysis pipeline accurately recovers input quantities. For temperature and luminosity we exclude the central $0.15\,R_{\rm 500c}$, where FLAMINGO suffers from low particle resolution and unrealistic metallicity profiles (Appendix \ref{subsec:inout}). The X-GAP data is treated identically. We find good agreement between input and recovered X-ray luminosities and gas masses. The \texttt{hydromass} pipeline estimates emissivities using radially dependent metallicity and temperature profiles derived from the spectral fits, enabling unbiased gas mass measurements (Appendix \ref{subsec:inout}, \ref{appendix:systematics}). Temperature measurements are less robust for strong feedback models, where disturbed systems violate assumptions such as spherical symmetry or a single-temperature structure. After excluding such cases through visual inspection, the recovered temperatures remain consistent with the spectroscopic-like weighting scheme \citep[][]{Mazzotta2004MNRAS_SL, Seppi2026arXiv260303440S}. Further details are provided in Appendix \ref{subsec:inout}.

We now focus on the comparison between the samples selected from FLAMINGO and the real X-GAP.

\subsection{Observable scaling relations}
In this section we study the relations linking core-excised X-ray luminosity and gas mass within 400 kpc to core-excised temperature. The values are tabulated in Table \ref{tab:xgap_meas}.

\subsubsection{Luminosity - Temperature}
For consistency with the core-excised temperature, we also consider core-excised X-ray luminosities. Similarly, we focus on the region between 0.15 and 1.0$\times$R$_{\rm 500c}$. Although the inconsistency between input and recovered temperatures primarily affects the strongest feedback simulations, we retain all systems in the comparison with X-GAP. From an observational standpoint, such discrepancies would not be identifiable, as the true temperatures are not known for real data.

\begin{table*}[]
    \centering
    \caption{Summary statistic about the comparison between X-GAP and various FLAMINGO flavours.}
    \begin{tabular}{|c|c|c|c|c|c|c|}
    \hline
    \hline
       \textbf{Model}  & \textbf{$B_{\rm LT}$} & \textbf{$B_{\rm MgasT}$} & Mean T [keV] & $N_{\rm grps}$ & Mean $\sigma_{\rm v}$ [km/s]& \textbf{Combined significance} \\
       \hline
       \rule{0pt}{2ex}    
        \textbf{X-GAP} & 42.43$_{-0.04}^{+0.04}$ & 12.20$_{-0.02}^{+0.02}$ & 1.30$_{-0.07}^{+0.06}$ & 44 & 433$_{-18}^{+18}$ & -- \\
        L1\_m8 & 42.63$_{-0.02}^{+0.03}$ & 12.26$_{-0.01}^{+0.01}$ & 1.45$_{-0.05}^{+0.05}$ & 47$_{-10}^{+10}$ & 431$_{-15}^{+14}$ & 3.1$\sigma$ \\
        L1\_m9 & 42.63$_{-0.02}^{+0.03}$ & 12.27$_{-0.01}^{+0.01}$ & 1.44$_{-0.06}^{+0.06}$ & 47$_{-11}^{+10}$ & 432$_{-16}^{+14}$ & 3.1$\sigma$ \\
        f$_{\rm gas}+2\sigma$ & 42.92$_{-0.02}^{+0.02}$ & 12.43$_{-0.01}^{+0.01}$ & 1.26$_{-0.05}^{+0.04}$ & 60$_{-11}^{+13}$ & 421$_{-14}^{+13}$ & 3.6$\sigma$ \\
        f$_{\rm gas}-2\sigma$ & 42.42$_{-0.03}^{+0.03}$ & 12.19$_{-0.01}^{+0.01}$ & 1.49$_{-0.06}^{+0.05}$ & 39$_{-9}^{+9}$ & 444$_{-13}^{+12}$ & 0.8$\sigma$ \\
        f$_{\rm gas}-4\sigma$ & 42.28$_{-0.03}^{+0.03}$ & 12.12$_{-0.02}^{+0.02}$ & 1.45$_{-0.06}^{+0.06}$ & 33$_{-7}^{+8}$ & 473$_{-19}^{+18}$ & 2.4$\sigma$ \\
        f$_{\rm gas}-8\sigma$ & 42.05$_{-0.05}^{+0.05}$ & 12.00$_{-0.03}^{+0.02}$ & 1.57$_{-0.09}^{+0.09}$ & 26$_{-6}^{+7}$ & 475$_{-18}^{+20}$ & 4.4$\sigma$ \\
        Jet & 42.83$_{-0.02}^{+0.02}$ & 12.38$_{-0.01}^{+0.01}$ & 1.24$_{-0.06}^{+0.06}$ & 64$_{-13}^{+13}$ & 410$_{-15}^{+13}$ & 3.7$\sigma$ \\
        Jet\_f$_{\rm gas}-4\sigma$ & 42.23$_{-0.05}^{+0.05}$ & 12.11$_{-0.03}^{+0.03}$ & 1.52$_{-0.06}^{+0.06}$ & 27$_{-7}^{+7}$ & 478$_{-14}^{+13}$ & 3.3$\sigma$ \\
        \hline
    \end{tabular}
    \tablefoot{The columns report: the group sample, the normalisation of the core-excised luminosity-temperature relation ($B_{\rm LT}$) and the gas mass-core-excised temperature ($B_{\rm MgasT}$), the mean temperature, the number of groups, the mean velocity dispersion, and the combined tension to X-GAP.}
    \label{tab:summary_stat}
\end{table*}

\begin{figure*}
    \centering
    \includegraphics[width=0.66\columnwidth]{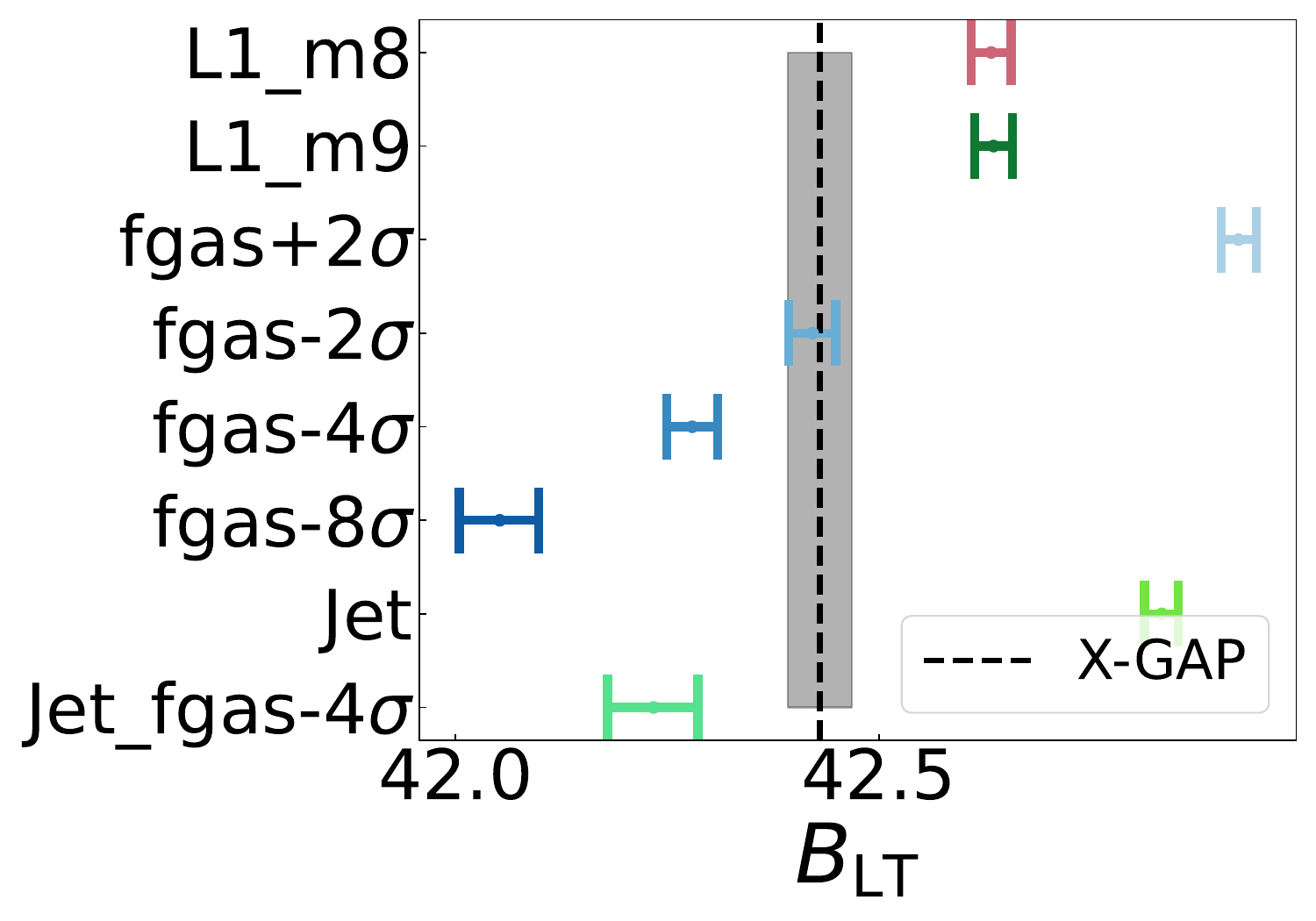}
    \includegraphics[width=0.66\columnwidth]{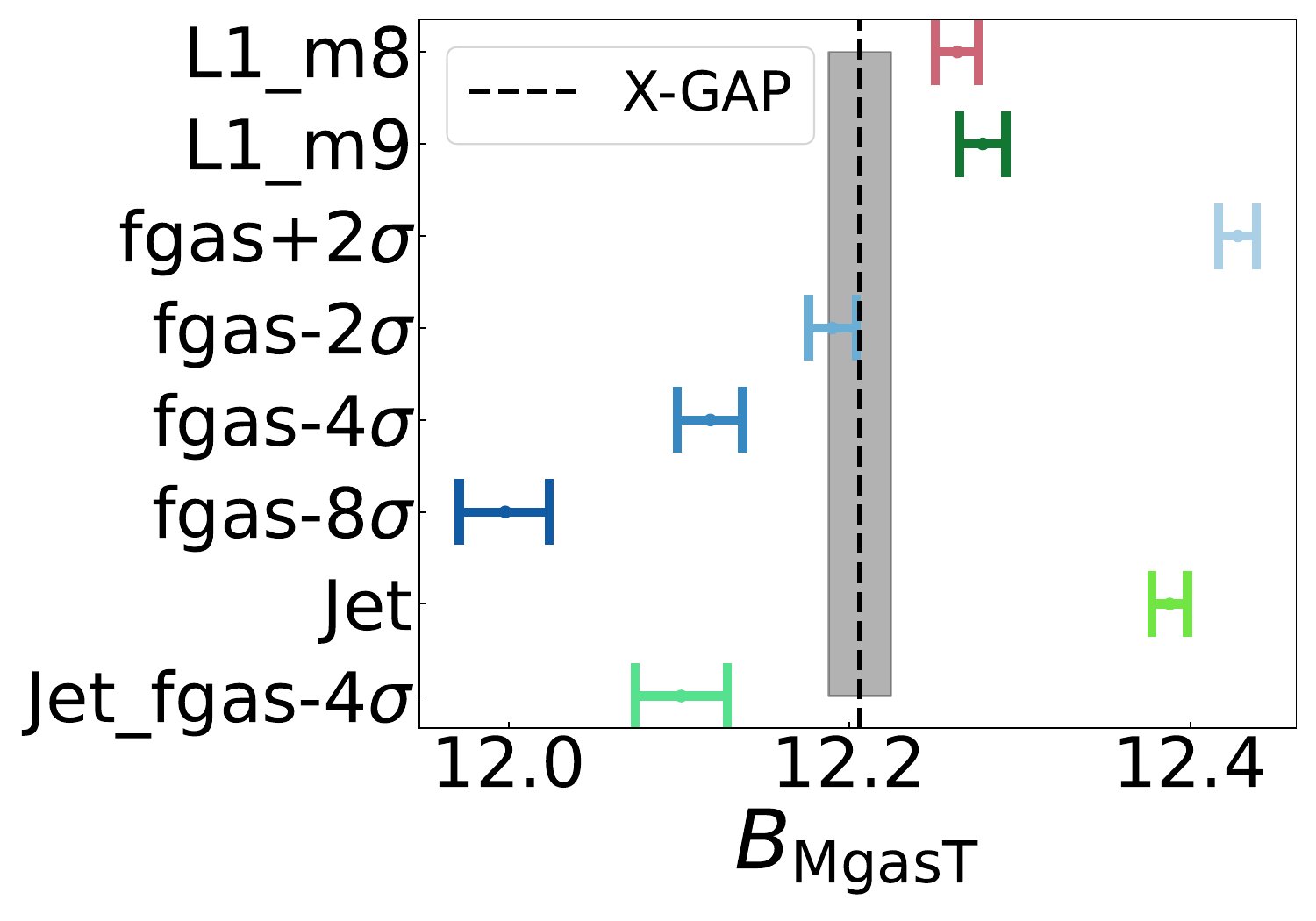}
    \includegraphics[width=0.66\columnwidth]{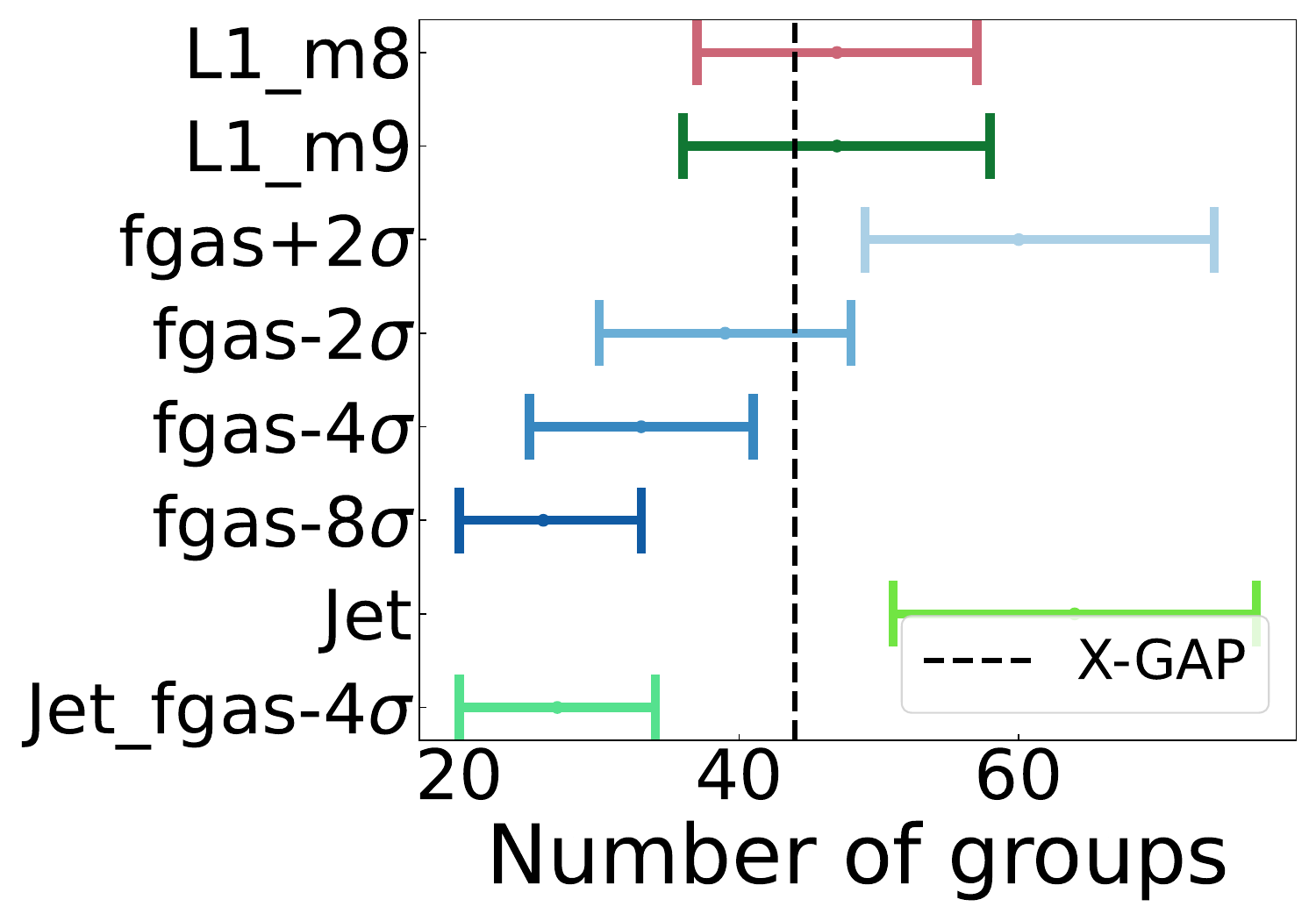}
    \includegraphics[width=0.66\columnwidth]{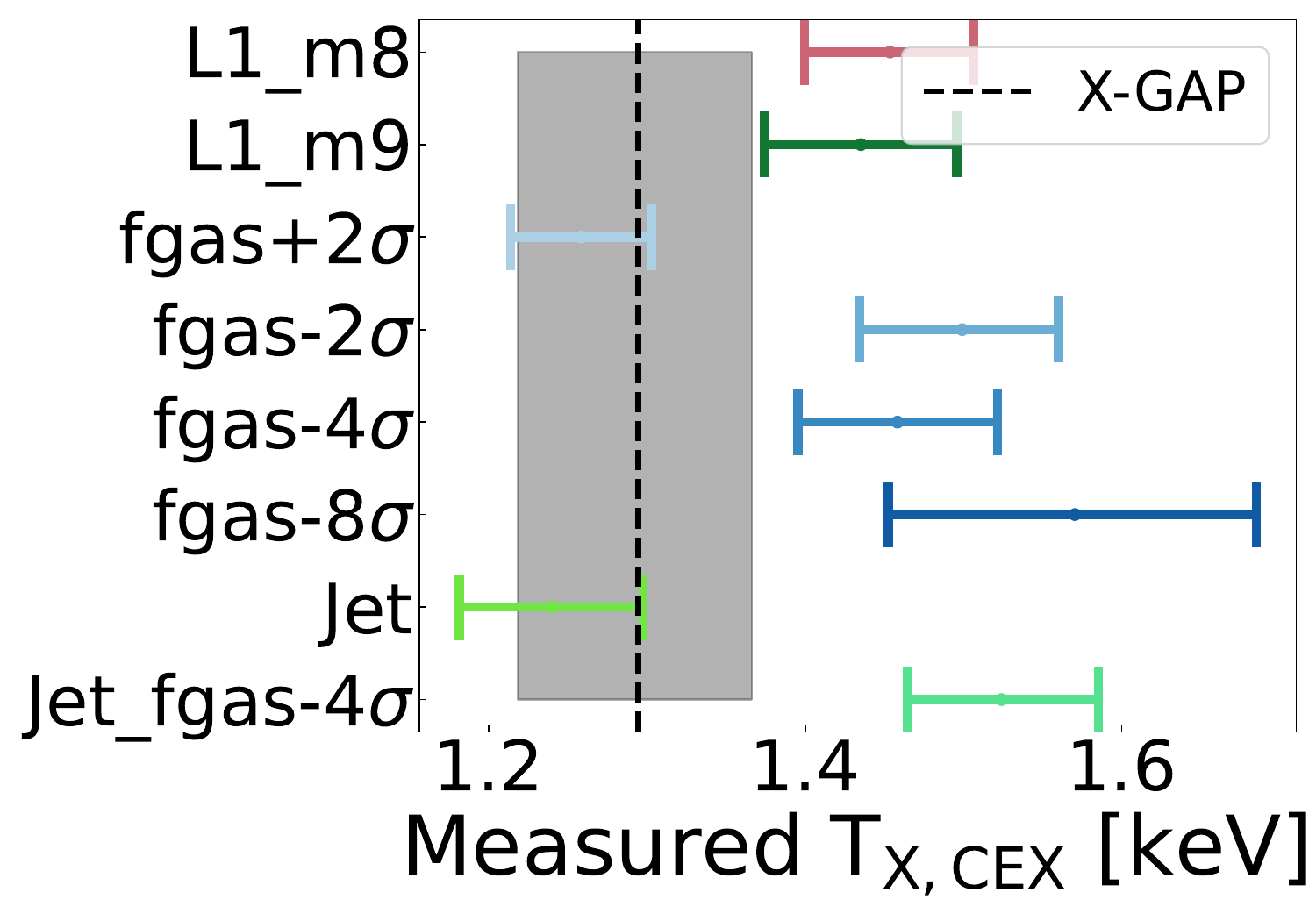}
    \includegraphics[width=0.66\columnwidth]{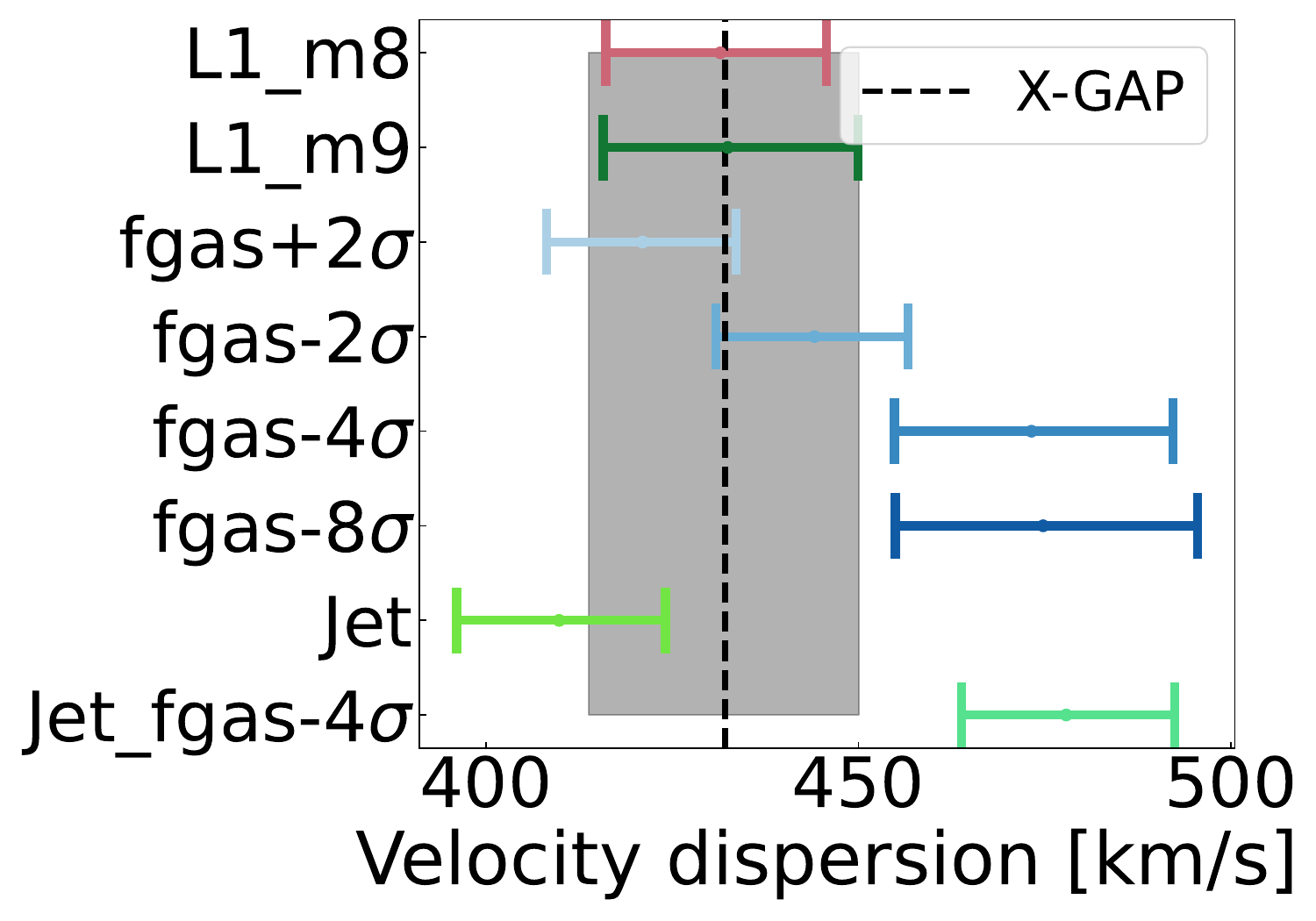}
    \caption{Observables used for the comparison between X-GAP and various FLAMINGO models: the normalisation of the scaling relation between X-ray luminosity and temperature (both core-excised), between gas mass within 400 kpc and core-excised temperature, the total number of groups, the mean temperature and galaxy member velocity dispersion. X-GAP shows the best agreement with the $f_{\rm gas}-2\sigma$ model.}
    \label{fig:observables_for_comparison}
\end{figure*}

The relation between the core-excised X-ray luminosity and temperature obtained in the previous sections is shown in the top panel of Fig. \ref{fig:LT_comparison}. Both quantities are direct observables obtained from XMM-Newton mocks, so that a direct comparison between the FLAMINGO models selected with our selection function and the real X-GAP is possible. All samples occupy a similar region of this parameter space, roughly between 5$\times$10$^{41}$ and 3$\times$10$^{43}$ erg/s and 0.5 and 3.0 keV. However, increasing the strength of AGN feedback shifts groups towards the bottom-right: stronger feedback reduces the gas content of galaxy groups, lowering their X-ray luminosity. As a result, at fixed observed flux the selected systems correspond to intrinsically more massive and hotter haloes. It is important to account for the correlation between these two observables: temperature regulates the mass scale related to the groups selection whereas luminosity encodes the gas retained by these haloes. 

We model the $L$--$T$ relation as a power law of the form $\log L = B_{\rm LT} + \alpha \log\left(\frac{T}{T_{\rm p}}\right)$, where $T_{\rm p}$ is the pivot temperature of 1.3 keV, equivalent to the mean core-excised temperature in X-GAP. In particular, the normalisation $B_{\rm LT}$ reflects the gas content as traced by the X-ray luminosity at a characteristic temperature typical of the systems considered here. Hereafter we will refer to $B_{\rm LT}$ involving core-excised quantities. 
We model the relation using a Bayesian hierarchical framework that accounts for errors in both variables by treating the true temperature as a latent quantity. The fit is performed in base-10 logarithmic space. We adopt uniform priors on the slope (1–4) and normalisation (40–45), and include an intrinsic scatter with a broad half-Gaussian prior up to 0.8 dex. As all samples follow the same selection by construction, the power-law relation can be directly inferred. Posterior sampling is performed using the No-U-Turn sampler implemented in \texttt{PyMC} \citep{pymc2023}.

For X-GAP we obtain $B_{\rm LT}=42.43_{-0.04}^{+0.03}$. The closest FLAMINGO run is $f_{\rm gas}-2\sigma$ with $B_{\rm LT}=42.43_{-0.02}^{+0.03}$. The most extreme cases are $f_{\rm gas}+2\sigma$ with $B_{\rm LT}=42.91_{-0.03}^{+0.03}$ and $f_{\rm gas}-8\sigma$ with $B_{\rm LT}=42.06_{-0.05}^{+0.05}$. We obtain similar results for L1\_m8 and L1\_m9 with $B_{\rm LT}=42.63_{-0.02}^{+0.02}$ (see Table \ref{tab:summary_stat}).

\subsubsection{Gas Mass - Temperature}
\label{subsubsec:MgasT}
We model the gas mass–temperature scaling relation similarly to the $L$–$T$ relation described above. However, the true R$_{\rm 500c}$ used in the input–output comparison (Sect. \ref{subsec:inout}) is not known observationally. We therefore measure the gas mass within a fixed aperture of 400 kpc, corresponding to the upper limit of the integral in Eq. \ref{eq:gas_mass}. This radius lies close to or within R$_{\rm 500c}$ for most selected systems, where the X-ray data provide good signal-to-noise ratio. The same quantity is measured for X-GAP using the best-fit \texttt{hydromass} models \citep[e.g.][and Table \ref{tab:xgap_meas}]{Eckert2025arXiv_letter}.

The result is shown in the bottom panel of Fig. \ref{fig:LT_comparison}. Similarly to the $L$--$T$ relation, increasing the strength of AGN feedback shifts the points to the bottom right direction, ejecting gas and increasing the overall temperature.
We fit the relation with the same method as for $L$--$T$, with priors on the slope (1-5) and normalisation (10-15). For X-GAP we obtain $B_{\rm MgasT}=12.20_{-0.02}^{+0.02}$. The closest FLAMINGO run is $f_{\rm gas}-2\sigma$ with $B_{\rm MgasT}=12.19_{-0.01}^{+0.01}$. The most extreme cases are $f_{\rm gas}+2\sigma$ with $B_{\rm MgasT}=12.42_{-0.01}^{+0.01}$ and $f_{\rm gas}-8\sigma$ with $B_{\rm MgasT}=12.00_{-0.03}^{+0.02}$. We find excellent agreement also for this relation between L1\_m8 ($B_{\rm MgasT}=12.27_{-0.01}^{+0.01}$) and L1\_m9 ($B_{\rm MgasT}=12.26_{-0.01}^{+0.01}$). Finally, we verified in Sect. \ref{appendix:Z03} that $B_{\rm MgasT}$ is recovered correctly independently of the gas particle metallicity, which is high in FLAMINGO \citep[][]{Braspenning2024MNRAS_flamingo}.

\subsection{Combining observables}
\label{subsec:combining_obs}

As we have shown in Sect. \ref{subsec:cosmic_variance}, the number of groups and the median temperature of the samples are potential discriminators between the FLAMINGO models. Here we combine the normalisation of the $L$--$T$ and $M_{\rm gas}$--$T$ relations obtained in the previous section with the number of selected groups, the mean of the measured core-excised temperature, and an estimate of the mean velocity dispersion of the groups.
For each observable, we construct its distribution across the simulation ensemble. We reference this process as the \texttt{MonteCarlo} procedure: for the $L$--$T$ and $M_{\rm gas}$--$T$ normalisations, $B_{\rm LT}$ and $B_{\rm MgasT}$, the distribution is obtained directly from the posterior samples of the \texttt{PyMC} fit described in the previous section.
The temperature distribution is estimated via 1000 bootstrap resamples of the values measured for the end-to-end simulations for each individual model.
Regarding the velocity dispersion, we adopt the calibrated relation between true and measured velocity dispersion derived in \cite{Seppi2025A&A_selfunc}, who performed an end-to-end simulation also in the optical band (see their Sect. 6 and Eqs. 17-18). For each object, a measured velocity dispersion is drawn from the corresponding log-normal distribution defined by that model, using the true velocity dispersion computed in Sect. \ref{subsubsec:flamingo_groupsel}. Similarly to temperature, the distribution is obtained via bootstrap sampling. These distributions include measurement uncertainties on a single mock X-GAP realization. For the number of groups we apply to all FLAMINGO runs the procedure described in Sect. \ref{subsec:cosmic_variance} for L1\_m8. We name this process the \texttt{N-light cone generation}. This provides, for each model, a distribution of the selected number of groups that self-consistently incorporates cosmic variance, Poisson noise, and uncertainties in the selection function.

The results are collected in Fig. \ref{fig:observables_for_comparison}. $B_{\rm LT}$ (top-left panel) shows a preference for the $f_{\rm gas}-2\sigma$ model, the only one that is compatible within 1$\sigma$ with X-GAP at 0.3$\sigma$. The deviations are largest for the $f_{\rm gas}+2\sigma$, Jet, and $f_{\rm gas}-8\sigma$, with tensions at the 11.6, 9.6, and 6.3$\sigma$ levels. $B_{\rm MgasT}$ (top-central panel) shows a preference for the $f_{\rm gas}-2\sigma$ model, again the only one that is compatible within 1$\sigma$ with X-GAP at 0.7$\sigma$. The largest deviations are in the $f_{\rm gas}+2\sigma$, Jet and $f_{\rm gas}-8\sigma$, with tensions at the 10.6, 8.8, and 6.7$\sigma$ levels. The distributions of the number of groups (top-left panel) are broad due to cosmic variance, therefore most models agree within 2$\sigma$ with X-GAP. The vertical dashed line marks the realised X-GAP catalogue size of 44 systems. Unlike the model distributions, which include Poisson fluctuations, cosmic variance, and selection-function uncertainties, this value is not assigned an additional error bar, since it corresponds to the fixed observed sample within the survey footprint. $f_{\rm gas}-8\sigma$ and Jet\_$f_{\rm gas}-4\sigma$ fall above this threshold by 2.7 and 2.4$\sigma$ respectively. The mean temperature of the sample (bottom-left panel) broadly shows a good agreement between models and data, with $f_{\rm gas}+2\sigma$ and Jet being the only ones under-predicting the mean X-GAP temperature, but agreeing with the data within 0.4 and 0.6$\sigma$. Instead $f_{\rm gas}-2\sigma$, $f_{\rm gas}-8\sigma$, Jet\_$f_{\rm gas}-4\sigma$ over-predicting it by 2.1, 2.2, and 2.4$\sigma$. We note that the measured temperature in Jet\_$f_{\rm gas}-4\sigma$ and $f_{\rm gas}-8\sigma$ tends to be underestimated as reported in Fig. \ref{fig:Tin_Tout}, which reduces the constraining power of temperature on its own. In addition, the underlying temperature distribution also depends on the slope of the temperature mass scaling relation in the group regime, which is below the mass range where the FLAMINGO models where calibrated \citep[see][and the discussion in Sect. \ref{sec:conclusion}]{Braspenning2024MNRAS_flamingo}.
Jet\_$f_{\rm gas}-4\sigma$ and $f_{\rm gas}-8\sigma$ show the largest tension for the mean velocity dispersion (bottom-right panel), in tension with the X-GAP measurement by 2.4 and 2.7$\sigma$. Finally, all the observables presented in Fig. \ref{fig:observables_for_comparison} are in excellent agreement between the high resolution L1\_m8 and L1\_m9. This is also a validation for the flagship FLAMINGO simulation L2p8\_m9 that uses the same resolution, but covers a larger box of 2.8 Gpc.

These observables are not statistically independent: for example, hotter systems tend to have higher velocity dispersions, potentially affecting the scaling relation as well as the total number of groups. Therefore, combining the individual p-values would overestimate the overall tension. Instead, we explicitly account for their covariance. In appendix \ref{appendix:statistical_comparison} we provide full details on our method. The final $\sigma_{\rm eq}$ quantifies the statistical significance of the difference between X-GAP and each FLAMINGO model, accounting for the covariance between the observables.

The final result is collected in Table \ref{tab:summary_stat} and shown in Fig. \ref{fig:final_comparison}. We conclude that $f_{\rm gas}-2\sigma$ provides the best comparison to X-GAP, with a global tension of 0.8$\sigma$. Given our framework, this means that there is a 52$\%$ probability of observing X-GAP if $f_{\rm gas}-2\sigma$ is the right model. We find a worse statistical agreement for all other models. $f_{\rm gas}-4\sigma$ is already too ejective, with lower scaling relation normalisations (Fig. \ref{fig:observables_for_comparison}) and at 2.4$\sigma$ global tension. All other models are ruled out with more than 3$\sigma$ confidence, with $f_{\rm gas}-8\sigma$ showing the largest discrepancy at 4.4$\sigma$. In particular, our result is in tension with recent kSZ measurements that prefer $f_{\rm gas}-8\sigma$ \citep[][]{Bigwood2025arXiv_kSZPk, Siegel2025arXiv_kszFGAS, McCarthy2025MNRAS_kSZGGL, Kovac2025JCAP_baryonification}. However, kSZ studies do not necessarily match the scales probed by our galaxy groups (see also discussion in Sect. \ref{subsubsec:comparison_to_lit}).

\section{Discussion}
\label{sec:discussion}
The impact of baryons on dark matter and the LSS is an open question in modern astronomy. In this work we investigated this problem through the impact of AGN feedback on galaxy groups. Here we further discuss our results.

\subsection{Scaling relations and selection}

In Sect. \ref{subsec:combining_obs} and Fig. \ref{fig:observables_for_comparison} we found that FLAMINGO generally tends to overestimate the X-GAP temperatures. The weak-feedback models $f_{\rm gas}+2\sigma$ and Jet are the only ones consistent with the data within 1$\sigma$, while models such as L1\_m8 and $f_{\rm gas}-2\sigma$ show discrepancies at the $\sim$2$\sigma$ level. This behaviour is likely driven by the interplay between the underlying $L_{\rm X}$–$M$ and $T$–$M$ scaling relations and the X-ray flux-limited selection function \citep{Seppi2025A&A_selfunc}. Because the selection effectively operates in the $L_{\rm X}$–$M$ plane, even small changes in the slope of this relation modify the mass distribution of the selected systems, which in turn shifts the mean temperature of the sample.

For instance, \citet{Braspenning2024MNRAS_flamingo} show (their Fig. 2) that the $L$–$M$ relation is slightly flatter in the Jet model compared to the fiducial case. A flatter slope increases the relative luminosity of lower-mass groups \citep[see also the larger gas fractions for the Jet model in Fig. 10 of][]{Schaye2023MNRAS_flamingo}, effectively lowering the mass scale probed by a flux-limited survey. Similarly, their Fig. 3 shows a modest steepening of the $T$–$M$ relation towards the low-mass regime, with most models lying below the fiducial model. The combined effect of these slope variations is therefore non-trivial: shifts in the mass distribution of the selected haloes propagate into shifts in the temperature distribution, even when the intrinsic scaling relations differ only mildly.

Importantly, FLAMINGO was not calibrated in the low-mass group regime, where observational constraints remain sparse. Small differences in the slopes of scaling relations can therefore be amplified by the flux-limited selection, producing systematic shifts in the mean temperature of an X-GAP-like sample.

\subsection{Comparison to other works}
\label{subsubsec:comparison_to_lit}
Our results are broadly consistent with \cite{Eckert2025arXiv_letter}, who compared FLAMINGO and X-GAP with population models and excluded the strongest feedback models. In particular $f_{\rm gas}-8\sigma$ showed a 5.7$\sigma$ tension, and the best agreement was provided by the fiducial feedback model. In this work we refined the analysis with a full end-to-end forward modelling of the observations and accounting for the covariance between observables. We find evidence for stronger feedback than the fiducial model, with $f_{\rm gas}-2\sigma$ showing the lowest tension to X-GAP, but still exclude the strongest model by more than 4$\sigma$. 

Both the results presented in this paper and the ones from \citet{Eckert2025arXiv_letter} are in tension with kSZ measurements combined with X-ray cluster gas masses presented in \cite{McCarthy2025MNRAS_kSZGGL, Kovac2025JCAP_baryonification, Bigwood2025arXiv_kSZPk, Siegel2025arXiv_kszFGAS}, which show a preference for the strongest FLAMINGO feedback schemes. 
A direct comparison with stacked kSZ measurements is, however, non-trivial. X-GAP is located in the local Universe at redshifts less than 0.05, while measurements of the stacked kSZ are sensitive to a much wider mass and redshift range. In addition, the X-ray cluster gas mass fractions may be subject to systematics relative to mass calibration and therefore aperture-related uncertainties on the integrated quantities. These differences suggest that the apparent tension may reflect a combination of redshift evolution, selection effects, and measurement systematics rather than a single discrepancy in feedback strength.

Altogether, these results show that we currently lack a complete understanding of AGN triggering throughout cosmic history. For example, \cite{Costello2025_BHflamingo} used the FLAMINGO simulations and found an anti-correlation between BH mass and gas fraction in the galaxy regime, however this reverses and becomes positive for galaxy groups: more massive haloes form early, host more massive BHs, eject quickly and reaccrete later, increasing the gas fractions. Conversely, \cite{Marini2025A&A_assemblyhistory} do not find evidence for such a trend in the Magneticum simulation. This hints to the possibility that the exact feedback evolution history remains unclear.

A variation of the AGN feedback efficiency across cosmic time, for example due to environmental effects, may change group properties as a function of redshift. In addition, the exact interplay between AGN feedback and the surrounding medium, both in thermal, kinetic, and jet modes, has an impact on the duty cycle \citep{Sotira2025A&A_AGNturbulence, Sotira2026arXiv_coldgas}. The shape of the f$_{\rm gas}$-M relation at the group regime likely explains the differences between the fiducial models and the Jet run. Although both models are calibrated to match the same gas fraction at cluster scales, the extrapolation toward the group regime probed in this work predicts higher gas fractions for the Jet model than for the fiducial case, while they converge at higher masses \citep[Fig. 10 in][]{Schaye2023MNRAS_flamingo}. This highlights that different prescriptions for AGN feedback as a function of halo mass can have distinct effects depending on the halo population under consideration (groups, clusters, or galaxies) which may also contribute to tensions with studies based on the kSZ effect.

\subsection{Impact on cosmology}
\label{subsubsec:cosmo_impact}
On top of all observables used in this work, together with kSZ and gas fractions, another useful feedback probe is the cross correlation between X-ray photons and cosmic shear, that can potentially discriminate between FLAMINGO models \citep[][]{McDonald2026arXiv_flamingo_xrayshear}. Disentangling all these effects in comparison to a local sample such as X-GAP requires deep, high-resolution observations of galaxy groups, a particularly relevant outlook in sight of NewAthena \citep[][]{Cruise2025NatAs_newAthena}. 

Finally, the relevance of group-scale feedback extends beyond halo astrophysics. The distance up to which one would find all the baryons ejected by feedback, effectively recovering the universal baryon fraction, depends on feedback intensity and efficiency. The closure radius, i.e. the radius at which the enclosed baryon fraction reaches the cosmic mean, was shown to vary from one to several R$_{\rm 200c}$ for group-scale systems  \cite{Ayromlou2023MNRAS_baryonsTNG}, which implies a substantial decoupling between baryons and dark matter. This can be probed by large scale X-ray observations of galaxy groups at low redshifts. The redistribution of baryons modifies the matter distribution and suppresses small-scale power in the matter power spectrum \citep[][]{vanDaalen2011MNRAS.415.3649V, Schneider2019JCAP_Pksupp, vanDaalen2020MNRAS.491.2424V, Gebhardt2026arXiv260106258G}. The suppression is mainly driven by the shape of the gas density profiles of galaxy groups, such that deep observations of local groups are highly complementary to LSS cosmological probes \citep[see][]{Debackere2020MNRAS_Pk, Mead2021MNRAS_HMcode, Grandis2024MNRAS_pk, Schneider2025JCAP_baryonification, Schaller2025MNRAS_FLA_Pk}. This has relevant consequences on cosmological experiments on the LSS \citep[][]{Semboloni2011MNRAS_WLbaryons, Arico'2023A&A_DES_cosmobaryons}, meaning that a deeper understanding of feedback paves the way towards a more accurate and robust knowledge of cosmology \citep[see e.g.][]{DESY62026arXiv260114559D}.

\section{Conclusion}
\label{sec:conclusion}

In this section, we provide a summary of this work and a future outlook.

\subsection{Summary}

In this work we investigated the impact of AGN feedback on the hot atmospheres of galaxy groups by comparing simulations and observations.  We used the FLAMINGO suite, which includes several AGN feedback prescriptions, and compared them with the X-GAP group sample. Selection effects were taken into account using the detection probability derived in \cite{Seppi2025A&A_selfunc}, which forward-modelled the RASS detection process. We found cosmic variance to be a large source of systematic for what concerns the number of selected groups, with relative variations larger than 20$\%$ (see Fig. \ref{fig:Number_of_groups}and also Appendix \ref{subsec:inout}). Therefore, the number of groups is not a driver in the comparison between data and models in this case, due to the relatively small volume (about 0.07 Gpc$^3$) probed by X-GAP. 

We further applied the X-GAP selection criteria of at least eight galaxy members and R$_{\rm 500c}$<15$'$ to construct simulated X-GAP analogues. We then generated end-to-end XMM-Newton mock observations, including instrumental effects, and analysed them with the same imaging and spectral procedures used for the real data (Fig. \ref{fig:workflow}).

We then analyse the X-ray simulation with the same method as for the real X-GAP sample \citep[][]{Eckert2025A&A_4436}. This framework enabled a direct comparison between simulations and observations in observable space, where temperature traces the halo mass scale and X-ray luminosity probes the gas content (Fig. \ref{fig:LT_comparison}). 

We used the $L$--$T$, $M_{\rm gas}$--$T$ normalisations, mean temperature, mean velocity dispersion, and the number of detected groups as discriminating observables, accounting for their covariance with a Gaussian copula approach (Fig. \ref{fig:final_comparison}). The $f_{\rm gas}-2\sigma$ model provides the best agreement with X-GAP, deviating by only $0.8\sigma$, while the strongest feedback model ($f_{\rm gas}-8\sigma$) is excluded at $4.4\sigma$. The $f_{\rm gas}-2\sigma$ model provides an X-GAP-like population with a typical gas fraction of 6.1\%, while the $f_{\rm gas}-8\sigma$ yields emptier groups with a typical gas fraction of 3.8\% (see Sect. \ref{subsec:hydro_runs}). Overall, our end-to-end framework provides a direct constraint on the fate of hot gas in galaxy groups, favouring a feedback scheme that is stronger than the fiducial FLAMINGO calibration but weaker than the most ejective scenarios.

\subsection{Outlook}
In future work, we plan to further expand the comparison to more simulations, to enrich the comparison including jets in this article. In this context, the coupling between the IGrM and AGN feedback is a function of the distance to the central SMBH: different recipes in hydrodynamical simulations may cause variations of the efficiency to which AGN feedback impacts the hot medium. This information is encoded in the shape of the entropy profiles, which provides a promising avenue for future tests.

Simulations at higher resolution such as COLIBRE \citep[][]{Schaye2025arXiv_colibre}, in addition to well resolved AGN jets \citep[][]{Bourne2023Galax_feedback} will help shed light on this topic. For example, \cite{Correa2026MNRAS_cSBFLAMINGO} showed that surface brightness concentration can be another discriminator between FLAMINGO models. This is directly related to the slope of the profile $\beta$. However, at the current resolution level the inner core of X-GAP-like groups is resolved with only a few tens of gas elements, and core measurements are not fully reliable from the observer perspective. This could be achieved with more massive clusters in FLAMINGO where the inner 0.1$\times$R$_{\rm 500c}$ is resolved with hundreds of gas elements, but the AGN impact on the hot ICM is milder.

Finally, our framework will allow us to correct group mass measurements in the real X-GAP and therefore also their relation to the gas fraction, accounting for their covariance, measurement uncertainties and systematics, and selection effects. 

\begin{acknowledgements}
RS and DE are supported by Swiss National Science Foundation project grant \#200021\_212576. This work used the DiRAC@Durham facility managed by the Institute for Computational Cosmology on behalf of the STFC DiRAC HPC Facility (www.dirac.ac.uk). The equipment was funded by BEIS capital funding via STFC capital grants ST/P002293/1, ST/R002371/1 and ST/S002502/1, Durham University and STFC operations grant ST/R000832/1. DiRAC is part of the National e-Infrastructure. M.A.B. is supported by a UKRI Stephen Hawking Fellowship (EP/X04257X/1). LL acknowledges support from INAF grant 1.05.24.05.15. The authors thank the anonymous referee for their helpful comments on this article
\end{acknowledgements}

\bibliographystyle{aa}
\bibliography{biblio}

\appendix

\section{Modelling the N$_{\rm gal}$ selection}
\label{appendix:Ngalsel}

\begin{figure}
    \centering
    \includegraphics[width=0.75\columnwidth]{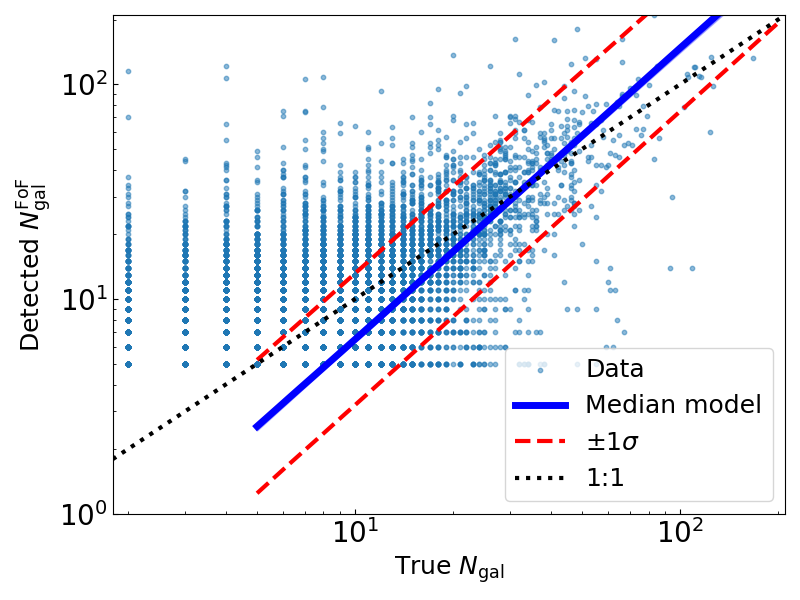}
    \includegraphics[width=0.75\columnwidth]{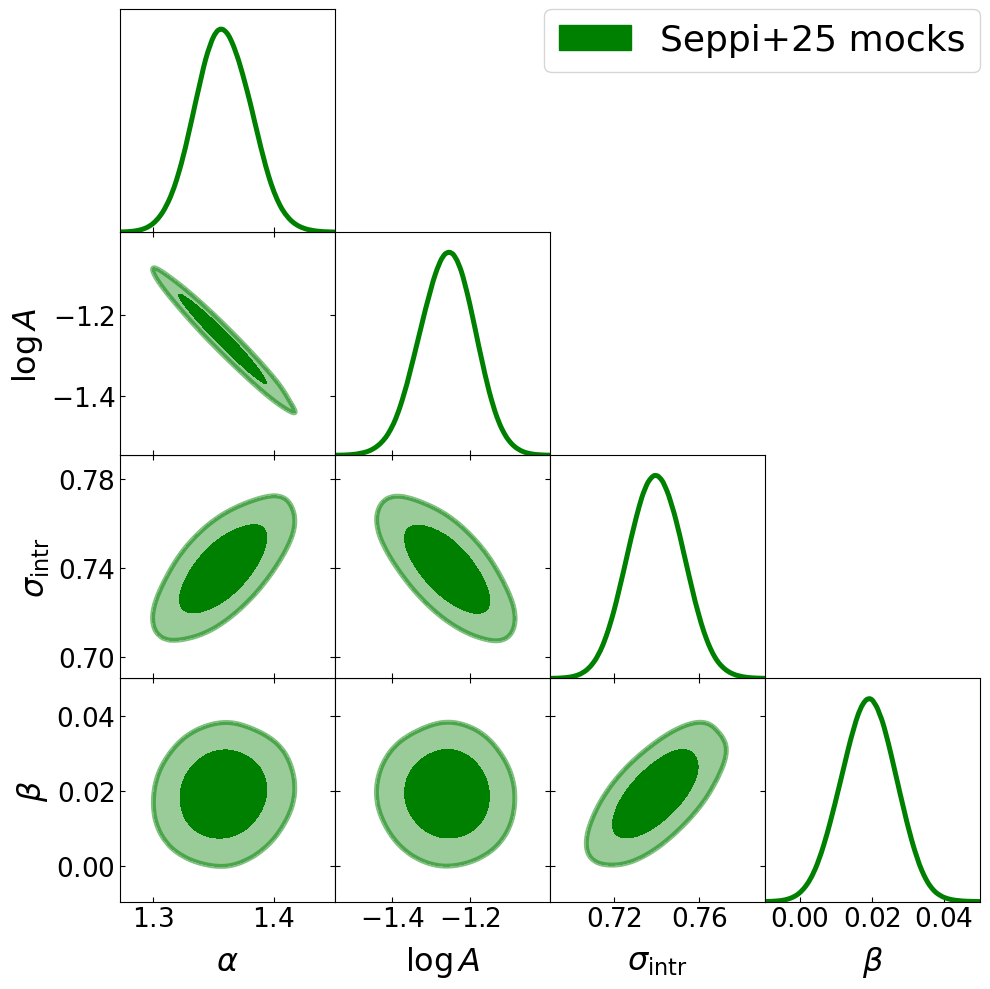}
    \caption{Relation between the number of true galaxy members and the ones identified by the optical finding process. \textbf{Top panel}: data from the simulations developed by \cite{Seppi2025A&A_selfunc} and the best fit model described in Eq. \ref{eq:Ngal_relation}. \textbf{Bottom panel}: Marginalized 1-D posterior distribution of the best fit parameters.}
    \label{fig:Ngal_corr}
\end{figure}

X-GAP is selected applying a lower cut equal to eight group members identified in SDSS. In this appendix we elaborate on the modelling of this cut, an ingredient necessary to complete the sample selection (Sect. \ref{subsubsec:flamingo_groupsel}).
To reproduce the $N_{\rm gal, SDSS}$ selection we retrieve the r-band luminosity and convert it to apparent magnitude according Eq. \ref{eq:rmag}:
\begin{align}
    M_r &= -2.5\times \log_{10}\frac{L_r}{L_{r0}} \nonumber \\
    K_r &= -2.5\times \log_{10}(\frac{1}{1+z}) \nonumber \\
    m_r &= M_r + 5\log_{10}[D_L/pc] - 5 + K_r, 
    \label{eq:rmag}
\end{align}
where $L_r$ is the optical r-band luminosity, and $L_r$ is the reference luminosity corresponding to 4.64 Mag, 
and $D_L$ is the luminosity distance in units of parsec.
Given the low redshift of our sample, the K-correction is small and we assume a flat galaxy SED in Eq. \ref{eq:rmag} \citep[see][Sect. 4 and Eq. 15]{Blanton2007AJ_Kcorr}. For each halo we count members with $m_r<17.77$. Although L1\_m8 misses galaxies below M$_\star\sim10^8$ M$_\odot$, these are mostly isolated, leading to a mild deficit in the total galaxy number density at $z<0.03$ compared to the UchuuSDSS mocks \citep{Seppi2025A&A_selfunc}. However, the number density of group members is consistent and requires no correction.  The simulated luminosities correspond to integrated stellar emission and therefore approximate, but do not necessarily exactly reproduce, SDSS Petrosian magnitudes. At the low redshifts considered here ($z<0.05$), such differences are expected to be subdominant relative to the empirical calibration between intrinsic and FoF-detected group membership described below.
We then account for the mismatch between true members and those identified by the FoF finder using the end-to-end mocks of \cite{Seppi2025A&A_selfunc}. These are used to calibrate the relation between the true number of members $N_{\rm gal}$ and the detected number $N_{\rm g,FoF}$. Since only systems with $N_{\rm g,FoF}\geq 5$ are retained, we model this relation with a truncated likelihood, i.e. the probability of observing $N_{\rm g,FoF}$ given $N_{\rm gal}$ and the selection cut. The model depends on four parameters: $A$, $\alpha$, $\sigma_{\rm intr}$, and $\beta$:
\begin{align}
    &\log \mu = \log A + \alpha \log N_{\rm gal}, \nonumber \\
    &\sigma_{\rm ToT} = \sigma_0 \cdot N_{\rm gal}^{-\beta}, \nonumber \\
    &P(\log N_{\rm g,FoF} | \log \mu) = \mathcal{N}(\log N_{\rm g,FoF} | \log \mu, \sigma_{\rm ToT}), \nonumber \\ 
    &P(N_{\rm g,FoF} \geq 5 | \log \mu) = \int_{\log 5}^{\infty} \mathcal{N}(\log y' | \log \mu, \sigma_{\rm ToT}) \, d\log y' = \nonumber \\
    &= 1 - \Phi\left( \frac{\log 5 - \log \mu}{\sigma_{\rm tot}} \right), \nonumber \\
    &\mathcal{L}_i = P(\log N_{\rm g,FoF}^{(i)}| \log \mu, N_{\rm g,FoF}^{(i)} \geq 5) = \frac{P(\log N_{\rm g,FoF}^{(i)} | \log \mu^{(i)})}{P(N_{\rm g,FoF}^{(i)} \geq 5 | \log \mu^{(i)})}, \nonumber \\
    &\log \mathcal{L}_i = \log \left[ P(\log N_{\rm g,FoF}^{(i)} | \log \mu^{(i)}) \right]
    - \log \left[ P(N_{\rm g,FoF}^{(i)} \geq 5 | \log \mu^{(i)}) \right], \nonumber \\ 
    &\log \mathcal{L} = \sum_{i=1}^{N_{\rm groups}} \log \mathcal{L}_i,
    \label{eq:Ngal_relation}
\end{align}
where $\mu$ is the mean expected $N_{\rm g,FoF}$ at fixed $N_{\rm gal}$, $\mathcal{N}$ is a normal distribution, $\Phi$ is the cumulative normal distribution function and we assumed that $P(\log N_{\rm g,FoF}^{(i)}, N_{\rm g,FoF} \geq 5 | \log \mu^{(i)}) = P(\log N_{\rm g,FoF}^{(i)} | \log \mu^{(i)})$, because since the group is in the catalogue we know that is has more than 5 detected members.
The relation between the true number of galaxy members, $N_{\rm gal}$, and the FoF-detected number, $N_{\rm g,FoF}$, is modelled probabilistically, accounting for intrinsic scatter and selection effects. We assume a log-normal distribution for $N_{\rm g,FoF}$ conditioned on $N_{\rm gal}$, with a scatter in log-space that varies with richness, reflecting larger relative fluctuations in poorer systems. 
We account for the detection threshold $N_{\rm g,FoF}\geq 5$ using a truncated log-normal likelihood, including the survival probability $P(N_{\rm g,FoF}^{(i)} \geq 5 \mid N_{\rm gal}^{(i)})$. The total log-likelihood is summed over all groups. This framework naturally incorporates both the selection cut and measurement scatter.

\begin{table}[]
    \centering
    \caption{Priors and posteriors of the relation between the number of true galaxy members and the ones recovered by the FoF.}
    
    \begin{tabular}{|c|c|c|}
    \hline
    \hline
       \textbf{Parameter}  & \textbf{Prior} & \textbf{Posterior} \\
       \hline
       \rule{0pt}{2ex}    
        $\alpha$ & $\mathcal{U}$(0, 2) & 1.36 $\pm$ 0.03  \\
        $\log A$ & $\mathcal{U}$(-5, 5) & -1.26 $\pm$ 0.07 \\
        $\sigma_{\rm intr}$ & $\mathcal{U}$(0.01, 5.0) & 0.74 $\pm$ 0.01 \\
        $\beta$ & $\mathcal{U}$(-5, 5) & 0.02 $\pm$ 0.01 \\
        \hline
    \end{tabular}
    \tablefoot{The symbol $\mathcal{U}(M,N)$ denotes a uniform prior between the values M and N. Posterior values are reported from the third column onward, for each case labelled in the top row.}
    \label{tab:ngal_rel_pars}
\end{table}

We apply the correction between number of true members and optical members using the best fit model in Eq. \ref{eq:Ngal_relation} using the parameters in Table \ref{tab:ngal_rel_pars}. This gives us a predicted number of optically identified galaxy members starting from the number of true galaxy members accounting for the full observational selection process \citep[][]{Damsted2024_axes, Eckert2024_xgap, Seppi2025A&A_selfunc}.

Fig. \ref{fig:Ngal_corr} shows the final result, with the relation between true and detected number of members, along with the best-fit model described above. The bottom panel shows the full marginalised posterior distributions for the parameters in Eq. \ref{eq:Ngal_relation}.

\section{X-GAP-FLAMINGO validation}
\label{subsec:inout}

In this appendix, we quantify the impact of cosmic variance on the selected X-GAP-like samples across the various FLAMINGO models and validate the recovery of intrinsic X-ray observables through our analysis pipeline.

\begin{figure}
    \centering
    \includegraphics[width=\columnwidth]{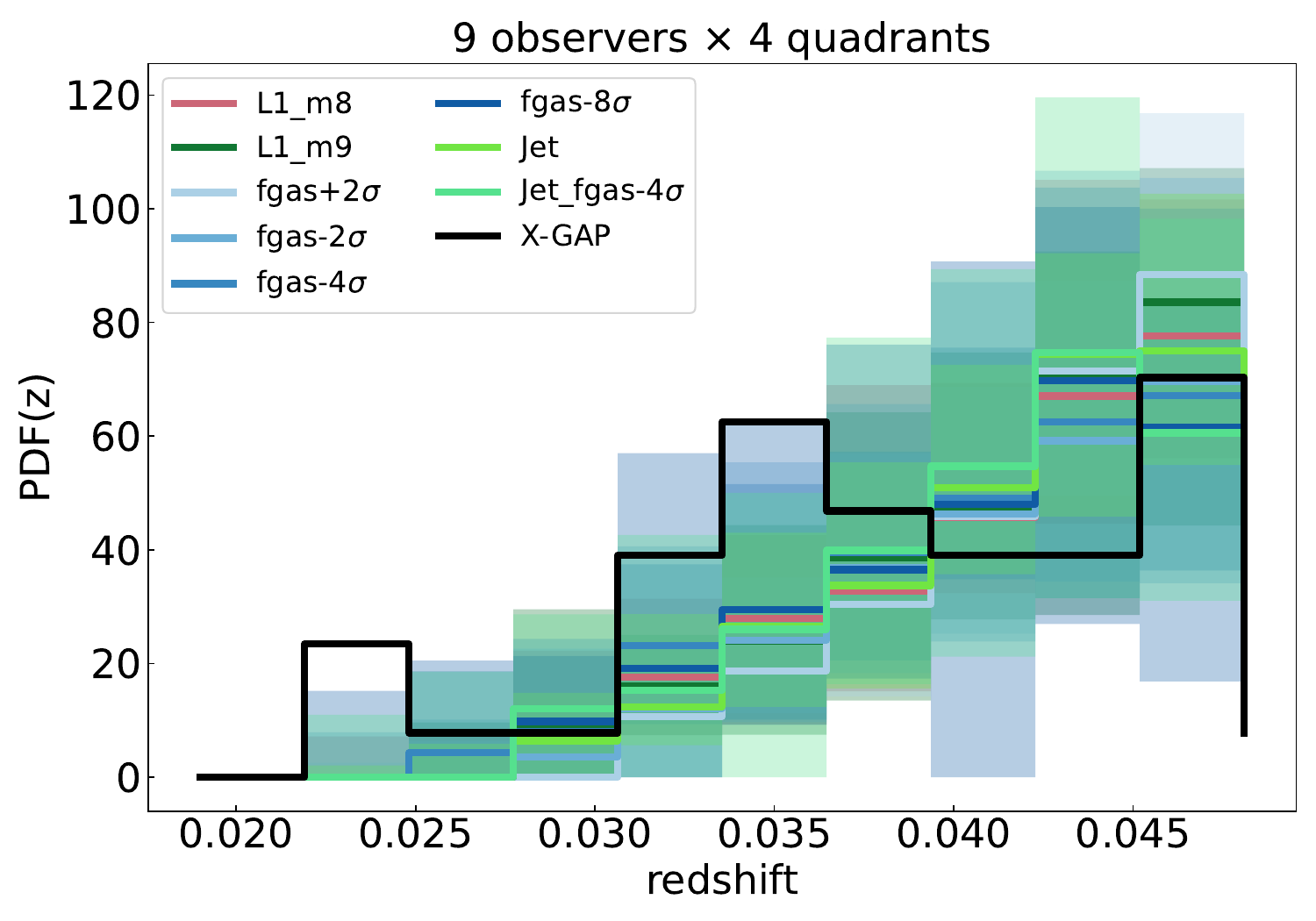}
    \includegraphics[width=\columnwidth]{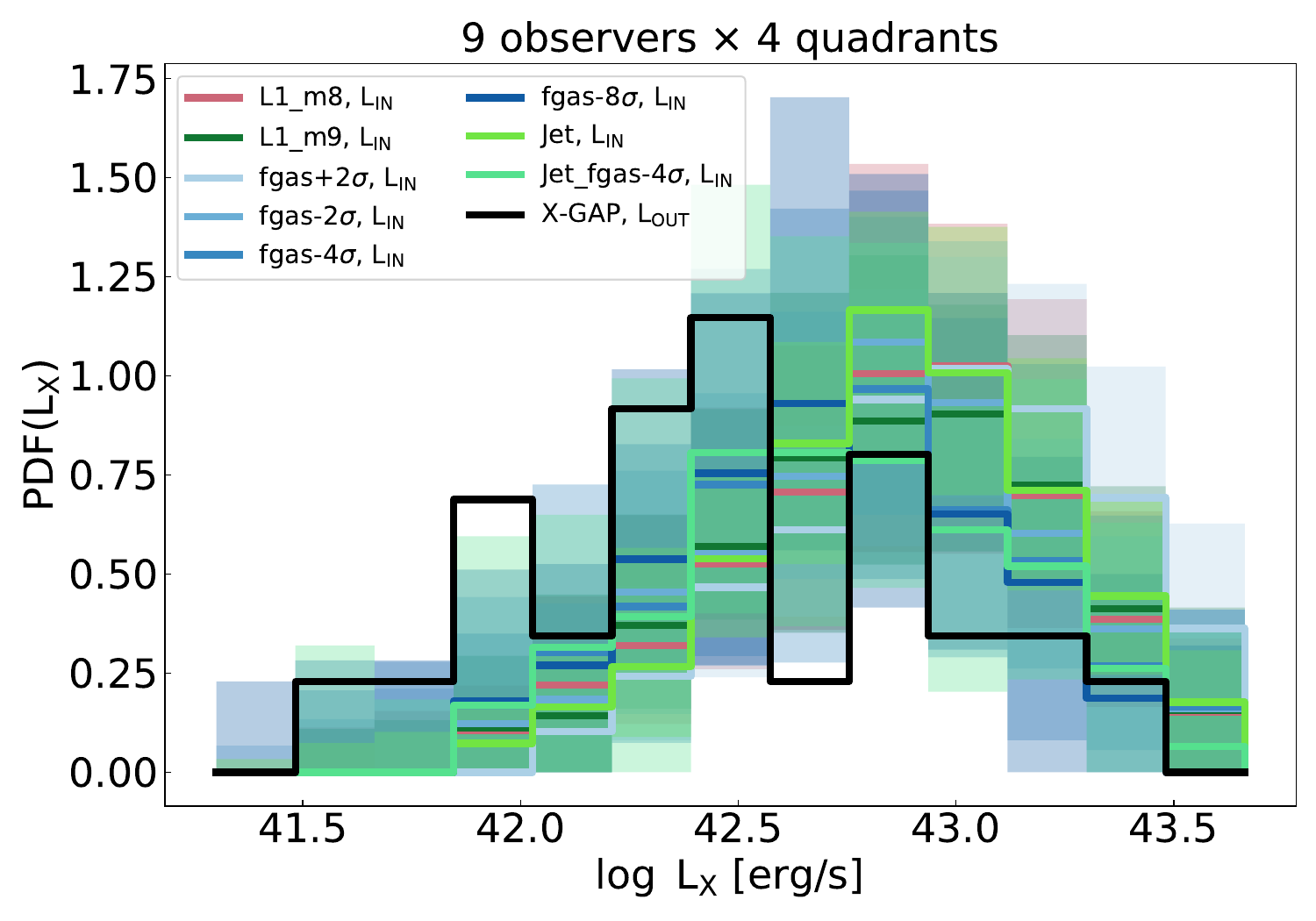}
    \caption{Redshift and luminosity distribution in the selected FLAMINGO samples and X-GAP.}
    \label{fig:PDF_Lx_z}
\end{figure}

\begin{figure}
    \centering
    \includegraphics[width=\columnwidth]{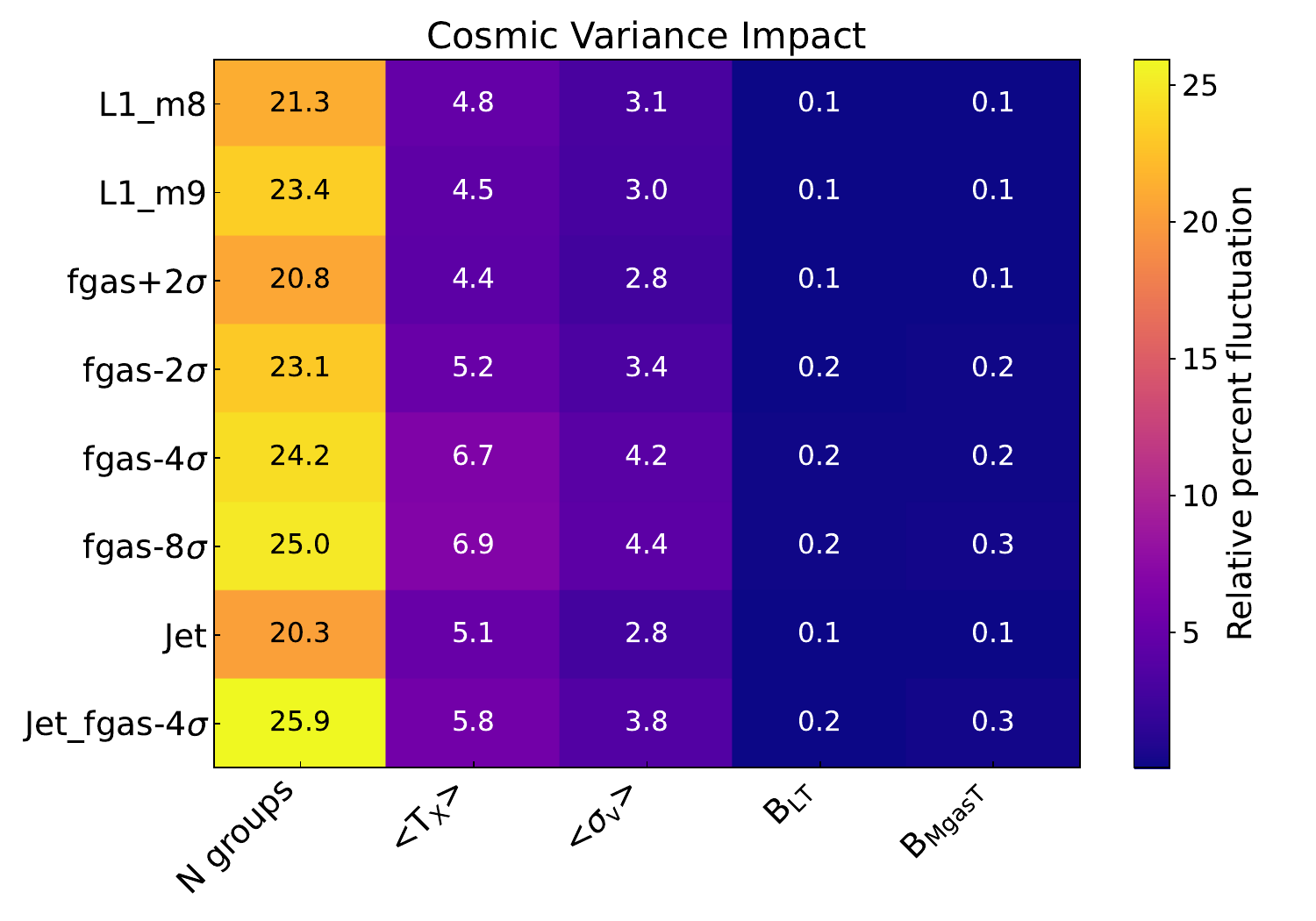}
    \caption{Relative fractional variation of each observable estimated from the cosmic variance tests.}
    \label{fig:cosmicvar_impact_OBS}
\end{figure}

After applying the selection described in Sect. \ref{subsubsec:flamingo_groupsel}, we compare the redshift and luminosity PDFs of the selected groups to X-GAP (Fig. \ref{fig:PDF_Lx_z}), where shaded regions indicate the 16--84th percentiles across observers and light cones.
X-GAP appears over(under)-dense at $z\sim0.035$ ($z\sim0.043$), leading to an excess (deficit) of systems around $L_{\rm X}\sim2\times10^{42}$ erg s$^{-1}$ ($\sim10^{43}$ erg s$^{-1}$), consistent with the flux-limited selection. As shown in Fig. \ref{fig:Number_of_groups}, such variations are dominated by cosmic variance, and it is unlikely that FLAMINGO reproduces the local Universe exactly. While large-scale structure fluctuations may affect observable distributions, this is mitigated by the copula approach (Sect. \ref{subsec:combining_obs}; Appendix \ref{appendix:statistical_comparison}). Moreover, the scaling-relation normalisations (M$_{\rm gas}$, L$_{\rm X,CEX}$), evaluated at fixed temperature, are largely unaffected and dominate the constraints.

We verify this using the cosmic variance and selection-function tests of Sect. \ref{subsec:cosmic_variance}. Relative variations, defined as $(p_{84}-p_{16})/(2\,p_{50})$, are shown in Fig. \ref{fig:cosmicvar_impact_OBS}. The number of groups varies by $>20\%$, while mean temperature and velocity dispersion vary at the 5\% and 4\% level, respectively. The scaling-relation normalisations at 1.3 keV are highly stable, with variations of 0.2\%.

\begin{figure}
    \centering
    \includegraphics[width=\columnwidth]{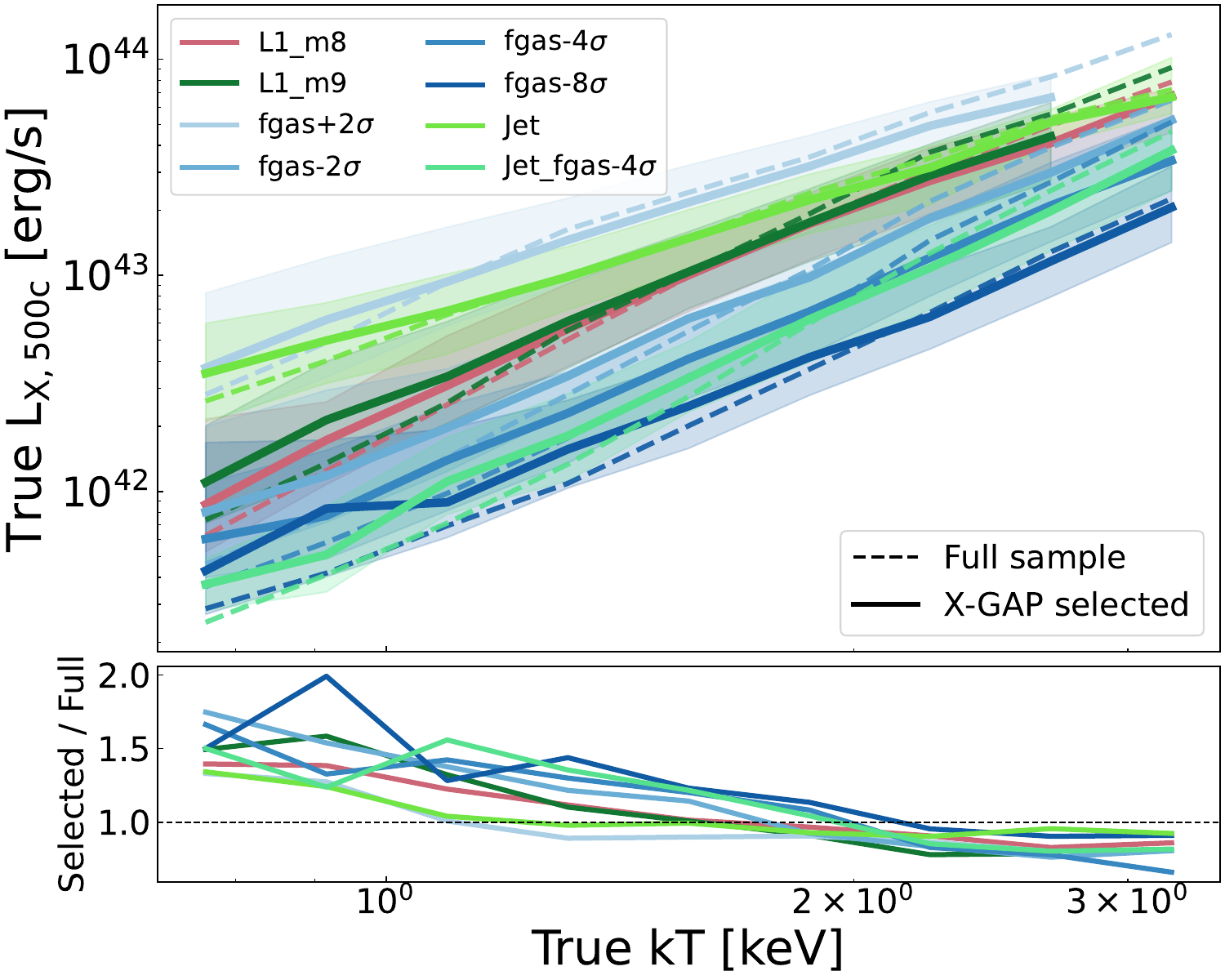}
    \caption{Luminosity-temperature relation for the selected systems (solid lines and shaded areas) and the full sample (dashed lines). Both are true input quantities.}
    \label{fig:LTselection}
\end{figure}

We also examine the impact of selection on the $L$--$T$ relation, analogous to the $f_{\rm gas}$--$M$ relation shown in Fig. \ref{fig:fgas}. The result is presented in Fig. \ref{fig:LTselection}. The selection bias follows a similar pattern: incompleteness at low masses and a down-scattered population at high masses caused by the upper radius cut. However, the effect is more pronounced for the $L$--$T$ relation because gas fraction is less directly tied to the selection observable (X-ray flux). We measure deviations of $\sim30\%$ around 1 keV and $\sim10\%$ near 3 keV.

\begin{figure}
    \centering
    \includegraphics[width=\columnwidth]{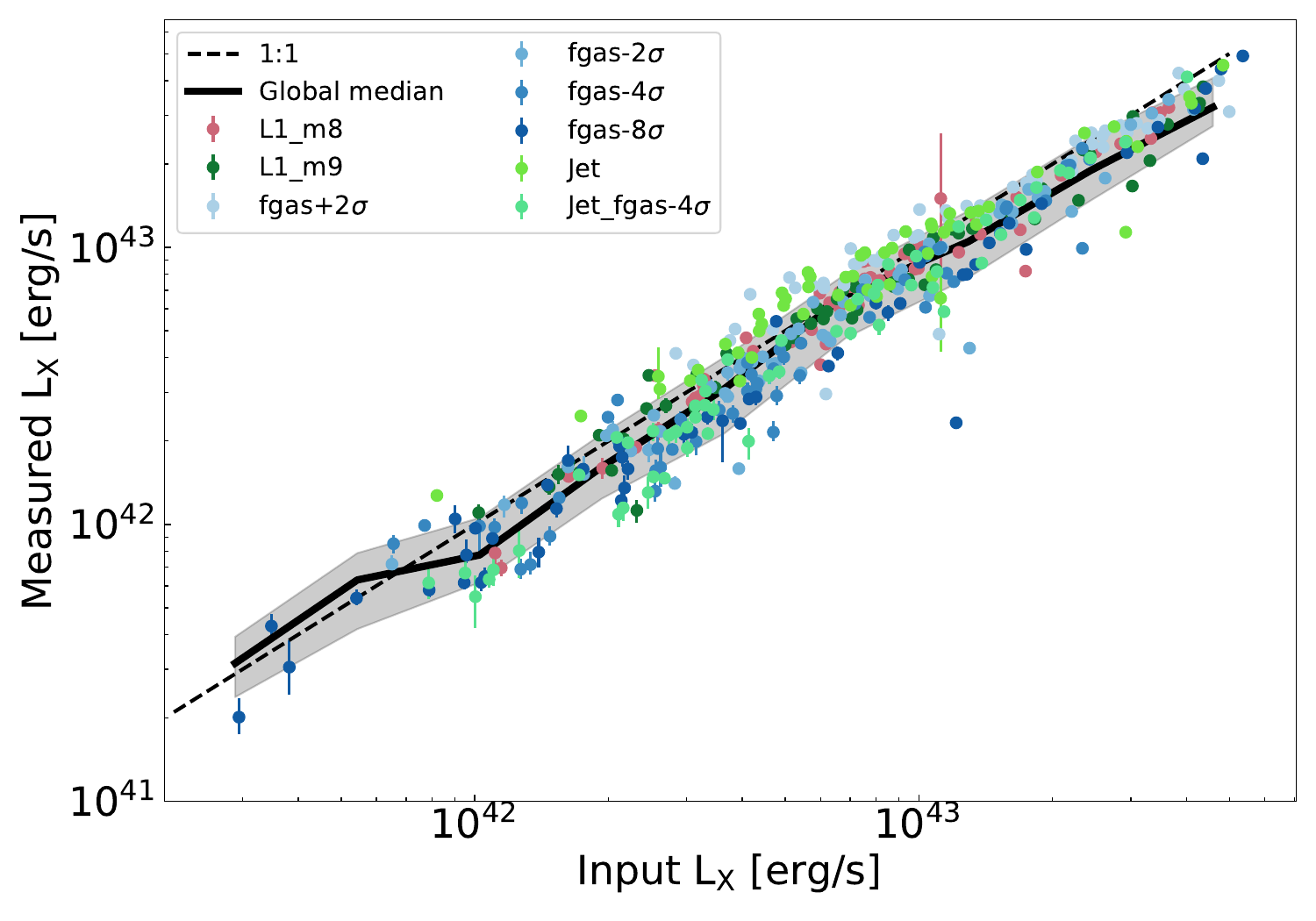}
    \caption{Comparison between the input X-ray luminosity and the output luminosity measured and modelled with \texttt{hydromass}.}
    \label{fig:Lin_Lout}
\end{figure}

We now focus the recovery of input X-ray luminosities, temperatures, and gas masses in the FLAMINGO groups.  

\subsection{X-ray luminosity}
\label{appendix:LinLout}

In our framework, X-ray luminosity is not computed as a simple photon count within an aperture. Instead, \texttt{hydromass} jointly models the gas density and temperature profiles to reproduce the observed surface brightness, which depends on the line-of-sight integral of $n_e^2$ and on temperature and metallicity through the cooling function \citep{Ploeckinger2020MNRAS_coolfunc}. 
Once the surface brightness and projected temperatures are fitted, the resulting 3D profiles yield the plasma emissivity, which is integrated along the line of sight to obtain the luminosity. The integration radius is critical, particularly for groups with flatter profiles, as it directly impacts the total luminosity. We therefore adapt \texttt{hydromass} to integrate out to the true R$_{\rm 500c}$ rather than the value inferred from the data.

We compare our measured luminosities to the ones tabulated by \citet{Braspenning2024MNRAS_flamingo} (see their Sect. 2.3, 2.4), and rescaled by the cylindrical correction $C_{\rm cyl}$ obtained in Sect.\ref{subsubsec:flamingo_groupsel}, i.e. the ones used for the selection function and as inputs to the \texttt{xmm\_simulator}. The result is shown in Fig. \ref{fig:Lin_Lout}. The black line shows the global measured luminosity at a given true value. The shaded area includes the 16th-84th percentiles. Although the solid lines appears to be just below the 1:1, with a global median that is 6$\%$ lower than the input value, the scatter on the median is well within the one-to-one line. We conclude that we do not find evidence for significant bias in the luminosity measurement. It means that any difference in the computation of the cooling rates using \texttt{cloudy} \citep[in][]{Braspenning2024MNRAS_flamingo} or \texttt{apec} (this work) does not impact the measurement of X-ray luminosity in our framework. In addition, a slight offset and a portion of the observed scatter may originate from the cylindrical correction (see Eq. \ref{eq:spherical_cylindircal_integrals}). The correction is applied to the reference true luminosities, whereas the reconstructed values are derived independently from the mock photon data. Therefore, any inaccuracy in the correction would manifest as a departure from the one-to-one relation. The close agreement observed here suggests that the cylindrical correction is indeed robust.

\begin{figure}
    \centering
    \includegraphics[width=\columnwidth]{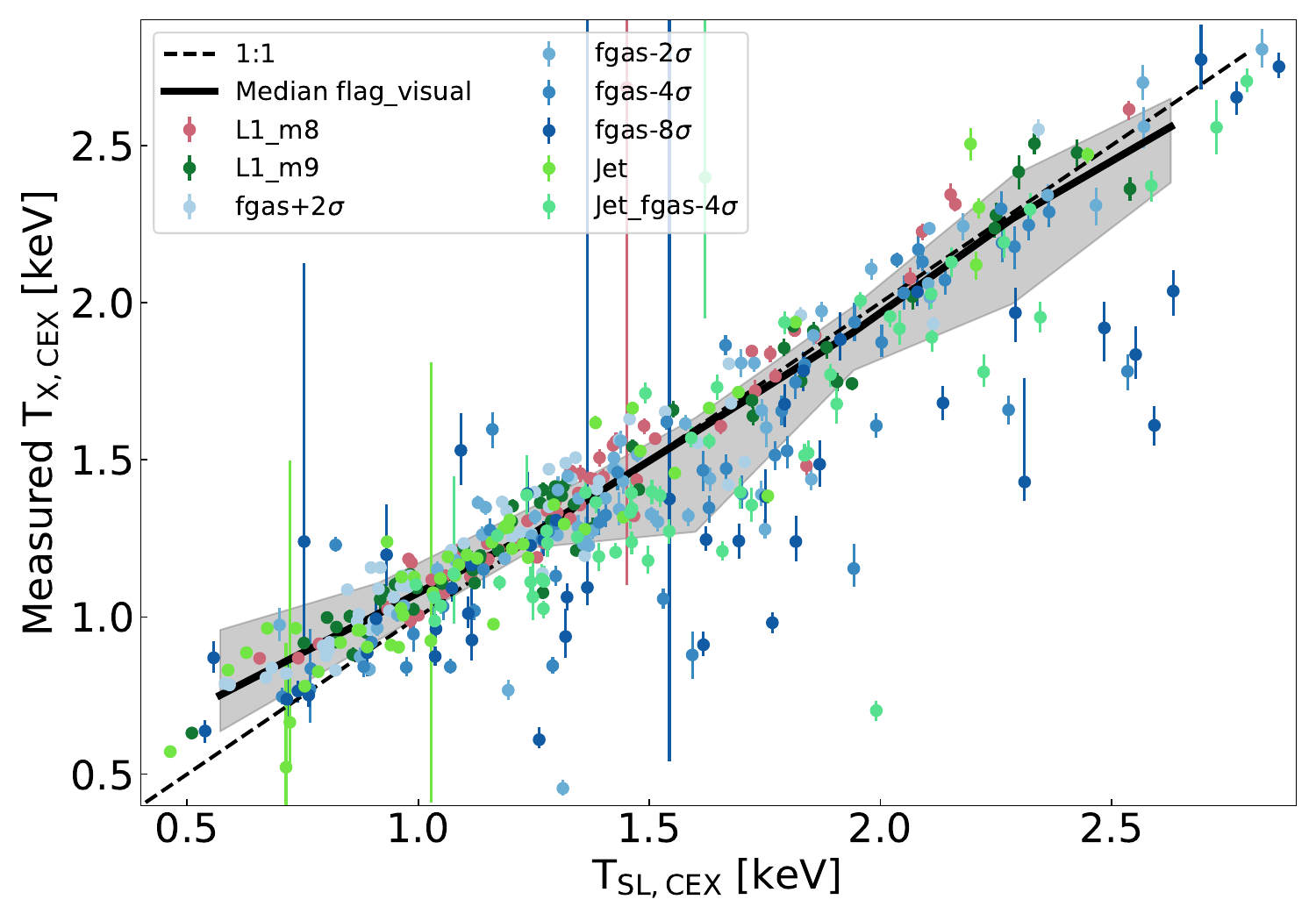}
    \caption{Comparison between the input spectroscopic-like temperature and the output temperature measured and modelled with \texttt{hydromass}. Both quantities are core-excised, using the estimated R$_{\rm 500c}$ as a reference.}
    \label{fig:Tin_Tout}
\end{figure}

\subsection{X-ray temperature}
A single definition of the true global hot gas temperature in simulations is not trivial and requires an assumption of weights for each gas element. The introduction of the spectroscopic-like weighting scheme bridged a gap between temperature definition in simulations and temperatures inferred from X-ray spectra \citep[][]{Mazzotta2004MNRAS_SL}, although some scatter is still expected \citep[][]{Seppi2026arXiv260303440S}. Dense and cool regions dominate the emissivity in X-ray spectra, which biases the temperature to such regions. An additional effect is given by metallicity: contribution from line emission is extremely relevant in groups where the temperature is not hot enough to fully ionize all elements \citep[e.g.,][]{Sanders2023arXiv_spectra}. The FLAMINGO simulations produced metal rich cores, with abundances that can reach super solar values in the inner region on average as reported by \citet{Braspenning2024MNRAS_flamingo}. The X-ray spectral analysis in such a regime that is extreme in comparison to observations is challenging. In addition, the particle resolution allows to resolve the core within the inner 0.1$\times$R$_{\rm 500c}$ with tens of particles at best in the L1\_m9 simulations. For these reasons, we compute global temperatures by excising the core and focus on the region between 0.15 and 1.0$\times$R$_{\rm 500c}$ (we do the same for observations). From the simulation perspective, we compute spectroscopic-like temperatures within this ring using the apertures defined with the estimated R$_{\rm 500c}$ to be consistent with the temperature estimated from the mock observations. 

The comparison is shown in Fig. \ref{fig:Tin_Tout}. We find the majority of the points to align close to the one-to-one line. However, we also obtain a cloud of points lying below the one-to-one line, made up mostly of systems in the strongest feedback simulations: $f_{\rm gas}-8\sigma$ and Jet\_$f_{\rm gas}-4\sigma$. From the visual inspection the majority of these systems showcase a complex IGrM, clearly disrupted by the strong AGN activity. We define a boolean flag, \textsc{flag\_visual}, which is set to True for systems that do not display obvious features likely to compromise the X-ray analysis, such as clear merger signatures, highly disturbed morphologies, very clumpy IGrM, strong asymmetries, or evident deviations from spherical symmetry.
The median temperature for these systems is shown by the black solid line in Fig. \ref{fig:Tin_Tout}. The black shaded area denotes the 16th-84th percentiles. It encompasses the one-to-one line for the majority of the temperature range considered here. A small disagreement at low temperature is likely due to the hydrogen column density, that absorbs the softer part of the spectrum, and the effective area, that is higher in the 1-2 keV range. Their combination may slightly bias our temperatures high in this regime, although the difference is not statistically significant (<2$\sigma$). Overall, Fig. \ref{fig:Tin_Tout} shows that for this subsample we find good agreement with the input spectroscopic-like temperature. For the global comparison to X-GAP, we use the full samples, otherwise we would bias the selection relative to the data.

\subsection{Gas Mass}

\begin{figure}
    \centering
    \includegraphics[width=\columnwidth]{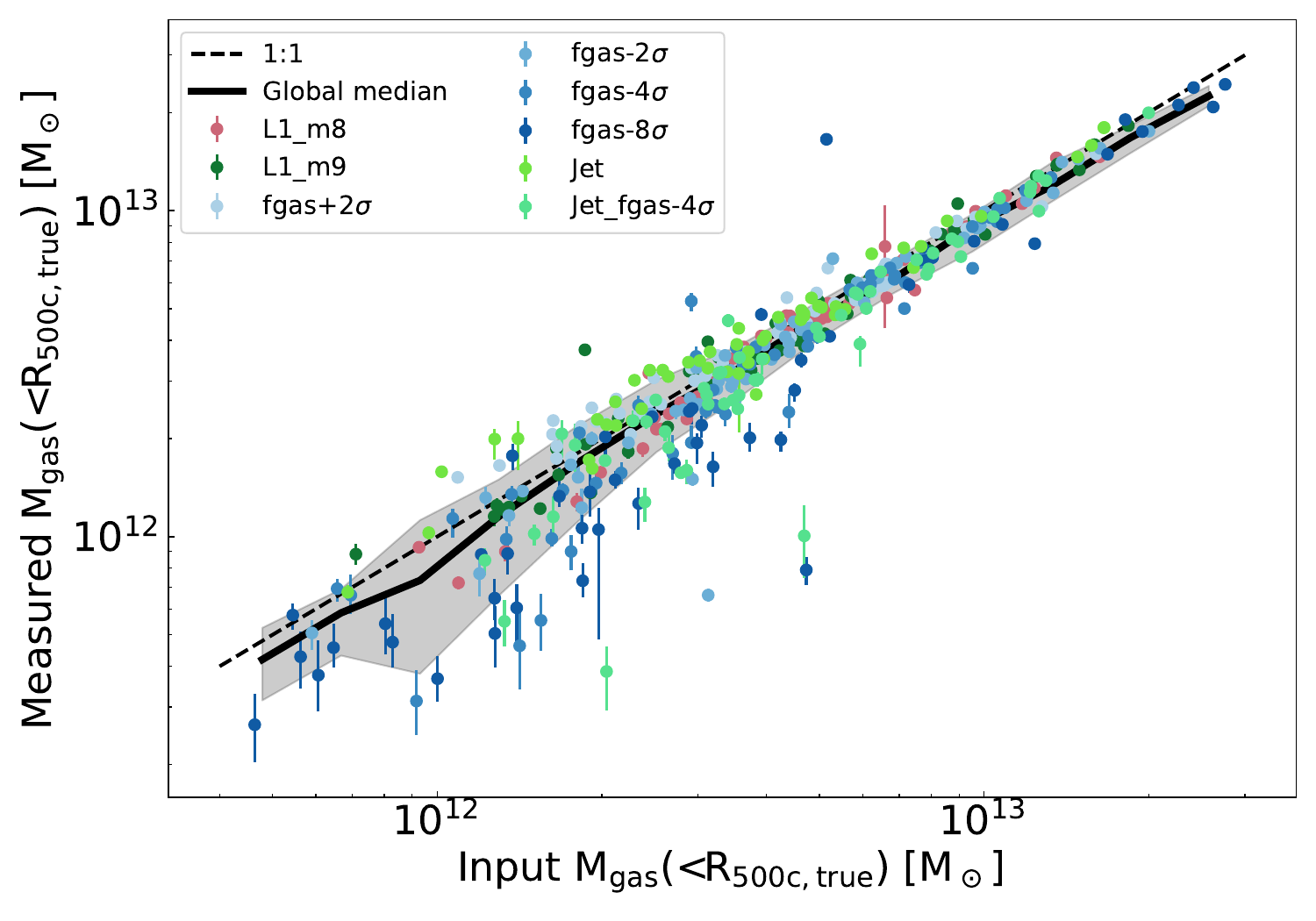}
    \caption{Comparison between the input and output gas mass measured and modelled with \texttt{hydromass}. Both quantities are integrated within the same radius (true R$_{\rm 500c}$).}
    \label{fig:Mgasin_Mgasout}
\end{figure}

The gas mass is measured by integrating the deprojected density profile following:
\begin{equation}
    M_{\rm gas} = 4\pi \mu_e m_p \int_0 ^{\rm R_{\rm 500c}} n_e(r)r^2 dr,
    \label{eq:gas_mass}
\end{equation}
where $\mu_e$ is the mean electron molecular weight and $m_p$ is the proton mass. In particular, the mean molecular weight per electron is computed in a flexible way based on the chosen abundance table \citep[][the one used for the X-GAP analysis]{Asplund2009ARA&Aabund} within \texttt{hydromass}. The pipeline computes a conversion factor that relates electron number density and emissivity, which therefore includes flexible dependencies on temperature and metallicity. 

The result is shown in Fig. \ref{fig:Mgasin_Mgasout}. The black line shows the global measured gas mass at a given true value. The shaded area includes the 16th-84th percentiles. We find excellent agreement with the one-to-one relation: our framework allows accurate gas mass measurements. Importantly, this agreement persists even though the FLAMINGO simulations produce inner metallicity profiles that are systematically higher than those observed \citep[][]{Schaye2023MNRAS_flamingo, Braspenning2024MNRAS_flamingo}. This indicates that our analysis pipeline is sufficiently flexible to recover unbiased gas masses, owing to the adaptive conversion between gas density and X-ray emissivity implemented in our modelling.

\section{The statistical comparison to X-GAP}
\label{appendix:statistical_comparison}

In this Appendix we detail the framework related to the statistical comparison between X-GAP and FLAMINGO in Sect. \ref{sec:results}.

\subsection{Observables correlation}
\begin{figure*}
    \centering
    \includegraphics[width=0.67\columnwidth]{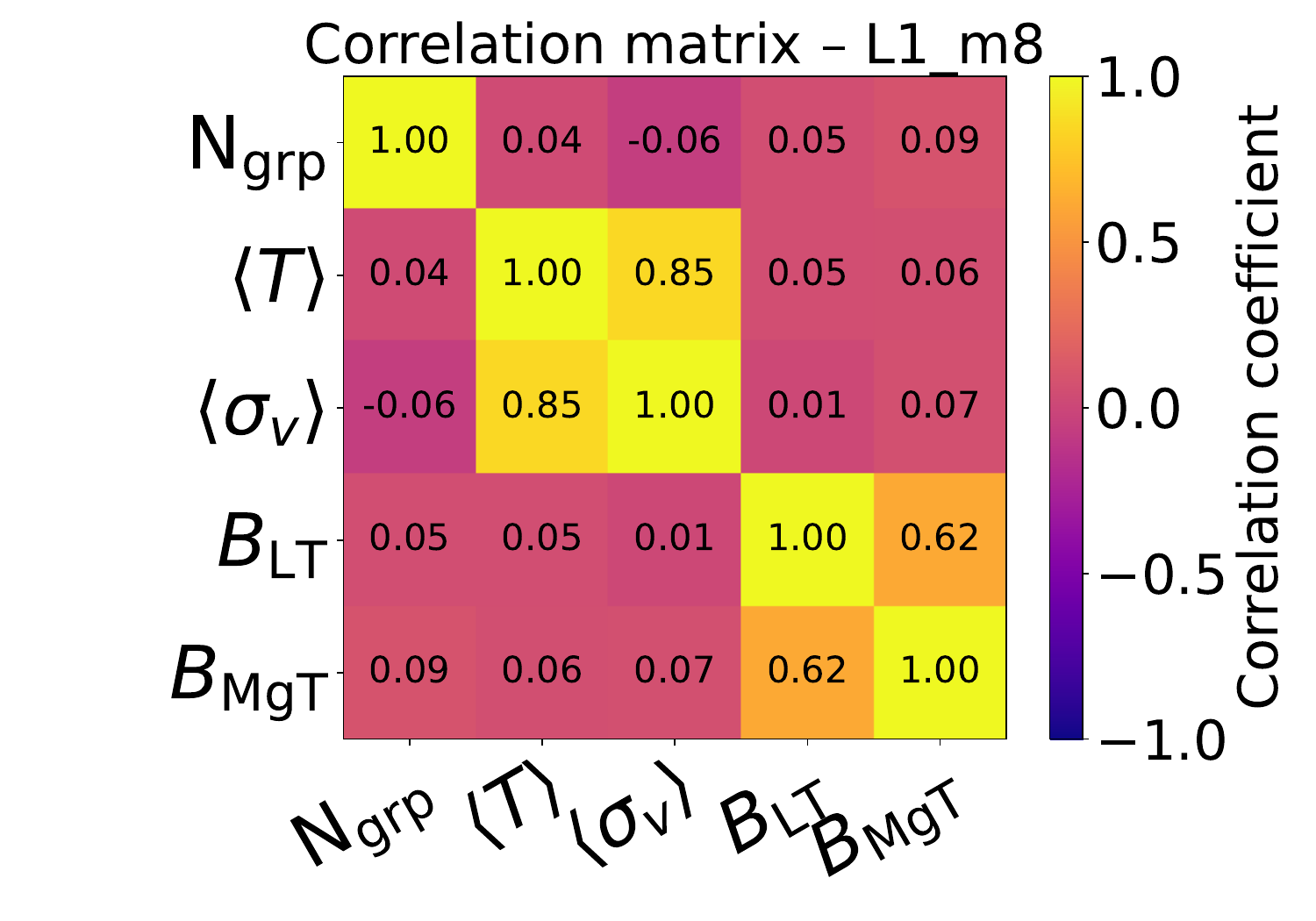}
    \includegraphics[width=0.67\columnwidth]{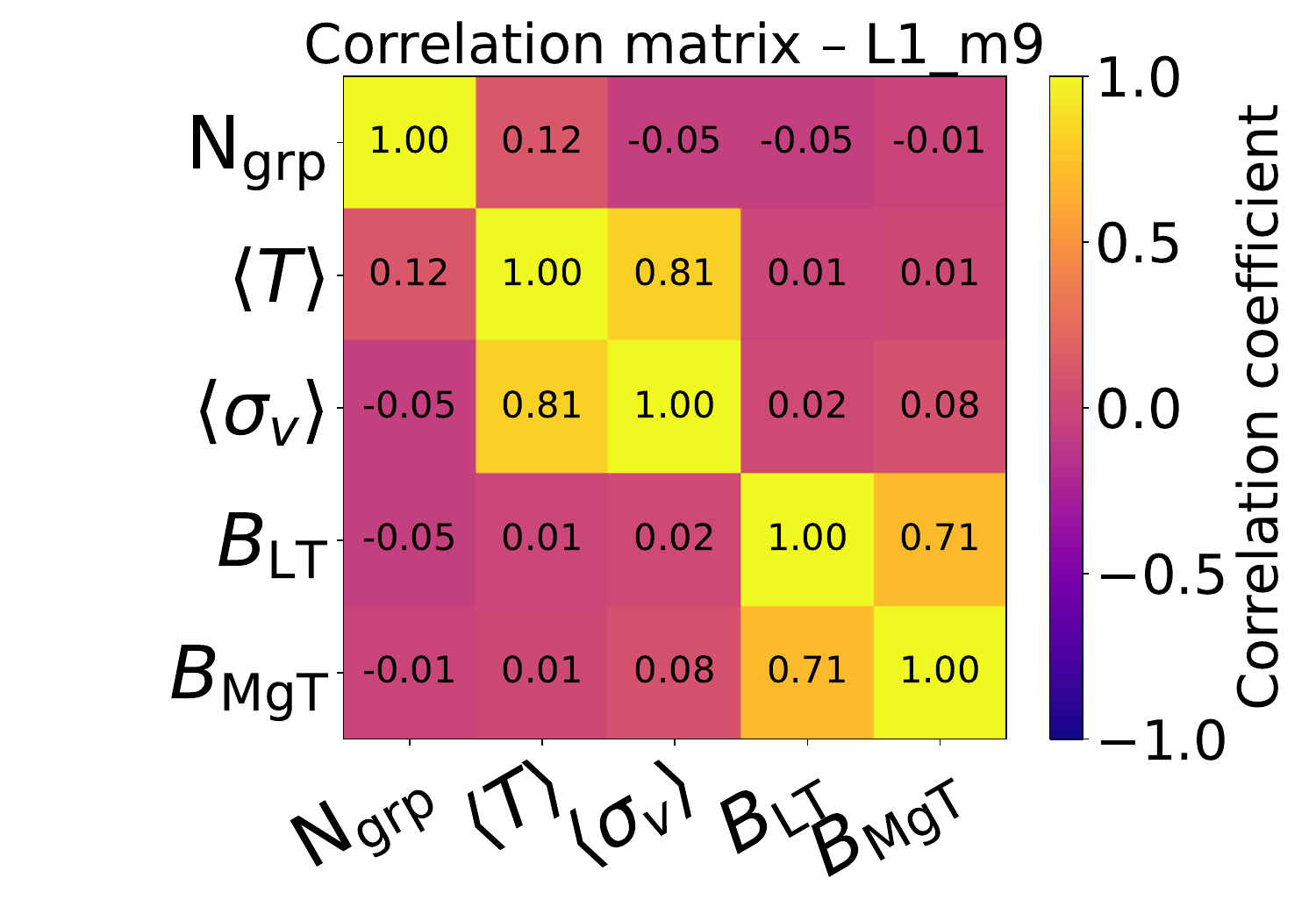}
    \includegraphics[width=0.67\columnwidth]{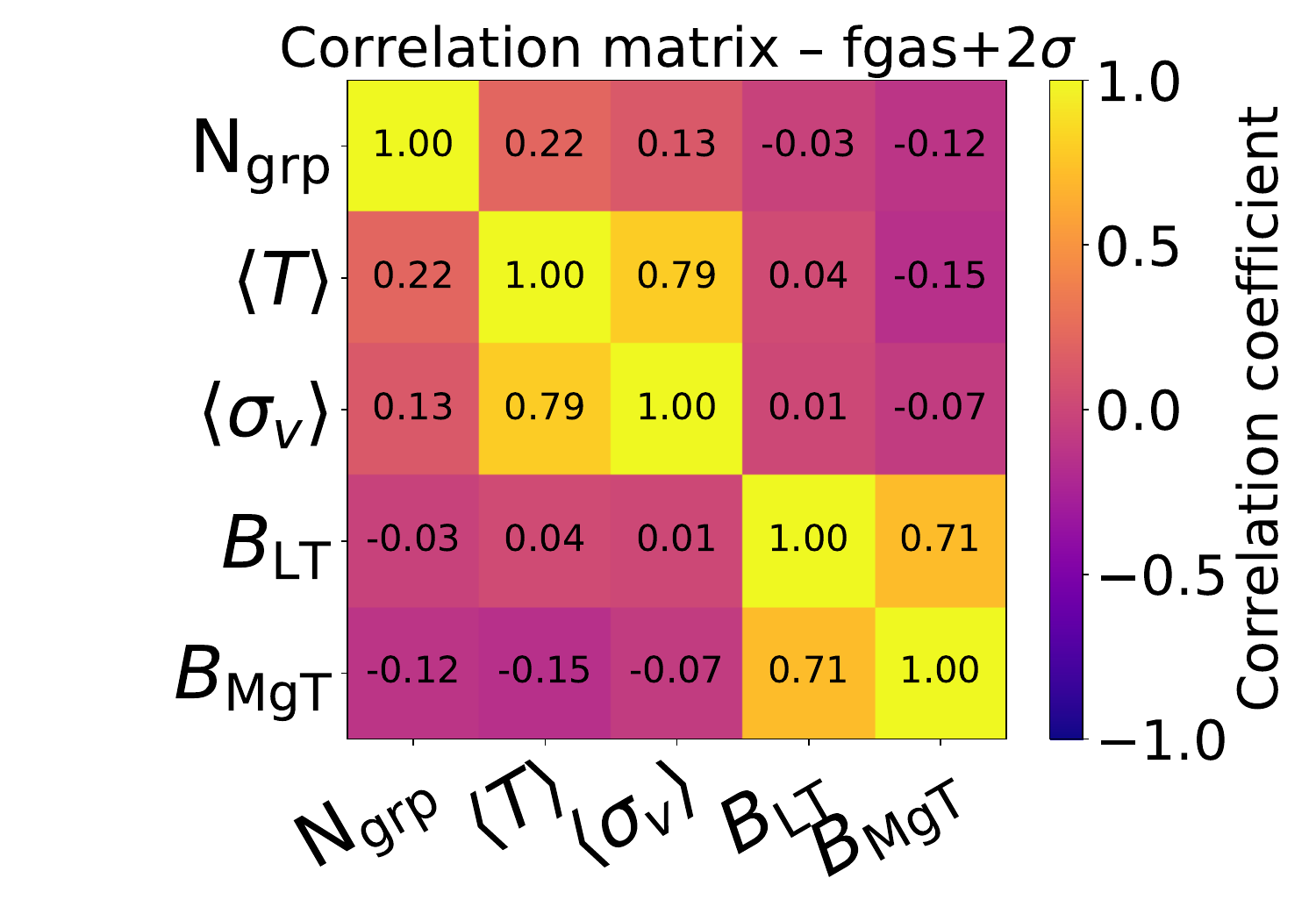}
    \includegraphics[width=0.67\columnwidth]{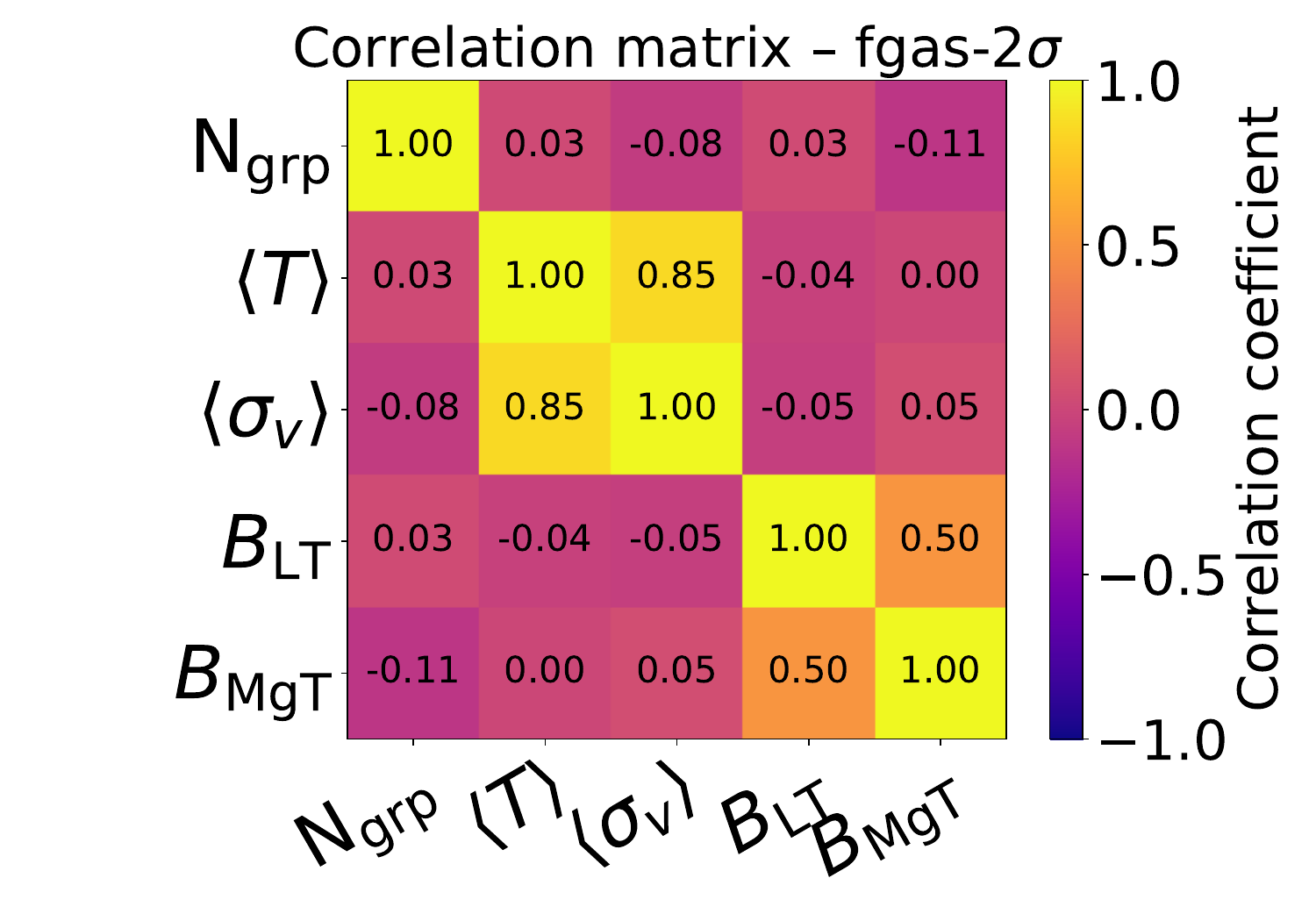}
    \includegraphics[width=0.67\columnwidth]{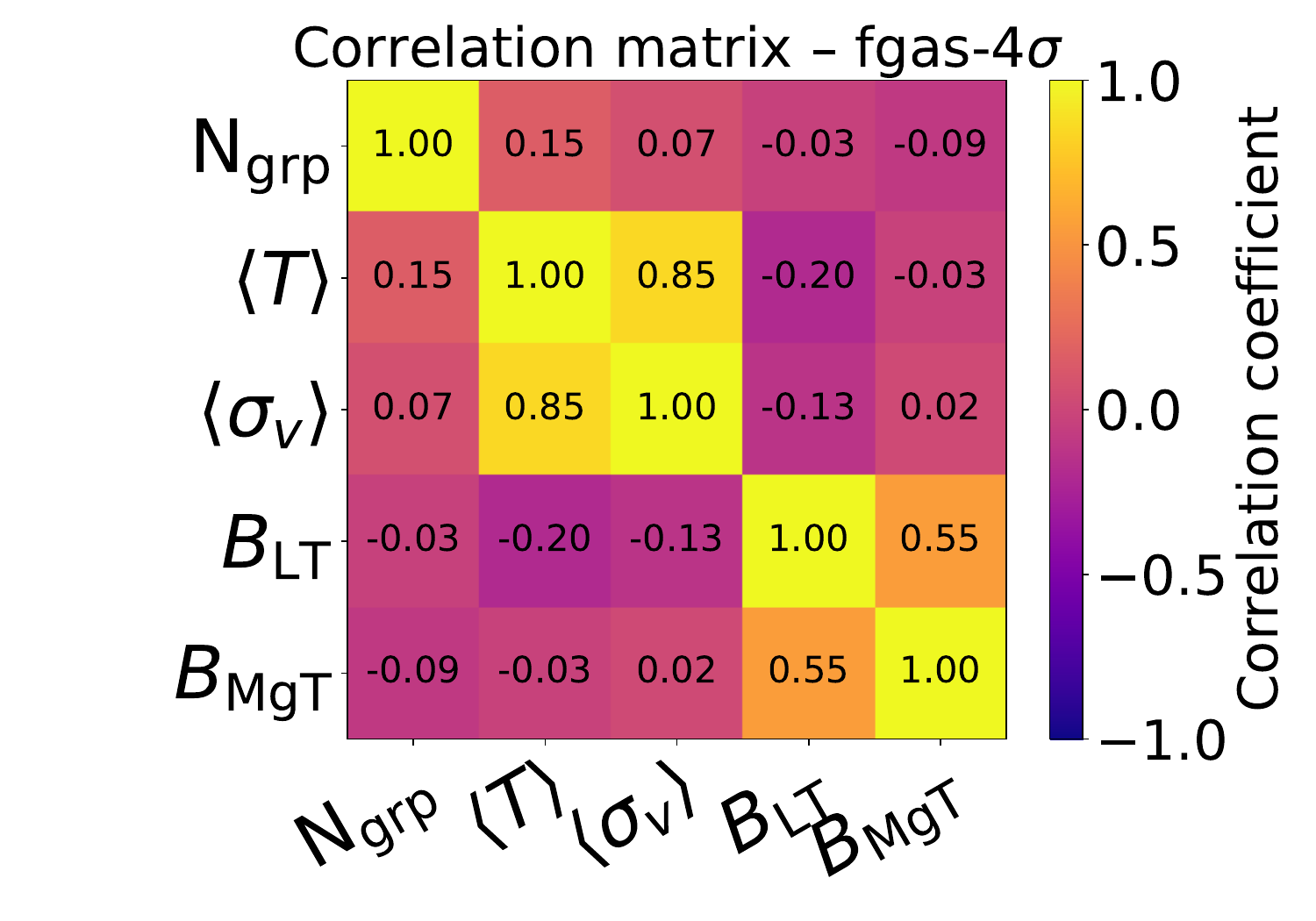}
    \includegraphics[width=0.67\columnwidth]{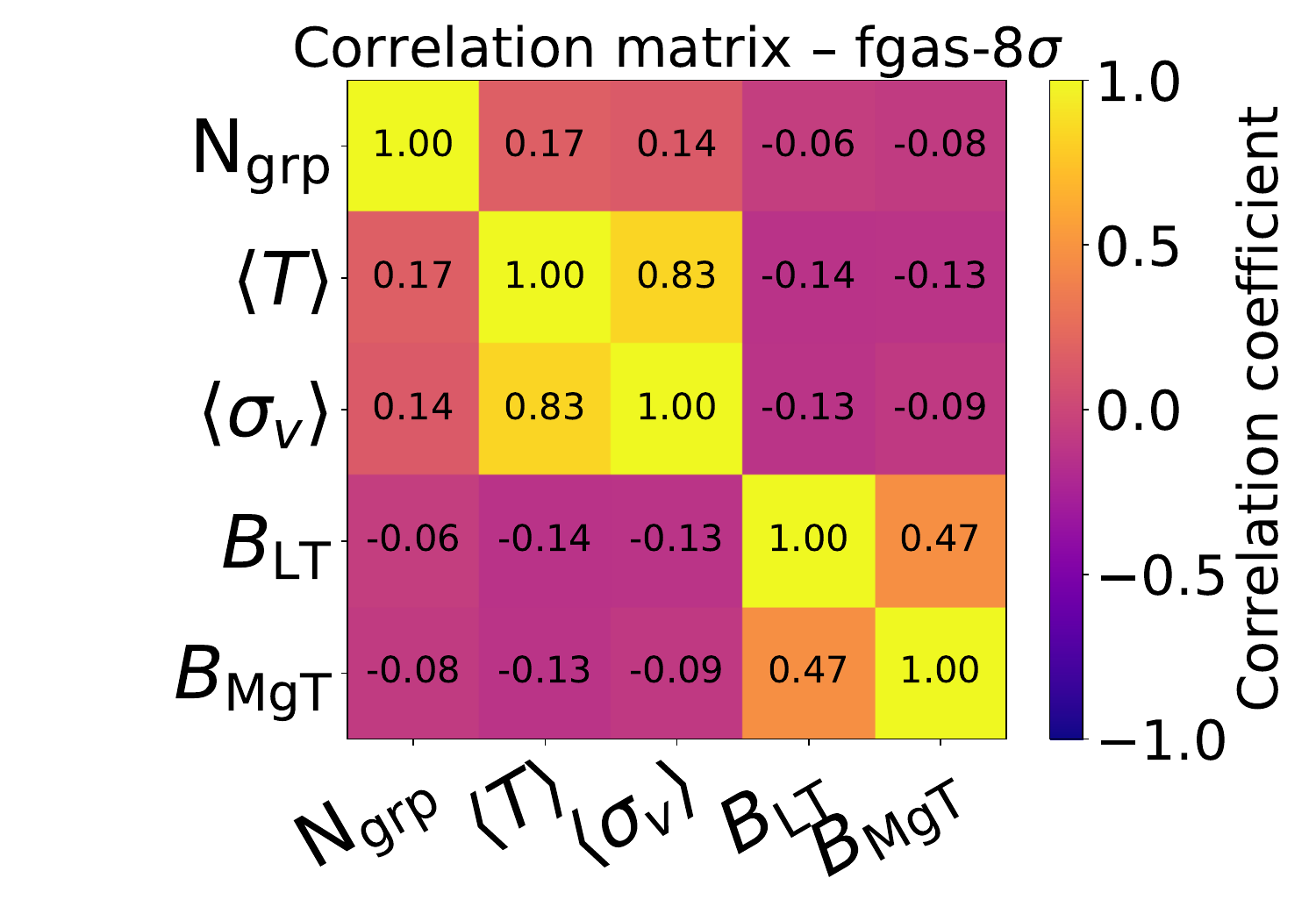}
    \includegraphics[width=0.67\columnwidth]{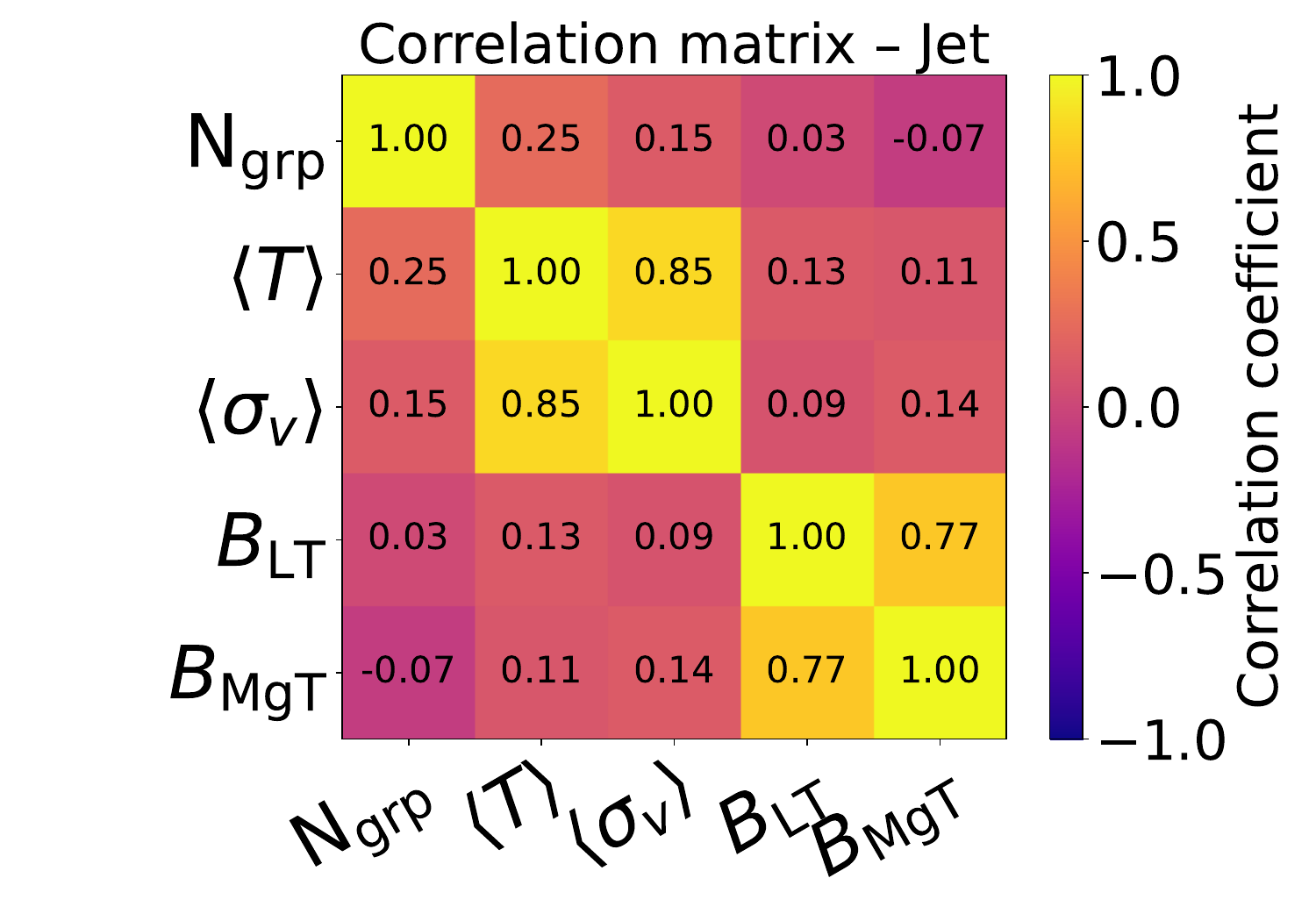}
    \includegraphics[width=0.67\columnwidth]{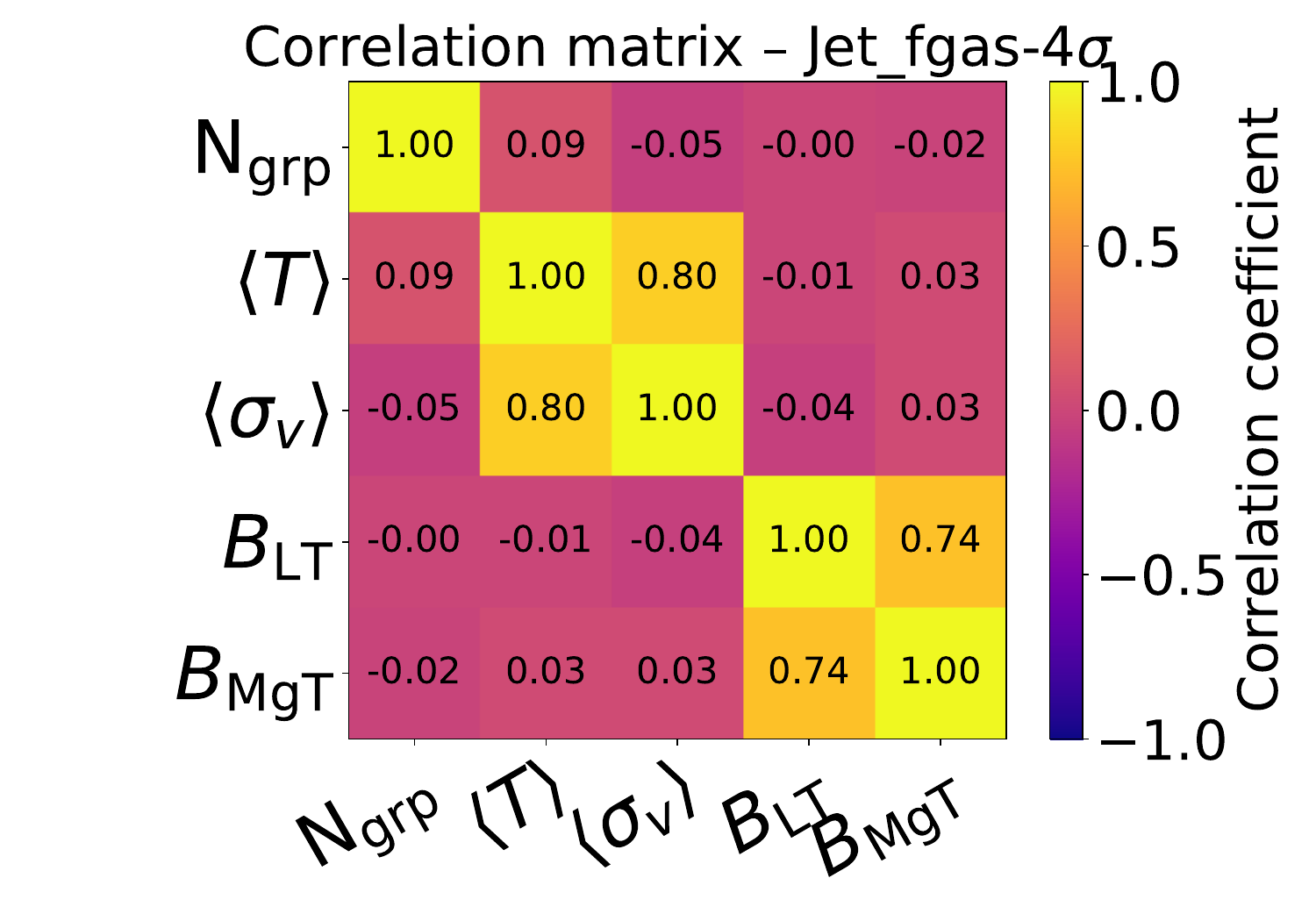}
    \caption{Correlation matrix between observables estimated from various light cones and selection function models for the various FLAMINGO models: the number of groups, the mean core-excised temperature, the mean velocity dispersion, and the normalisation of the scaling relations.}
    \label{fig:corrmat}
\end{figure*}

From the \texttt{N-light cone generation} procedure (explained in Sect. \ref{subsec:combining_obs}) we also store the input temperature and input velocity dispersion. Although these are not identical to the measured quantities, they provide a physically motivated baseline to quantify correlations between observables driven by halo population and selection effects. We also fit the normalisation of the scaling relation between input luminosity and temperature as explained in the previous section. This provides a data vector with $B_{\rm LT}$, $B_{\rm MgasT}$, $N_{\rm groups}$, mean temperature, and mean velocity dispersion for each experiment. From the ensemble we compute the correlation matrix $C_{i,j}$ between the observables, shown in Fig. \ref{fig:corrmat}.
Results are consistent across models.
We find a strong covariance between mean temperature and velocity dispersion (correlation $\sim$0.8), as both trace halo mass. The scaling-relation normalisations are also correlated (0.5--0.7), reflecting the link between gas mass and X-ray luminosity \citep[e.g.][]{Eckert2020_pyproffit}. In contrast, the number of groups shows no significant correlation with other observables, being dominated by cosmic variance. Similarly, the $L$--$T$ normalisation is also largely independent: it describes luminosity variations at fixed temperature, so changes in temperature or velocity dispersion do not directly affect its normalisation. The same holds for the $M_{\rm gas}$--$T$ normalisation.

\subsection{The copula approach}
For each observable individually, we construct its empirical cumulative distribution function (CDF) $\mathcal{F}_i$ from the \texttt{Monte Carlo} procedure, without assuming any analytic form. For temperature, velocity dispersion, and $B_{\rm LT}$, the \texttt{MonteCarlo} distributions primarily capture measurement and modelling uncertainties (bootstrap scatter and posterior sampling). However, the light-cone ensemble shows that field-to-field (cosmic variance and selection-function) fluctuations can also contribute to the scatter of sample-averaged quantities.
To ensure a consistent comparison with X-GAP, we add this additional variance in quadrature to the measurement-driven scatter when it is not already dominant. 
The final empirical distributions $\mathcal{F}_i$ therefore represent the total expected uncertainty for each observable, combining measurement effects and sample-to-sample fluctuations.

We use a Gaussian copula to model the observables covariance. A Gaussian copula provides a flexible framework to model the joint distribution of multiple variables by separating the individual properties of each observable (often named marginal distributions) from the way they co-vary (often named dependence structure). This approach is motivated by the need to combine observables with diverse, empirical distributions while preserving their intrinsic covariance. By mapping the marginals onto a latent multivariate Gaussian space, we can generate synthetic realizations that honour the measured correlation matrix regardless of the underlying shape of the individual 1D distributions. Operationally, this is implemented as follows:
we draw random Gaussian vectors $z_j$ from the correlation matrix $C_{i,j}$. Each component is transformed into a uniform variable using the standard normal CDF: $u_j = \Phi(z_j)$. The $u_j$ values span the range between 0 and 1 but retain the expected correlation from $C_{i,j}$. For each observable, $u_j$ is mapped onto a physical value $x_j = \mathcal{F}_j^{-1}(u_j)$ using the inverse empirical CDF of that observable. This procedure produces synthetic realizations that follow the empirical 1D distributions of each observable, and reproduce the measured correlation structure between them.
Finally, since the empirical CDF only contain scatter due to the measurement process, we add the scatter estimated from the \texttt{N-light cone generation} to also account for systematics due to cosmic variance and selection function uncertainties (see also Fig. \ref{fig:cosmicvar_impact_OBS} about the relative impact on each observable).
These samples represent the distribution of outcomes expected from that model under the same survey selection and cosmic variance as X-GAP.
\begin{figure}
    \centering
    \includegraphics[width=\columnwidth]{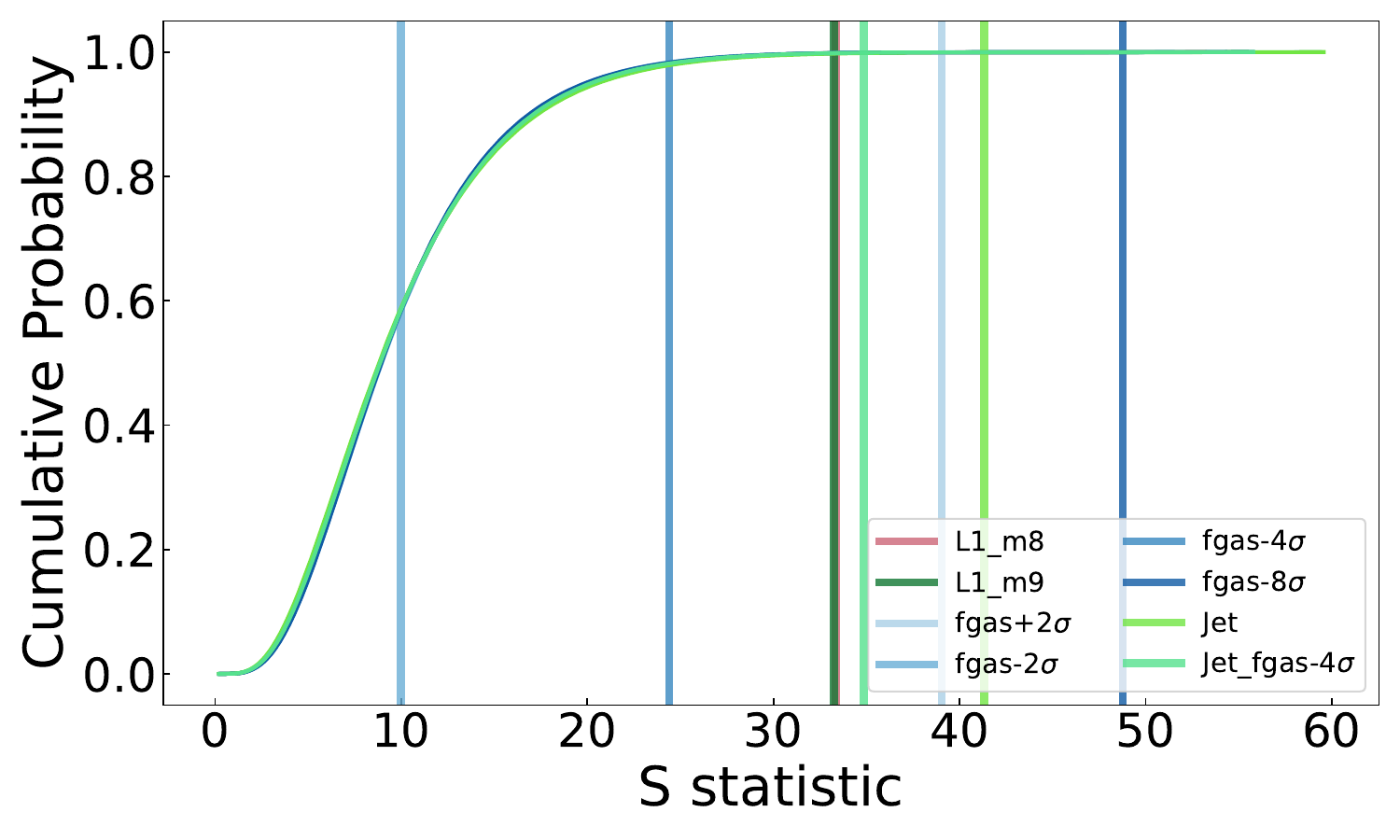}
    \caption{Cumulative distribution functions of the summary statistic S for the different FLAMINGO models and for the X-GAP sample. The curved lines represent the distribution of S generated via a Gaussian copula to account for the covariance between observables in the FLAMINGO realisations. The vertical lines indicate the position of the observed X-GAP data within each model distribution. A model is in good agreement if the vertical line falls near the median of the CDF (like $f_{\rm gas}-2\sigma$), while it is statistically disfavoured if the vertical line falls in the extreme tail (like $f_{\rm gas}-8\sigma$).}
    \label{fig:Sstat}
\end{figure}
\begin{figure}
    \centering
    \includegraphics[width=\columnwidth]{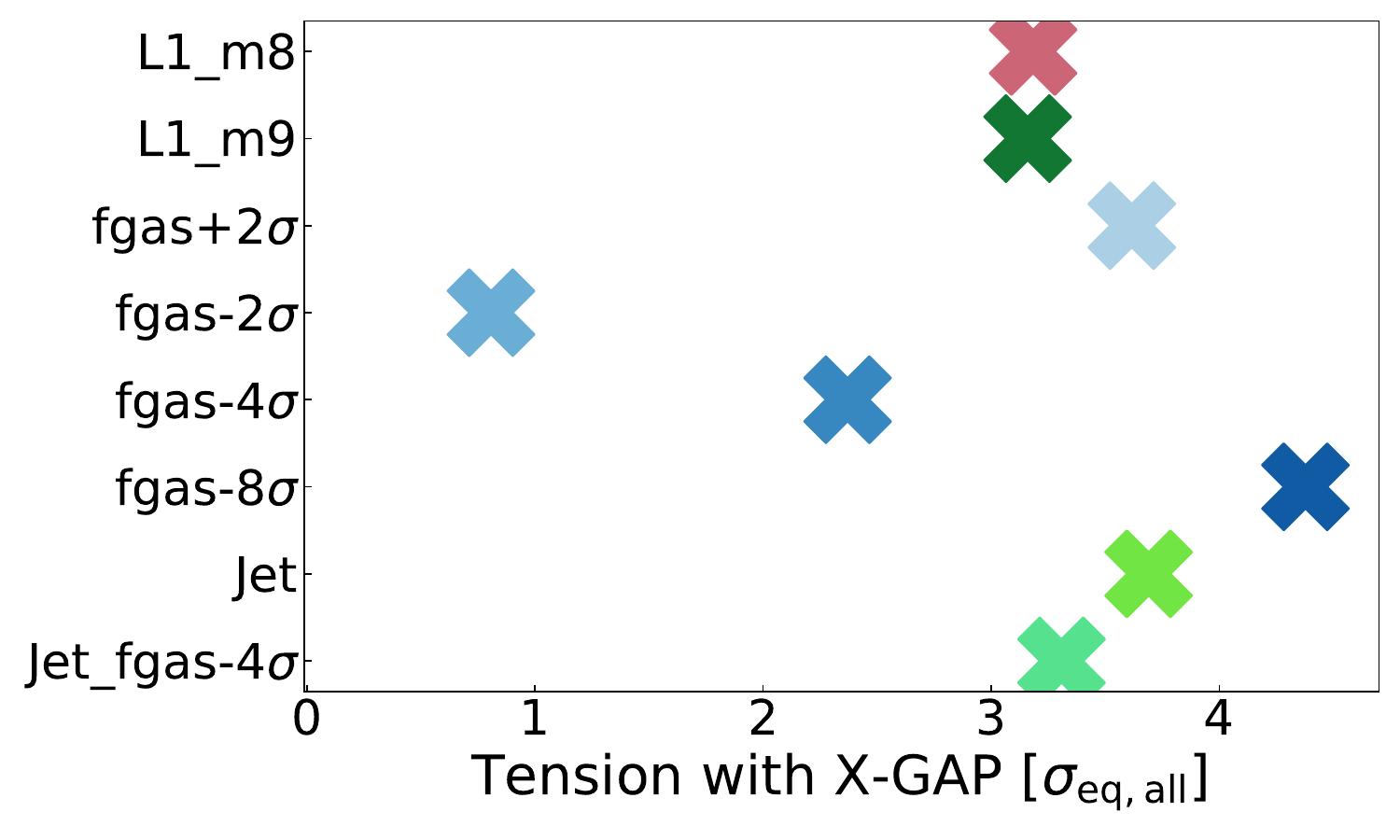}
    \caption{Final comparison to X-GAP. It includes the normalisation of the $L$--$T$ (both core-excised) and $M_{\rm gas}$(<400 kpc)--$T$ (core-excised) scaling relations, the expected number of groups, the mean core-excised temperature, and velocity dispersion. The tension expressed in Gaussian equivalent significance accounts for the intrinsic correlation between observables as explained in Sect. \ref{subsec:combining_obs}. The values are also reported in Table \ref{tab:summary_stat}.}
    \label{fig:final_comparison}
\end{figure}
Now, for each observable we compute a two-sided tail probability $p_j$ and define the summary statistic $S$ following the Fisher method such that:
\begin{align}
    p_j &= 2 \times \text{min}(\mathcal{F}_j(x_j), 1-\mathcal{F}_j(x_j)), \nonumber \\
    S &= -2\sum_{i=1}^5\log p_j.
    \label{eq:summary_probSTAT}
\end{align}
Because the $p_j$ are correlated, we do not expect $S$ to necessarily follow the standard $\chi^2$ distribution. We show the distribution of the S statistic for each model in Fig. \ref{fig:Sstat} \citep[a method similar to][]{Regamey2026A&A_SBI}. We also calculate the S statistic for the X-GAP sample itself and indicate its position with a vertical line of the corresponding colour. The discriminating power of the test lies in the percentile at which the X-GAP value intersects the model's CDF. The CDFs are very similar across the FLAMINGO runs: the discriminating power arises primarily from the location of the X-GAP statistic within each distribution. The model $f_{\rm gas}-2\sigma$ places the X-GAP value close to the median of the distribution and provides the best agreement with the data, whereas the fiducial runs (L1\_m8 and L1\_m9) lie at higher percentiles and are mildly disfavoured. Models with very strong feedback (e.g. $f_{\rm gas}-8\sigma$) place the X-GAP value in the extreme tail of the distribution and are strongly inconsistent with the observations. The joint probability $p_{\rm joint}$ is given empirically by the fraction of $S$ larger than the one computed for X-GAP, i.e. the fraction of simulated realizations that are at least as extreme as X-GAP in the combined 4-D space. Finally, we convert the joint probability into an equivalent two-sided Gaussian significance
\begin{equation}
    \sigma_{\rm eq} = \Phi^{-1}(1-\frac{p_{\rm joint}}{2}),
\end{equation}
where $\Phi^{-1}$ is the inverse standard normal cumulative distribution function.
The $\sigma_{\rm eq}$ quantifies how unusual the X-GAP measurements are relative to each model, while fully accounting for covariance between observables (Fig. \ref{fig:final_comparison} and Table \ref{tab:summary_stat}).

\section{Potential systematics}
\label{appendix:systematics}
Here we address three potential sources of systematics: the gas particle metallicity the sample size, and the cosmological model.

\subsection{Gas particle metallicity}
\label{appendix:Z03}

\begin{figure}
    \centering
    \includegraphics[width=\columnwidth]{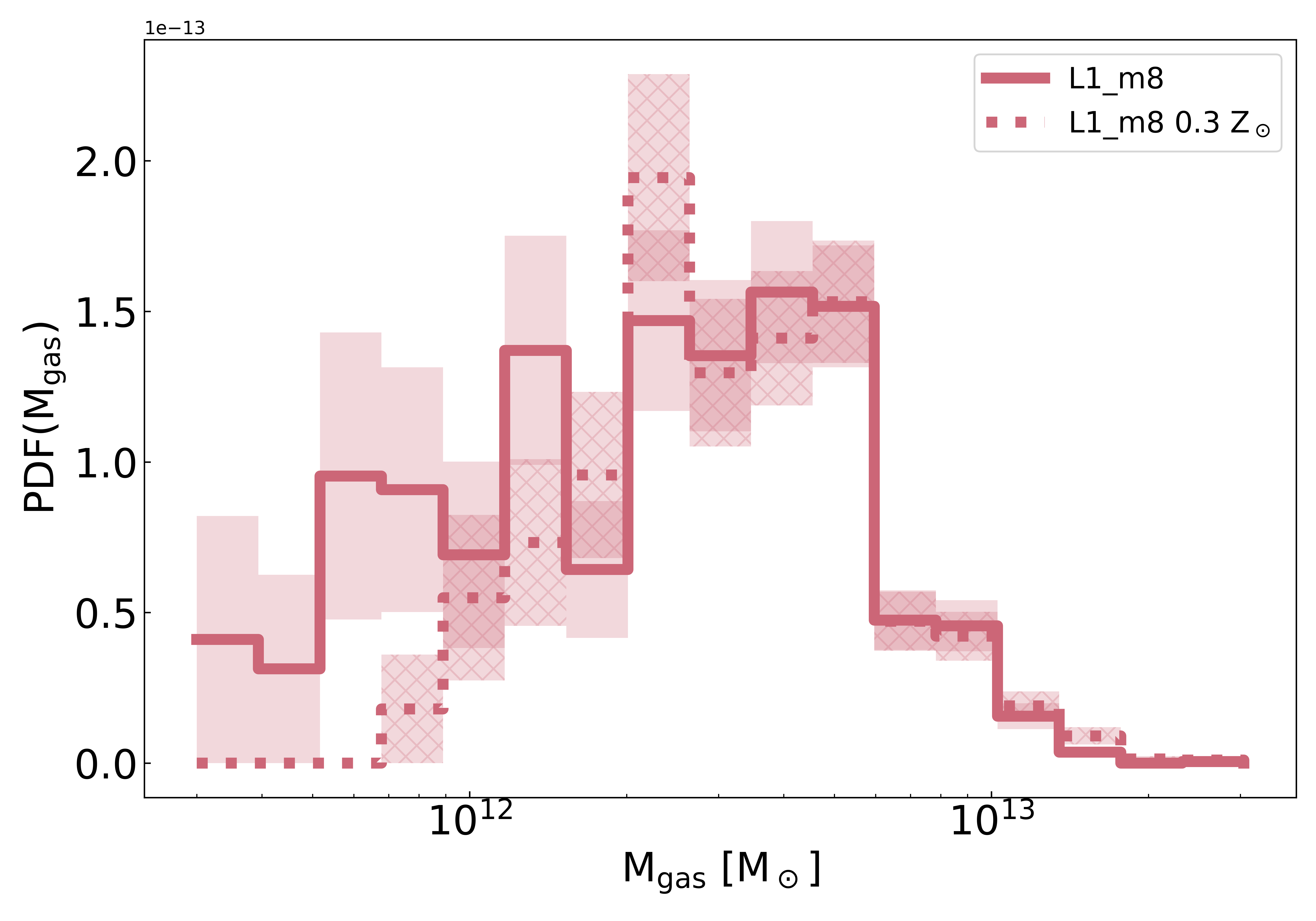}
    \caption{Comparison of the gas mass distribution between the sample selected from L1\_m8 as described in Sect. \ref{subsubsec:flamingo_groupsel} and the variation assuming Z=0.3 for all gas elements.}
    \label{fig:PDF_Mgas}
\end{figure}

FLAMINGO predicts metal abundances higher than observed, reaching super-solar values in group cores \citep{Braspenning2024MNRAS_flamingo}. In this mass regime, where the gas is not fully ionised, line emission contributes significantly to the total luminosity, making metallicity a key driver of $L_{\rm X}$. Higher abundances can therefore boost the luminosity at fixed gas mass.
This may affect the results in Fig. \ref{fig:observables_for_comparison}. The $L$--$T$ normalisation is largely unaffected, as luminosity drives the selection. In contrast, the $M_{\rm gas}$--$T$ relation is more sensitive: an enhanced $L_{\rm X}$ would require lower $M_{\rm gas}$ to match the observations.

We recompute the input X-ray luminosities with \texttt{pyXSIM} by fixing the metallicity of gas particles within R$_{\rm 500c}$ to 0.3 Z$_\odot$ for haloes with M$_{\rm 500c}>5\times10^{12}$ M$_\odot$. This requires collecting the gas particles around each halo and recalculating their emissivity at fixed abundance. Because the procedure is computationally expensive, we restrict the analysis to a single L1\_m8 light cone at $z\leq0.05$. We then apply the selection described in Sect. \ref{subsubsec:flamingo_groupsel} using these recomputed luminosities instead of the catalogue values. Since Appendix \ref{subsec:inout} shows that our pipeline accurately recovers gas masses, we do not generate new mock XMM-Newton observations but directly compare the gas-mass distributions of the original L1\_m8-selected sample and the variant computed with 0.3 Z$_\odot$.

The result is shown in Fig. \ref{fig:PDF_Mgas}. The solid line with the shaded area denotes the standard X-GAP-like sample from L1\_m8, while the dotted line with hatched shaded area corresponds to the Z=0.3 Z$_\odot$ variation. Overall we find good agreement between the two, especially for average to higher gas masses where the distributions are almost indistinguishable. The agreement is within 1$\sigma$ down to about M$_{\rm gas}$=10$^{12}$ M$_\odot$. Below such a threshold the standard L1\_m8 contains more objects, which skews the median gas mass to lower values compared to the Z=0.3 variation. This is in agreement with fainter groups being boosted by the higher metallicity. The median gas mass in the two samples is 4.5$\pm$0.2$\times$10$^{12}$ and 4.7$\pm$0.2 $\times$10$^{12}$ M$_\odot$. The uncertainties are estimated from 10000 bootstrap samples. The two values differ by about 5$\%$, although they are compatible at the 0.9$\sigma$ level. Therefore, we do not find a statistical disagreement between the two.

In addition, the normalisation of the $M_{\rm gas}$--$T$ is even less affected overall, because it is computed at fixed temperature. We find $B_{\rm MgasT}$=12.45$\pm$0.01 for L1\_m8 and $B_{\rm MgasT}$=12.46 for the Z=0.3 variation. We conclude that even though a particle selection with lower metallicity would not allow selecting a few low gas mass objects, the normalisation of the $M_{\rm gas}$--$T$ relation, which brings constraining power to our framework, in not significantly affected by this systematic.

\subsection{Sample size}
\label{subsec:sample_size}

\begin{figure}
    \centering
    \includegraphics[width=0.95\columnwidth]{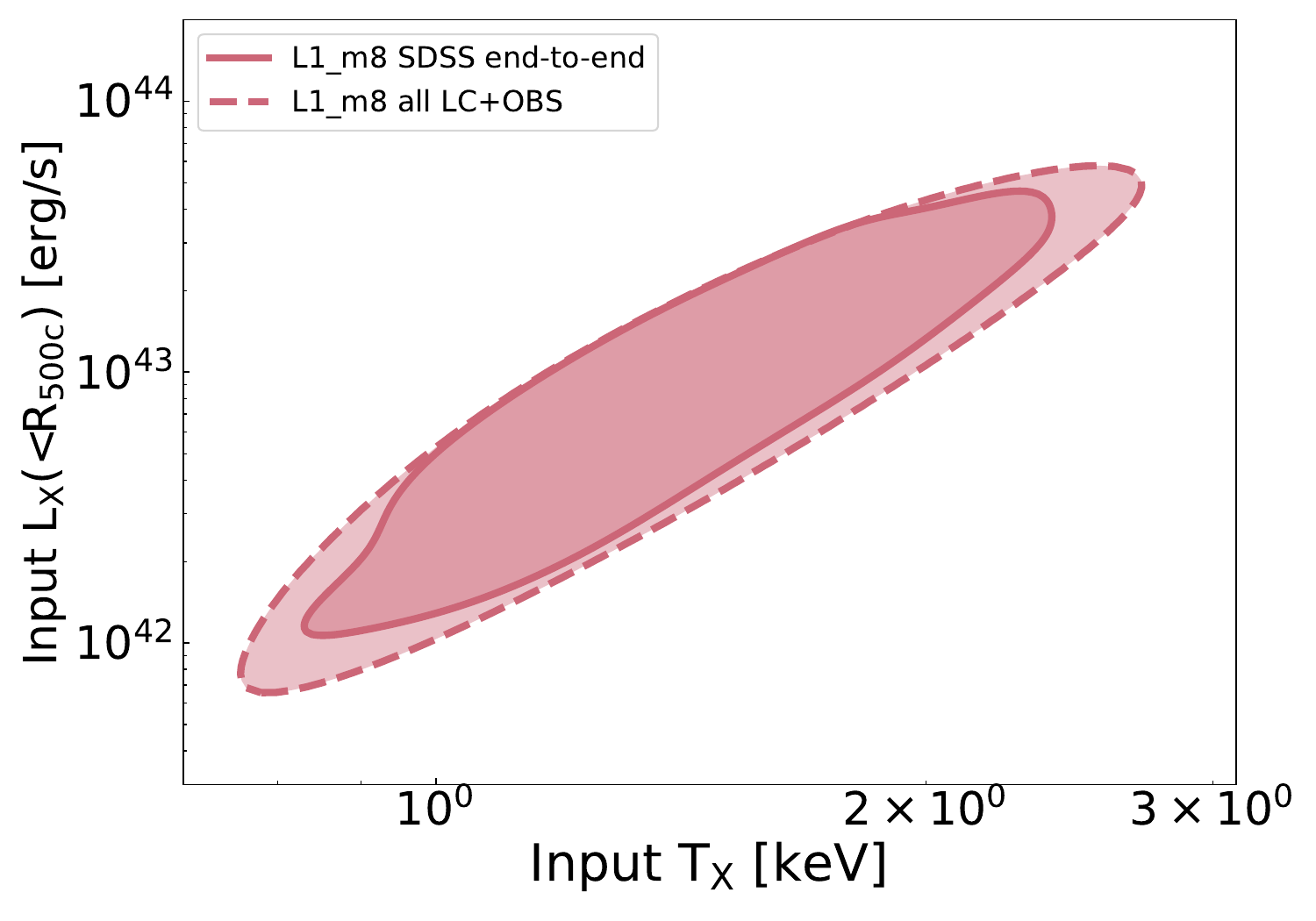}
    \includegraphics[width=0.95\columnwidth]{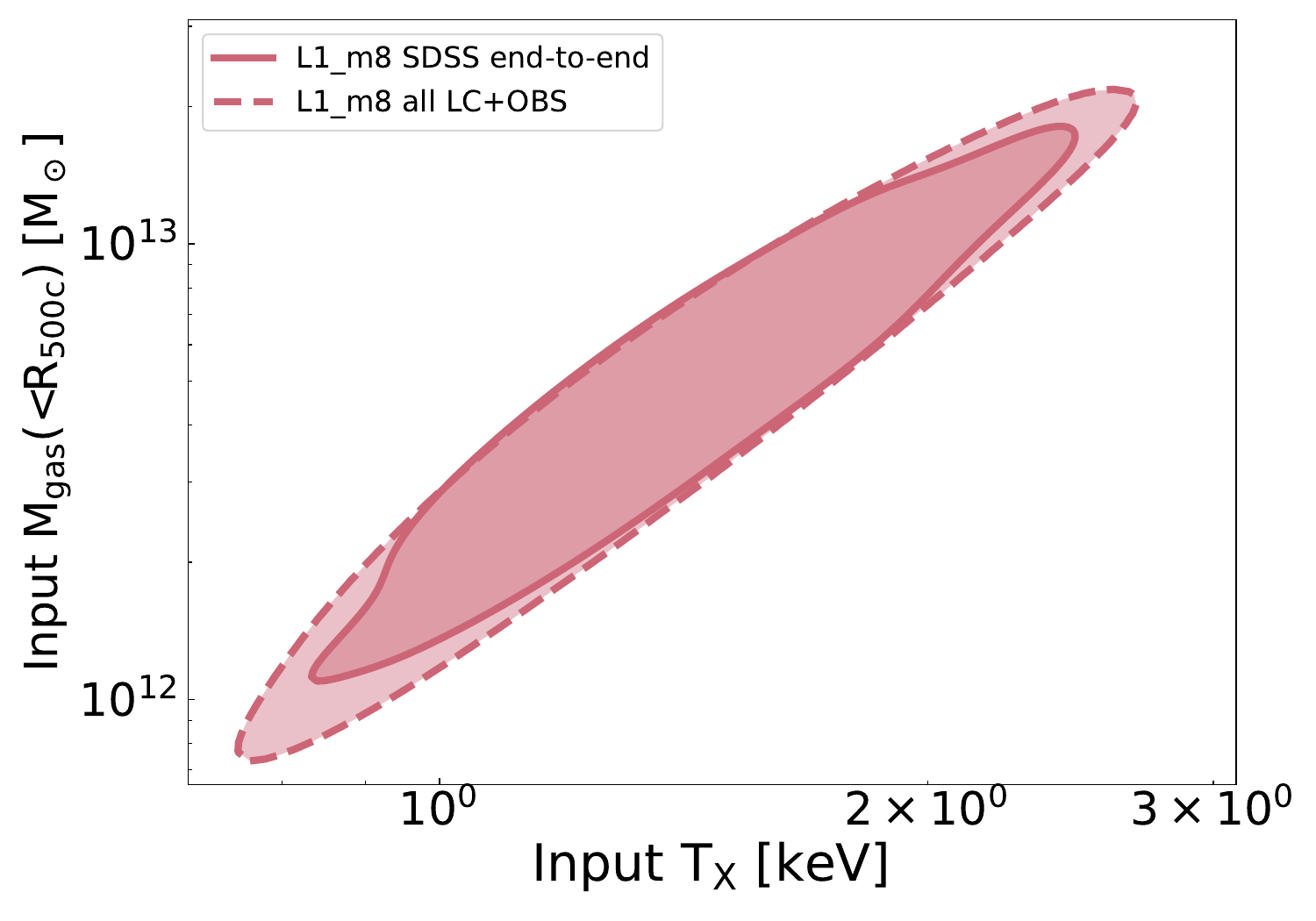}
    \caption{Comparison of the input $M_{\rm gas}$--$T$ and $L$--$T$ relations between the L1\_m8 sample (solid lines) used for comparison to X-GAP and the full combination of all light cones and observers used for the cosmic variance tests (dashed lines).}
    \label{fig:sample_size}
\end{figure}

We study whether the relatively limited sample size of about 50 groups in X-GAP and the one used in this work for comparison to the models has an impact on our results. We investigate the 2D distribution of gas mass and temperature, and X-ray luminosity and temperature. We compare the sample from L1\_m8 used for comparison to X-GAP to the full combination of all light cones and observers used for the cosmic variance tests. We use the input values, because we do not have the full end-to-end pipeline for all light cones and observers. In total we collect 2368 haloes. 

The results are shown in Fig. \ref{fig:sample_size}. The two populations occupy a very similar region of the parameter space. The larger sample combining all light cones and observers is more likely to contain rarer objects that deviate from the mean relations, therefore the dashed contours extend to slightly larger regions of the parameter space. 

The median values of each quantity are very close in the two samples ($T$ of 1.40 and 1.41 keV, $M_{\rm gas}$ of 4.65 and 4.48$\times$10$^{13}$ M$_\odot$, $L_{\rm X}$ of 7.48 and 7.39$\times$10$^{42}$ erg/s). The 1D 16th percentiles move from 1.79 to 1.81 keV (7.59 to 8.41$\times$10$^{13}$ M$_\odot$, 3.1 to 2.7$\times$10$^{42}$ erg/s). The 1D 84th percentiles move from 1.15 to 1.07 keV (2.5 to 2.1$\times$10$^{13}$ M$_\odot$, 1.71 to 1.73$\times$10$^{43}$ erg/s). 

Finally, the normalisation of the two scaling relations move from 12.56 to 12.55 for $M_{\rm gas}$--$T$ and from 42.76 to 42.75 for $L$--$T$. We conclude that the sample size does not significantly affect our results.

\subsection{Cosmology}
\label{appendix:cosmo}

The number of selected groups is expected to also depend on cosmology. To study this effect, we follow a methodology similar to \cite{Regamey2026A&A_SBI} and integrate the halo mass function from \cite{Tinker2008} up to z=0.05 and in the 10$^{13}$--10$^{14}$ M$_\odot$ range. We account for an SDSS-like sky coverage. We rescale the total number of generated haloes by the ratio between the X-GAP selected systems and the number of all haloes in the L1\_m8 light cone, that is equal to about 6$\%$. The result is a rough estimate of the total number of selected groups. We then vary the cosmological model and study the impact on the final number of groups. We use the best fit parameters from Planck \citep{Planck2020A&A}, DES-Y3 \citep{Abbott2022PhRvD_DESY3}, eRASS1 clusters \citep{Ghirardini2024_erass1cosmo}, and SPT clusters \citep{Bocquet2025PhRvD.111f3533B}. 

\begin{figure}
    \centering
    \includegraphics[width=\columnwidth]{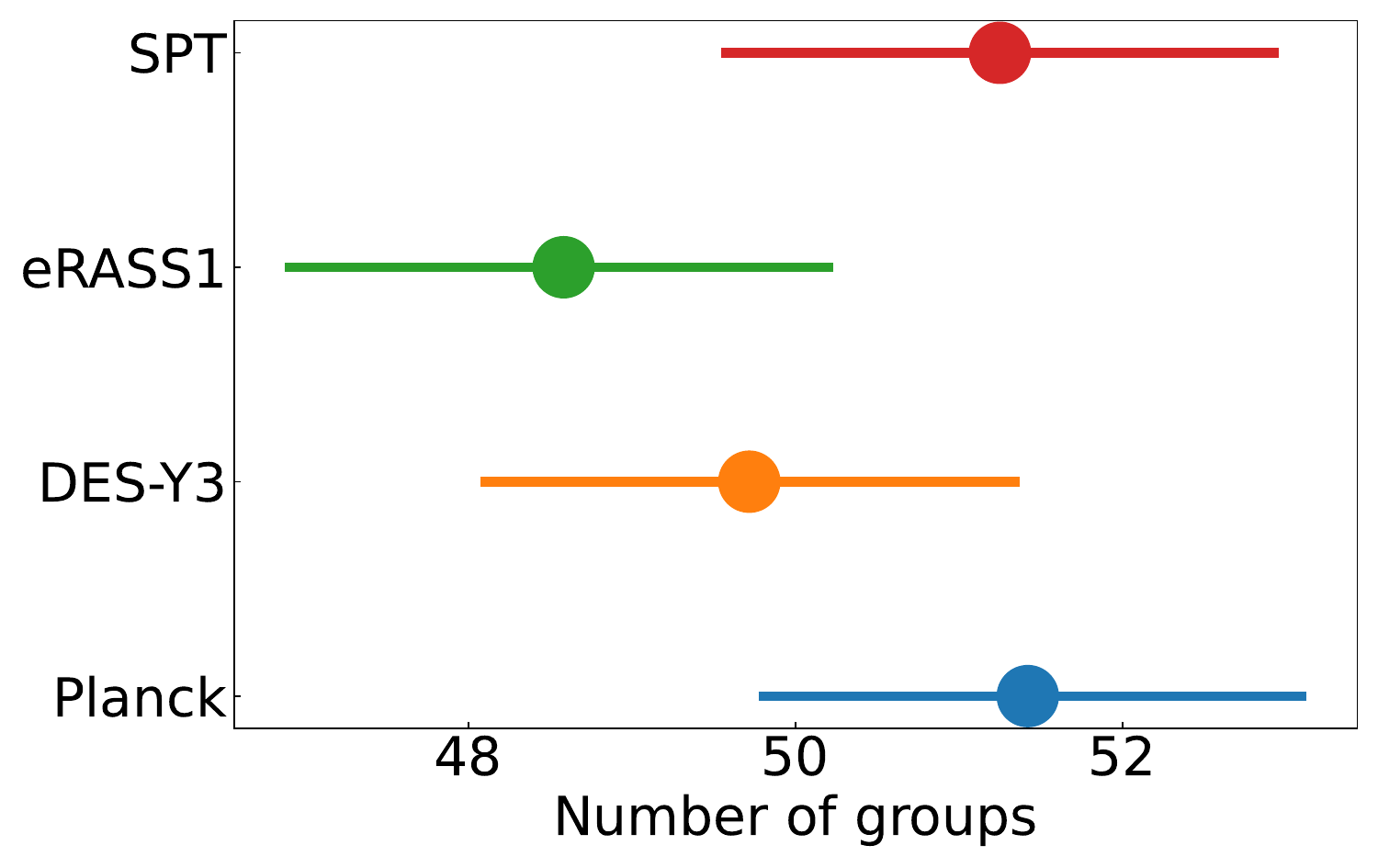}
    \caption{Comparison of the predicted number of X-GAP-like selected groups for different cosmological models.}
    \label{fig:Ngrps_cosmo}
\end{figure}

The result is shown in Fig. \ref{fig:Ngrps_cosmo}. The errorbars are estimated from 1000 samples generated from a different random seed for each cosmology. We obtain a median number of 51.4$\pm$1.8 for Planck, 49.7$\pm$1.7 for DES-Y3, 48.5$\pm$1.7 for eRASS1, and 51.3$\pm$1.7 for SPT. All results are compatible within 1$\sigma$. In addition, the 
statistical uncertainty and the variation between different cosmological models are about one order of magnitude lower than the uncertainty due to cosmic variance estimated in Sect. \ref{subsec:cosmic_variance}. We conclude that our results are not significantly affected by the choice of a specific cosmological model.

\section{X-GAP measurements}

Table \ref{tab:xgap_meas} collects the individual values of gas mass within 400 kpc, core-excised X-ray luminosity and temperature measured for each individual X-GAP group. These three observables have been measured for this work and have not been published before. We refer the reader to \cite{Eckert2025A&A_4436} for full details about the X-ray analysis pipeline, and \cite{Eckert2024_xgap} for the X-GAP main master table. 

\begin{table*}[]
    \centering
    \caption{Measurements of gas mass within 400 kpc, core-excised temperature, and core-excised X-ray luminosity for X-GAP groups. See \cite{Eckert2024_xgap} for full properties.}
    
    \begin{tabular}{|c|c|c|c|c|}
    \hline
    \hline
       \textbf{Group ID} & \textbf{redshift} & \textbf{$\log_{\rm 10}$M$_{\rm gas}$(<400 kpc) [M$_\odot$]} & \textbf{$\log_{\rm 10}$L$_{\rm X, CEX}$ [erg/s]} & \textbf{T$_{\rm X, CEX}$ [keV]} \\
       \hline
       \rule{0pt}{2ex}    
        8050 & 0.047 & 12.38$_{-0.01}^{+0.01}$ & 42.85$_{-0.01}^{+0.01}$ & 1.76$_{-0.06}^{+0.04}$ \\
10842 & 0.040 & 12.33$_{-0.01}^{+0.01}$ & 42.74$_{-0.01}^{+0.01}$ & 1.62$_{-0.03}^{+0.03}$ \\
885 & 0.047 & 12.55$_{-0.01}^{+0.01}$ & 43.07$_{-0.01}^{+0.01}$ & 2.50$_{-0.08}^{+0.08}$ \\
1011 & 0.046 & 12.36$_{-0.01}^{+0.01}$ & 42.76$_{-0.01}^{+0.01}$ & 1.56$_{-0.03}^{+0.03}$ \\
9695 & 0.038 & 12.38$_{-0.00}^{+0.01}$ & 42.78$_{-0.00}^{+0.00}$ & 1.74$_{-0.02}^{+0.02}$ \\
4654 & 0.022 & 11.84$_{-0.02}^{+0.02}$ & 41.64$_{-0.02}^{+0.01}$ & 0.86$_{-0.02}^{+0.01}$ \\
1695 & 0.039 & 12.53$_{-0.02}^{+0.02}$ & 43.03$_{-0.03}^{+0.02}$ & 1.97$_{-0.13}^{+0.15}$ \\
15641 & 0.027 & 12.10$_{-0.00}^{+0.00}$ & 42.21$_{-0.01}^{+0.01}$ & 0.95$_{-0.01}^{+0.01}$ \\
3128 & 0.032 & 12.03$_{-0.01}^{+0.01}$ & 42.25$_{-0.01}^{+0.01}$ & 1.10$_{-0.01}^{+0.02}$ \\
2620 & 0.039 & 12.42$_{-0.01}^{+0.01}$ & 42.89$_{-0.01}^{+0.01}$ & 1.57$_{-0.02}^{+0.02}$ \\
12349 & 0.036 & 12.25$_{-0.01}^{+0.01}$ & 42.53$_{-0.01}^{+0.00}$ & 1.04$_{-0.01}^{+0.01}$ \\
16150 & 0.032 & 11.79$_{-0.04}^{+0.05}$ & 41.71$_{-0.08}^{+0.06}$ & 1.10$_{-0.02}^{+0.02}$ \\
3460 & 0.043 & 12.18$_{-0.02}^{+0.02}$ & 42.35$_{-0.02}^{+0.02}$ & 1.25$_{-0.03}^{+0.02}$ \\
9771 & 0.044 & 12.17$_{-0.02}^{+0.02}$ & 42.28$_{-0.03}^{+0.02}$ & 1.38$_{-0.07}^{+0.08}$ \\
1601 & 0.034 & 12.08$_{-0.01}^{+0.01}$ & 42.20$_{-0.02}^{+0.01}$ & 1.07$_{-0.01}^{+0.01}$ \\
10094 & 0.031 & 11.81$_{-0.02}^{+0.02}$ & 41.86$_{-0.02}^{+0.02}$ & 1.08$_{-0.03}^{+0.03}$ \\
6159 & 0.024 & 11.79$_{-0.05}^{+0.05}$ & 41.42$_{-0.04}^{+0.04}$ & 0.72$_{-0.03}^{+0.04}$ \\
8102 & 0.033 & 12.13$_{-0.01}^{+0.01}$ & 42.34$_{-0.01}^{+0.01}$ & 1.10$_{-0.02}^{+0.02}$ \\
9178 & 0.040 & 11.89$_{-0.17}^{+0.11}$ & 41.56$_{-0.18}^{+0.10}$ & 0.84$_{-0.07}^{+0.09}$ \\
39344 & 0.028 & 12.01$_{-0.01}^{+0.01}$ & 42.19$_{-0.01}^{+0.01}$ & 0.85$_{-0.01}^{+0.01}$ \\
5742 & 0.034 & 12.06$_{-0.03}^{+0.02}$ & 41.93$_{-0.03}^{+0.02}$ & 0.79$_{-0.02}^{+0.03}$ \\
6058 & 0.045 & 11.87$_{-0.07}^{+0.04}$ & 41.76$_{-0.07}^{+0.04}$ & 0.89$_{-0.04}^{+0.05}$ \\
1162 & 0.044 & 12.27$_{-0.01}^{+0.01}$ & 42.64$_{-0.02}^{+0.01}$ & 1.42$_{-0.02}^{+0.02}$ \\
2424 & 0.040 & 12.15$_{-0.02}^{+0.02}$ & 42.26$_{-0.02}^{+0.02}$ & 1.17$_{-0.02}^{+0.03}$ \\
46701 & 0.042 & 12.13$_{-0.02}^{+0.02}$ & 42.24$_{-0.03}^{+0.02}$ & 1.05$_{-0.05}^{+0.05}$ \\
35976 & 0.036 & 12.29$_{-0.02}^{+0.02}$ & 42.64$_{-0.02}^{+0.02}$ & 1.28$_{-0.04}^{+0.05}$ \\
9399 & 0.035 & 12.18$_{-0.02}^{+0.02}$ & 42.29$_{-0.02}^{+0.02}$ & 1.10$_{-0.03}^{+0.03}$ \\
4436 & 0.046 & 12.29$_{-0.01}^{+0.01}$ & 42.59$_{-0.01}^{+0.02}$ & 1.58$_{-0.07}^{+0.06}$ \\
9647 & 0.023 & 12.06$_{-0.02}^{+0.02}$ & 41.89$_{-0.02}^{+0.02}$ & 0.80$_{-0.02}^{+0.02}$ \\
4936 & 0.042 & 11.77$_{-0.14}^{+0.12}$ & 41.42$_{-0.10}^{+0.11}$ & 0.86$_{-0.06}^{+0.08}$ \\
828 & 0.046 & 12.62$_{-0.01}^{+0.01}$ & 43.28$_{-0.01}^{+0.01}$ & 2.47$_{-0.06}^{+0.06}$ \\
15776 & 0.036 & 12.14$_{-0.04}^{+0.04}$ & 42.29$_{-0.04}^{+0.03}$ & 1.01$_{-0.05}^{+0.07}$ \\
9370 & 0.038 & 12.12$_{-0.04}^{+0.03}$ & 42.27$_{-0.03}^{+0.03}$ & 1.09$_{-0.04}^{+0.05}$ \\
1398 & 0.046 & 12.56$_{-0.01}^{+0.01}$ & 43.25$_{-0.01}^{+0.01}$ & 2.58$_{-0.07}^{+0.07}$ \\
40241 & 0.049 & 12.15$_{-0.02}^{+0.02}$ & 42.34$_{-0.02}^{+0.01}$ & 0.89$_{-0.01}^{+0.01}$ \\
11844 & 0.038 & 12.12$_{-0.06}^{+0.05}$ & 42.15$_{-0.04}^{+0.04}$ & 0.69$_{-0.04}^{+0.04}$ \\
10159 & 0.031 & 11.73$_{-0.11}^{+0.07}$ & 41.36$_{-0.09}^{+0.06}$ & 0.59$_{-0.04}^{+0.07}$ \\
11320 & 0.045 & 12.39$_{-0.01}^{+0.01}$ & 42.93$_{-0.02}^{+0.02}$ & 2.46$_{-0.23}^{+0.50}$ \\
11631 & 0.046 & 12.42$_{-0.02}^{+0.02}$ & 42.95$_{-0.01}^{+0.01}$ & 1.57$_{-0.09}^{+0.12}$ \\
16393 & 0.046 & 12.09$_{-0.02}^{+0.02}$ & 42.30$_{-0.03}^{+0.02}$ & 1.24$_{-0.04}^{+0.07}$ \\
22635 & 0.034 & 12.21$_{-0.03}^{+0.02}$ & 42.48$_{-0.03}^{+0.02}$ & 1.17$_{-0.06}^{+0.11}$ \\
28674 & 0.037 & 12.14$_{-0.01}^{+0.01}$ & 42.36$_{-0.02}^{+0.01}$ & 1.07$_{-0.05}^{+0.06}$ \\
3513 & 0.036 & 11.90$_{-0.09}^{+0.07}$ & 41.84$_{-0.07}^{+0.07}$ & 1.25$_{-0.17}^{+0.28}$ \\
3669 & 0.048 & 12.49$_{-0.01}^{+0.01}$ & 43.11$_{-0.02}^{+0.02}$ & 1.85$_{-0.16}^{+0.23}$ \\
    \hline
    \end{tabular}
    \label{tab:xgap_meas}
\end{table*}

\end{document}